\documentclass[fleqn,twoside]{article}
\usepackage{epsfig}
\usepackage{xspace}

\usepackage{graphicx}
\usepackage[figuresright]{rotating}

\newcommand{\GeV}{\ensuremath{\mbox{GeV}}\xspace}

\newcommand{\GeVc}{\ensuremath{\mbox{GeV}/c}\xspace}
\newcommand{\MeVc}{\ensuremath{\mbox{MeV}/c}\xspace}

\newcommand{\cm}{\ensuremath{\mbox{cm}}\xspace}
\newcommand{\mm}{\ensuremath{\mbox{mm}}\xspace}
\newcommand{\micron}{\ensuremath{\mu \mbox{m}}\xspace}
\newcommand{\mrad}{\ensuremath{\mbox{mrad}}\xspace}
\newcommand{\ns}{\ensuremath{\mbox{ns}}\xspace}
\newcommand{\m}{\ensuremath{\mbox{m}}\xspace}

\newcommand{\ps}{\ensuremath{\mbox{ps}}\xspace}

\newcommand{\bfGeVc}{\ensuremath{\mathbf {\mbox{\bf GeV}/c}}\xspace}

\begin{document}

\title{\bf Measurement of the production cross-section of positive
  pions in p--Al collisions at 12.9~\bfGeVc}

\author{HARP Collaboration}

\maketitle

\begin{abstract}
A precision measurement 
of the double-differential production cross-section,
${{d^2 \sigma^{\pi^+} }}/{{d p d\Omega }}$,
for pions of positive charge, performed in the HARP experiment is
presented.
The incident particles are protons of 
12.9~\GeVc momentum impinging on an aluminium target of 5\% nuclear
interaction length.
The measurement of this cross-section has a direct application
to the calculation of the neutrino flux of the K2K experiment.
After cuts, 210~000 secondary tracks reconstructed in the forward
spectrometer were used in this analysis.
The results are given for secondaries within a momentum range from
0.75~\GeVc to 6.5~\GeVc, and within an angular range from 30~\mrad to
210~\mrad. 
The absolute normalization was performed using 
prescaled beam triggers counting protons on target.
The overall scale of the cross-section 
is known to better than 6\%,
while the average point-to-point error is 8.2\%.

\vspace{1pc}
\end{abstract}

\vspace{6cm}
\begin{center}
Accepted for publication in Nucl. Phys. B
\end{center}
\newpage
\begin{center}
{\large HARP collaboration}\\
\newcommand{\afdoct}{{3}\xspace}
\vspace{0.1cm}
{\small
M.G.~Catanesi, 
M.T.~Muciaccia, 
E.~Radicioni,
S.~Simone
\\ 
{\bf Universit\`{a} degli Studi e Sezione INFN, Bari, Italy} 
\\
R.~Edgecock, 
M.~Ellis$^1$, 
S.~Robbins$^{2,3}$, 
F.J.P.~Soler$^{4}$
\\
{\bf Rutherford Appleton Laboratory, Chilton, Didcot, UK} 
\\
C.~G\"{o}\ss ling,
M.~Mass  
\\
{\bf Institut f\"{u}r Physik, Universit\"{a}t Dortmund, Germany} 
\\
S.~Bunyatov, 
A.~Chukanov,  
D.~Dedovitch,
A.~Elagin,
M.~Gostkin, 
A.~Guskov, 
D.~Khartchenko, 
O.~Klimov, 
A.~Krasnoperov, 
D.~Kustov, 
K.~Nikolaev,
B.~Popov$^5$, 
V.~Serdiouk, 
V.~Tereshchenko, 
A.~Zhemchugov
\\
{\bf Joint Institute for Nuclear Research, JINR Dubna, Russia} 
\\
E.~Di~Capua, 
G.~Vidal--Sitjes$^{6,1}$  
\\
{\bf Universit\`{a} degli Studi e Sezione INFN, Ferrara, Italy}  
\\
A.~Artamonov$^7$, 
P.~Arce$^8$, 
S.~Giani, 
S.~Gilardoni$^{6}$, 
P.~Gorbunov$^{7}$,
A.~Grant,  
A.~Grossheim$^{6}$, 
P.~Gruber$^{6}$, 
V.~Ivanchenko$^9$, 
A.~Kayis-Topaksu$^{10}$,
L.~Linssen, 
J.~Panman, 
I.~Papadopoulos,  
J.~Pasternak$^{6}$, 
E.~Tcherniaev, 
I.~Tsukerman$^7$, 
R.~Veenhof, 
C.~Wiebusch$^3$, 
P.~Zucchelli$^{11,12}$ 
\\
{\bf CERN, Geneva, Switzerland} 
\\
A.~Blondel, 
S.~Borghi, 
M.~Campanelli, 
A.~Cervera--Villanueva, 
M.C.~Morone, 
G.~Prior$^{6,13}$, 
R.~Schroeter
\\
{\bf Section de Physique, Universit\'{e} de Gen\`{e}ve, Switzerland} 
\\
\newcommand{\afkek}{{14}\xspace}
\newcommand{\afkyot}{{14}\xspace}
I.~Kato$^{\afkyot}$,
T.~Nakaya$^{\afkyot}$,
K.~Nishikawa$^{\afkyot}$,
S.~Ueda$^{\afkyot}$
\\
{\bf University of Kyoto, Japan} %
\\
V.~Ableev, 
U.~Gastaldi
\\
{\bf Laboratori Nazionali di Legnaro dell' INFN, Legnaro, Italy} 
\\
\newcommand{\aflanl}{{15}\xspace}
G.~B.~Mills$^{\aflanl}$  
\\
{\bf Los Alamos National Laboratory, Los Alamos, USA} %
\\
J.S.~Graulich$^{16}$,  
G.~Gr\'{e}goire 
\\
{\bf Institut de Physique Nucl\'{e}aire, UCL, Louvain-la-Neuve,
  Belgium} 
\\
M.~Bonesini, 
M.~Calvi,   
A.~De~Min, 
F.~Ferri, 
M.~Paganoni, 
F.~Paleari
\\
{\bf Universit\`{a} degli Studi e Sezione INFN, Milano, Italy} 
\\
M.~Kirsanov
\\
{\bf Institute for Nuclear Research, Moscow, Russia} 
\\
A. Bagulya, 
V.~Grichine,  
N.~Polukhina
\\
{\bf P. N. Lebedev Institute of Physics (FIAN), Russian Academy of
Sciences, Moscow, Russia} 
\\
V.~Palladino
\\
{\bf Universit\`{a} ``Federico II'' e Sezione INFN, Napoli, Italy} 
\\
\newcommand{\afclmb}{{15}\xspace}
L.~Coney$^{\afclmb}$, 
D.~Schmitz$^{\afclmb}$
\\
{\bf Columbia University, New York, USA} %
\\
G.~Barr, 
A.~De~Santo$^{17}$, 
C.~Pattison, 
K.~Zuber$^{18}$  
\\
{\bf Nuclear and Astrophysics Laboratory, University of Oxford, UK} 
\\
F.~Bobisut, 
D.~Gibin,
A.~Guglielmi, 
M.~Laveder, 
A.~Menegolli, 
M.~Mezzetto
\\
{\bf Universit\`{a} degli Studi e Sezione INFN, Padova, Italy} 
\\
J.~Dumarchez, 
S.~Troquereau, 
F.~Vannucci 
\\
{\bf LPNHE, Universit\'{e}s de Paris VI et VII, Paris, France} 
\\
V.~Ammosov, 
V.~Gapienko,  
V.~Koreshev, 
A.~Semak, 
Yu.~Sviridov,  
V.~Zaets
\\
{\bf Institute for High Energy Physics, Protvino, Russia} 
\\
U.~Dore
\\
{\bf Universit\`{a} ``La Sapienza'' e Sezione INFN Roma I, Roma,
  Italy} 
\\
D.~Orestano, 
M.~Pasquali,  
F.~Pastore, 
A.~Tonazzo, 
L.~Tortora
\\
{\bf Universit\`{a} degli Studi e Sezione INFN Roma III, Roma, Italy}
\\
C.~Booth, 
C.~Buttar$^{4}$,  
P.~Hodgson, 
L.~Howlett
\\
{\bf Dept. of Physics, University of Sheffield, UK} 
\\
M.~Bogomilov, 
M.~Chizhov, 
D.~Kolev, 
R.~Tsenov
\\
{\bf Faculty of Physics, St. Kliment Ohridski University, Sofia,
  Bulgaria} 
\\
S.~Piperov, 
P.~Temnikov
\\
{\bf Institute for Nuclear Research and Nuclear Energy, 
Academy of Sciences, Sofia, Bulgaria} 
\\
M.~Apollonio, 
P.~Chimenti,  
G.~Giannini, 
G.~Santin$^{19}$  
\\
{\bf Universit\`{a} degli Studi e Sezione INFN, Trieste, Italy} 
\\
Y.~Hayato$^{\afkek}$, 
A.~Ichikawa$^{\afkek}$, 
T.~Kobayashi$^{\afkek}$
\\
{\bf KEK, Tsukuba, Japan} %
\\
J.~Burguet--Castell, 
J.J.~G\'{o}mez--Cadenas, 
P.~Novella,
M.~Sorel,
A.~Tornero
\\
{\bf  Instituto de F\'{i}sica Corpuscular, IFIC, CSIC and Universidad de Valencia,
Spain} 
}
\end{center}
\vfill
\rule{0.3\textwidth}{0.4mm}
\newline
$^{~1}${Now at Imperial College, University of London, UK.}
\newline
$^{~2}${Now at Bergische Universit\"{a}t Wuppertal, Germany.}
\newline
$^{~3}$Jointly appointed by Nuclear and Astrophysics Laboratory,
            University of Oxford, UK.
\newline
$^{~4}${Now at University of Glasgow, UK.}
\newline
$^{~5}${Also supported by LPNHE (Paris).}
\newline
$^{~6}${Supported by the CERN Doctoral Student Programme.}
\newline
$^{~7}${Permanently at ITEP, Moscow, Russian Federation.}
\newline
$^{~8}${Permanently at Instituto de F\'{\i}sica de Cantabria,
            Univ. de Cantabria, Santander, Spain.} 
\newline
$^{~9}$On leave of absence from the Budker Institute for Nuclear Physics,
Novosibirsk, Russia.
\newline
$^{10}${On leave of absence from \c{C}ukurova University, Adana, Turkey.}
\newline
$^{11}$On leave of absence from INFN, Sezione di Ferrara, Italy.
\newline
$^{12}${Now at SpinX Technologies, Geneva, Switzerland.}
\newline
$^{13}${Now at Lawrence Berkeley National Laboratory, Berkeley, USA.}
\newline
$^{14}${K2K Collaboration.}
\newline
$^{15}${MiniBooNE Collaboration.}
\newline
$^{16}${Now at Section de Physique, Universit\'{e} de Gen\`{e}ve, Switzerland, Switzerland.}
\newline
$^{17}${Now at Royal Holloway, University of London, UK.}
\newline
$^{18}${Now at University of Sussex, Brighton, UK.}
\newline
$^{19}${Now at ESA/ESTEC, Noordwijk, The Netherlands.}

\newpage

\section{Introduction}
\label{sec:intro}

The objective of the HARP experiment is a systematic study of hadron
production for beam momenta from 1.5~\GeVc to 15~\GeVc for a large range of target
nuclei~\cite{harp-prop}. 
The main motivations are: a) to measure pion yields for a quantitative
design of the proton driver of a future neutrino factory, b) to improve
substantially the calculation of the atmospheric neutrino flux and
c) to provide input for the flux calculation of accelerator neutrino
experiments, such as K2K and MiniBooNE.

The measurement described in 
this paper is of particular relevance in the context of the recent results
presented by the K2K experiment~\cite{k2k, k2k-last}, which have shown
evidence for neutrino oscillations at a confidence level
of four standard deviations. The K2K experiment uses an accelerator-produced
$\nu_\mu$ beam with an average energy of 1.3~\GeV directed at 
the Super-Kamiokande detector. The K2K analysis compares the observed
$\nu_\mu$ spectrum in Super-Kamiokande, located at a distance of about
250~km from the neutrino source, with the predicted
spectrum in the absence of oscillations. This, in turn,
is computed by multiplying the observed spectrum at the near detector 
(located at 300~\m from the neutrino source) by
the so-called `far--near ratio', $R$, defined as the ratio between the
predicted flux at the far and near detectors. This factor corrects for the
fact that at the near detector, the neutrino source is not point-like, but
sensitive to effects such as the finite size of the decay tunnel,
etc., whereas
at the Super-Kamiokande site the neutrino source can be considered as point-like.
According to the neutrino oscillation 
parameters measured in atmospheric neutrino
experiments~\cite{Ashie:2005ik} the distortion of the spectrum 
measured with the far detector is predicted to be maximal 
in the energy range between 0.5 and 1~\GeV.
The determination of $R$ is the leading
energy-dependent systematic error in the K2K
analysis~\cite{k2k,k2k-last}. 

The HARP experiment has a large acceptance in the momentum and angular
range relevant for K2K neutrino flux. It covers 80\% of the
total neutrino flux in the near detector and in the relevant region
for neutrino oscillations. Thus, it can provide an independent, and
more precise, measurement of the pion yield needed as input to the
calculation of the K2K far--near ratio than that currently available. 

The neutrino beam of the K2K experiment originates from
the decay of light hadrons, produced by exposing an aluminium target 
to a proton beam of momentum 12.9~\GeVc.  
In this paper, the measurement of
the double-differential cross-section, 
$
{{d^2 \sigma^{\pi^+ }}}/{{dpd\Omega }}
$
of positive pion production for 
protons of 12.9~\GeVc momentum impinging on a thin Al target 
of 5\% nuclear interaction length ($\lambda_{\mathrm{I}}$) is presented,
{\em i.e.} reproducing closely the conditions of the K2K beam-line for
the production of secondaries.

%
%

The HARP apparatus~\cite{harp-prop,harpTech}
is a large-acceptance spectrometer consisting of a
forward and large-angle detection system. 
The forward spectrometer covers polar angles up to 250~\mrad which
is well matched to 
the angular range of interest for the K2K beam line.

The results reported here are based on data taken in 2002 in
the T9 beam of the CERN PS.
About 3.4 million incoming protons were selected.
After cuts, 
209~929
secondary tracks reconstructed in the forward
spectrometer were used in this analysis.
The results are given in the region relevant for K2K, that is the
momentum range from 
0.75~\GeVc to 6.5~\GeVc and within an angular range from 30~\mrad to
210~\mrad. 
The absolute normalization was performed using 
280~542 
`minimum-bias' triggers.

This paper is organized as follows. The experimental apparatus is
outlined in Section~\ref{sec:apparatus}.
Section~\ref{sec:forward-tracking} describes
tracking with the forward spectrometer. 
Section~\ref{sec:tracking-eff} discusses the 
calculation of the
reconstruction efficiency. 
Section~\ref{sec:pid} summarizes the particle identification
(PID) capabilities of the spectrometer and describes the PID algorithm.
Sections~\ref{sec:xsec} and \ref{sec:atlantic} give details of the cross-section calculation.  
Results are discussed in Section~\ref{sec:results}. 
A comparison with previous data is presented in Section~\ref{sec:compare}.
An illustrative calculation of the K2K far--near ratio is shown in 
Section~\ref{sec:far-near}.
A summary is given in Section~\ref{sec:summary}.

\section{Experimental apparatus}
\label{sec:apparatus}
\begin{figure}[tbp]
\begin{center}
\hspace{0mm} \epsfig{file=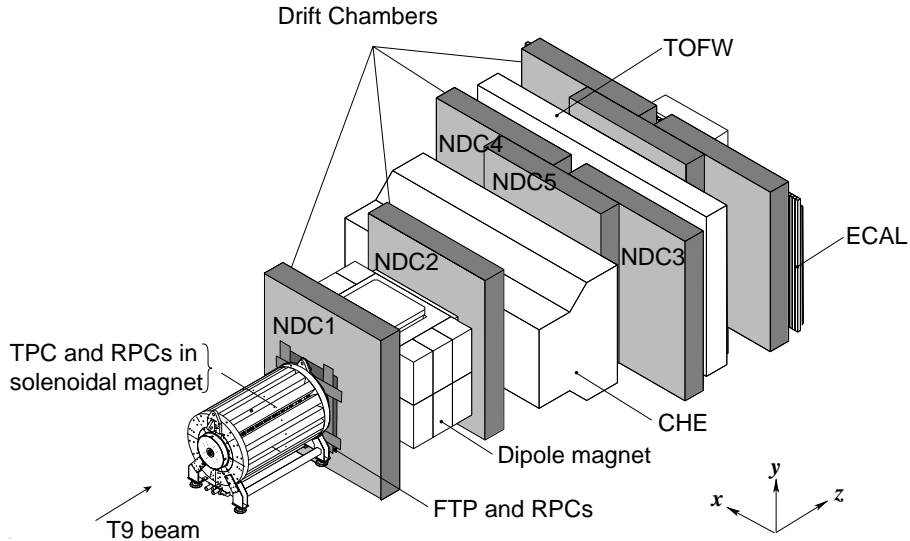,width=12cm}
\end{center}
\caption{Schematic layout of the HARP spectrometer. 
The convention for the coordinate system is shown in the lower-right
corner. 
The three most downstream (unlabelled) drift chambers are only partly
equipped with electronics and not used for tracking.
}
\label{fig:harp}
\end{figure}

The HARP detector, shown in Fig.~\ref{fig:harp}, consists of forward
and large-angle detection systems. 
The convention used for the coordinate system is also given in the
Figure. 
In the large-angle region a TPC
positioned in a solenoidal magnet is used for tracking. The forward
spectrometer is built around a dipole magnet 
with an integral field of $\int{B_{y}dL}$=0.66~T\,m
for momentum analysis, 
with large planar drift chambers (NDC) for particle tracking, and three
detectors used for particle identification: 
a time-of-flight wall (TOFW),
a threshold Cherenkov detector (CHE), 
and an electromagnetic calorimeter (ECAL).
The target itself is located inside the TPC. Beam instrumentation, including
three timing detectors (BTOF) and two threshold Cherenkov detectors
(BCA and BCB), provides
identification of the incoming particle and the determination of the
interaction time at the target.  The impact point of the beam particle
on the target and its direction are measured by a set of multi-wire
proportional chambers (MWPCs).  Several trigger detectors are
available to select events with an interaction and to define the
normalization.
 
Data were taken with several beam momenta and target configurations.
In addition to the data taken with the thin aluminium target of
5\%~$\lambda_{\mathrm{I}}$ at an incident proton momentum of 12.9~\GeVc,
runs were also taken with an empty target holder.
These data allow a subtraction to be made of the interactions
occurring in the material on the path of the incident beam.
Other relevant configurations for the measurement described here are
the data taken with and without target with other beam momenta (1.5,
3.0, 5.0, 8.0, 8.9 and 15~\GeVc) with electrons, pions and protons.  
These settings have been used to determine the response of the 
spectrometer to these particles in terms of efficiency, momentum
resolution and particle identification capability.
Data with thick Al targets, such as a replica of the K2K target, have
also been taken, but are not yet used in the present analysis.
The momentum definition of the T9 beam is known with a precision of
the order of 1\%~\cite{ref:t9}. 

A detailed description of the HARP experiment is
given in Ref.~\cite{harpTech}. 
In this analysis we utilize primarily the detector components
of the forward spectrometer and the beam instrumentation.
Below, the elements which are important for this analysis will be mentioned.

\subsection{Beam and trigger detectors}
\label{sec:beamtrigger}

\begin{figure}[bt]
\begin{center}
\epsfig{file=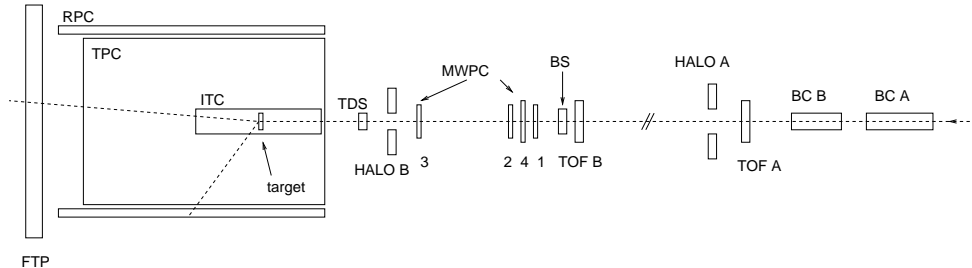,width=0.8\textwidth}
\caption{ Schematic view of the trigger and beam
equipment. The description is given  in the text. The beam
enters from the right. The MWPCs are numbered: 1, 4, 2, 3 from right to
left. On the left, the position of the target inside the inner field cage of
the TPC is shown.}
\label{fig:beamline}
\end{center}
\end{figure}

\begin{figure}[bt]
\begin{center}
\epsfig{file=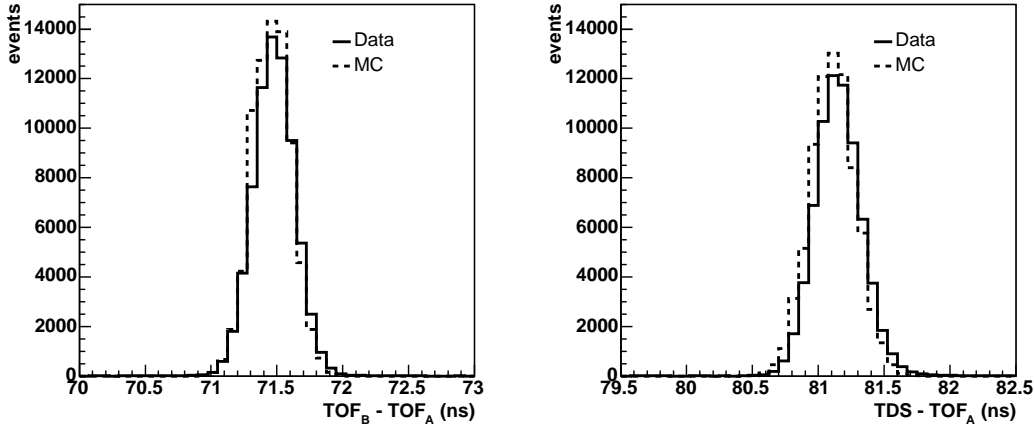,width=0.9\textwidth}
\caption{The timing resolution of the beam TOF detectors. The left
  hand panel shows the time difference measured between TOFA and TOFB,
  the right hand panel the time difference between TDS and TOFA.  
}
\label{fig:btof}
\end{center}
\end{figure}

\begin{figure}[bt]
\begin{center}
\epsfig{file=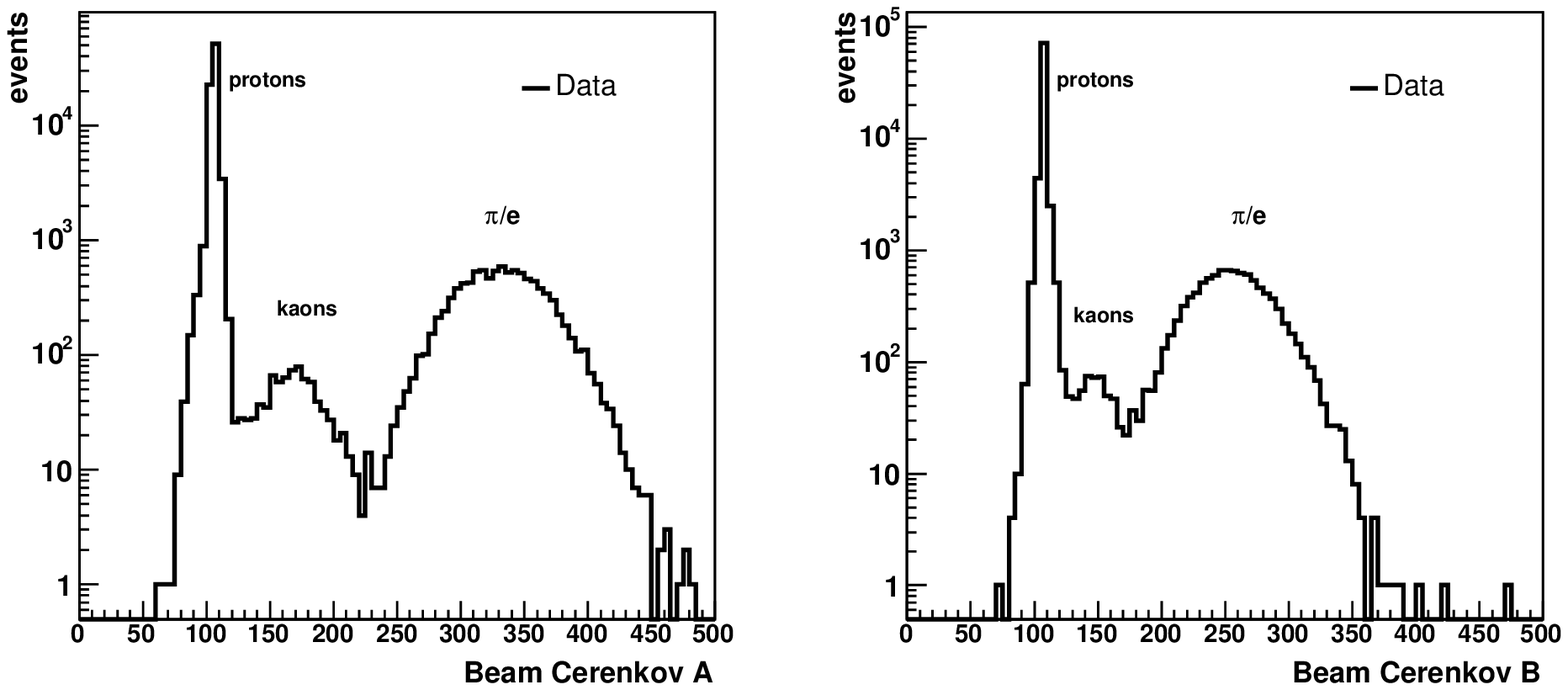,width=0.9\textwidth}
\caption{Pulse height spectra expressed in ADC
counts from beam Cherenkov counters BCA (left) and BCB (right) at 12.9~\GeVc. 
The pedestal (marked `protons' representing particles with a velocity
below threshold) is around channel 110 in both cases. } 
\label{fig:cheb}
\end{center}
\end{figure}

A schematic picture of the equipment in the beam line is shown in
Fig.~\ref{fig:beamline}. It is instrumented with the following systems:

\begin{itemize} 
\item A set of four multi-wire
proportional chambers (MWPCs) measures the position and direction of
the incoming beam particles,
with an accuracy of $\approx$1~\mm in position and $\approx$0.2~\mrad
in angle per projection.
\item A beam time-of-flight system (BTOF)
measures time difference over a distance of $21.4$~m. It is made of two
identical scintillation hodoscopes, TOFA and TOFB (originally built
for the NA52 experiment~\cite{ref:NA52}),
which, together with a small target-defining trigger counter (TDS,
also used for the trigger and described below), provide particle
identification at low energies. This allows separation of pions, kaons
and protons up to 5~\GeVc and provides the initial time at the
interaction vertex ($t_0$). The resolution is shown in
Fig.~\ref{fig:btof}.  The $t_0$, {\it i.e.} the time at which the
incident beam particle is predicted to cross the mid-plane of the
target ($z=0$), is calculated after particle identification.  The
weighted average of the individual measurements of $t_0$ from the
three timing detectors is calculated, taking into account the
velocity, $\beta$, of the particle using the known beam momentum and
the particle mass deduced after identification.  The timing resolution
of the combined BTOF system is about 70~\ps.
\item A system of two N$_2$-filled Cherenkov detectors (BCA and BCB) is
used to tag electrons at low energies
and to tag pions at higher energies. 
The electron and pion tagging efficiency is found to be close to
100\%.
At momenta larger than 12~\GeVc it is also possible to tag kaons as can 
be seen in Fig.~\ref{fig:cheb} which shows the pulse height spectrum 
of BCA and BCB for a 12.9~\GeVc beam. 
This spectrum displays raw channel counts without pedestal subtraction
(the pedestal is around channel 110).
The kaon and pion peaks can be clearly distinguished from the
pedestal
peak at low pulse-height which is due to heavier
particles below Cherenkov threshold such as protons.
The electrons are part of the pion peak.

\end{itemize} 

The target is positioned inside the inner field cage of the TPC.
It has a cylindrical shape with a nominal diameter of 30~\mm.
The aluminium (99.999\% pure) target used for the 
measurement described here has a
nominal thickness of 5\%~$\lambda_{\mathrm{I}}$.
Precise measurements of the thickness have been performed at different
locations on its surface and show a maximum variation between
19.73~\mm and 19.85~\mm.  

A set of trigger detectors completes the beam instrumentation: a
thin scintillator slab covering the full aperture of the last
quadrupole magnet in the beam line to start the trigger logic
decision (BS); a small scintillator disk, TDS mentioned above, positioned
upstream of the target to ensure that only 
particles hitting the target cause a trigger; and `halo' counters
(scintillators with a hole to let the beam particles pass) to veto
particles too far away from the beam axis. 

The TDS is designed to have a very high efficiency (measured to be 99.9\%).
It is located as near as possible to the entrance of
the TPC and has a 20~mm diameter,
smaller than the target which has a 30~mm diameter.  
Its time resolution ($\sim 130 $~\ps) is sufficiently good to be used
as an additional detector for the BTOF system.

A double plane of scintillation counters (FTP),
positioned upstream of the dipole magnet, is 
used to select events with an interaction in the target 
and outgoing charged particles in the forward region. 
The plane covers the full aperture of the dipole
magnet, with the exception of a central hole with a diameter of
60~\mm to let the beam particles pass.  
The efficiency of the FTP was measured using events which had been
taken simultaneously using triggers which did not
require the FTP and amounts to $>$99.8\%.

\subsection{Drift chambers}
\label{sub:ndc} 
The main tracking device of the HARP forward spectrometer is
a set of large drift chambers (NDC) placed upstream and
downstream of the dipole magnet. These chambers were
originally built for the NOMAD experiment~\cite{NOMAD_NIM}, where they served both as a 
target for neutrino interactions and as a tracker for the produced
charged particles. 

The spectrometer contains five NDC modules, each of which is made of four
chambers.  The chambers consist of three wire planes, with one plane ($x$)
of wires oriented vertically; the other 
two ($u$ and $v$) are rotated with respect to
the vertical by $\pm 5$~degrees. 
They have been described
elsewhere~\cite{NOMAD_NIM_DC} and we refer to~\cite{harpTech} for 
a detailed description of their performance under HARP conditions. 
This performance can be summarized in terms of two quantities,
spatial resolution and hit efficiency per plane. After internal
alignment of the individual wires, 
the spatial resolution of the chamber is about 340~\micron. 
The hit efficiency is smaller in HARP than it 
was in NOMAD due to the use of a different, 
non-flammable but less efficient gas mixture. 
The hit efficiency varies between 80\% and 85\% in the central
NDC modules.

\subsection{PID detectors}

Particle identification is performed in the forward spectrometer 
through the combination of several detectors
downstream of the dipole magnet (CHE, TOFW and ECAL). We refer
to~\cite{harpTech} for a detailed description of the three
systems. 

A large scintillator wall (TOFW) covering the full acceptance of the
downstream tracking system is used 
in conjunction with the timing information from the beam detectors 
to measure the time-of-flight of the secondary particles 
for momenta up to 5~\GeVc.
The TOFW measures the time-of-flight of particles emanating
from the target, and this, 
together with the charged track trajectory length, $l$,
determines the velocity, $\beta$, of the particle.

The single scintillator counters are BC408 bars from Bicron, 2.5~\cm
thick and 21~\cm wide.  The counters are grouped into three mechanical
structures (palisades).   In the left and right palisades, 
scintillators are 250~\cm long and are mounted vertically, while
in the central palisade scintillators are 180~\cm long and are mounted
horizontally. 
The counters overlap partially by 2.5~\cm to ensure full coverage. 
The scintillator slabs are viewed by one photo-multiplier tube (Philips
XP2020) at each side.
The TOFW and its performance is described in detail in
Ref.~\cite{ref:tofw}.
With an intrinsic resolution of the individual counters of $160~\ps$,
a time-of-flight resolution better than $180~\ps$ is achieved using
this detector in combination with the BTOF system.

The threshold Cherenkov detector (CHE) separates pions from protons
for momenta above the pion threshold (2.6~\GeVc) and identifies
electrons below the pion threshold. 
The radiator gas (perfluorobutane C$_{4}$F$_{10}$) is chosen for 
its high refractive index, which allows the detector to be operated at
atmospheric pressure.
The particles traverse about 2~m of the radiating medium and generate
photons that are deflected by about 135$^\circ$ upward or downward
by two large cylindrical mirrors 6~m long with a radius of  
curvature of 2.4~m. 
Thirty-eight EMI 9356-KA photo-multipliers were used for
their very low noise and high gain characteristics.  
In order to increase their useful light-collection area to a diameter
of 340~\mm, the photo-multipliers were matched to aluminized Winston
cones. 

Finally, the electromagnetic calorimeter (ECAL) provides electron rejection.
It is segmented longitudinally into two planes.
The two calorimeter planes were assembled from existing calorimeter
modules of the CHORUS experiment~\cite{ref:chorus-cal}.
These planes consist of 62 and 80 modules, covering a total active width of
4.96~m and 6.4~m, respectively. 
Each module is composed of 
scintillating fibres (1~\mm diameter) embedded in extruded lead sheets 
with a volume ratio 1/4.
The ratio of the energy deposition in the two planes is different for
electrons compared to hadrons.  In addition, the comparison of the
momentum of the particle measured by the curvature of its trajectory
and the energy 
deposition in the calorimeter provides another way to identify
electrons.  The ECAL complements the electron rejection of the
Cherenkov above the pion Cherenkov threshold.

\section{Tracking with the forward spectrometer}
\label{sec:forward-tracking}

\subsection{Tracking algorithm}
\label{sub:tracka} 

The track reconstruction algorithm starts by building 
two-dimensional (2D) segments
per NDC module. 
Those are later combined to create 3D track segments (also per module). 
The requirements are the following:

\begin{itemize}
\item {\em Plane (2D) segment}: At least three hits out of four in the 
same projection ($u$, $x$ or $v$) compatible
with being aligned. The drift sign associated to each hit
is decided during the plane segment reconstruction phase.
\item {\em Track (3D) segment}: Two or three plane 
segments of different projections, whose intersection defines a 3D
straight line. In the case where only two plane segments are found, an
additional hit in the remaining projection is required. This hit must
intersect the 3D straight line defined by the other two projections.
\end{itemize}

Consequently, to form a track segment at least seven hits
(from a total of 12 measurement planes) are needed  within the same NDC
module.  
Once track segments are formed in the individual modules
they are combined (downstream of the dipole magnet) to 
obtain longer track segments. Finally, downstream tracks 
are connected
with either the interaction vertex or a 3D track segment in NDC1 
(the NDC module upstream the dipole magnet, see Fig.~\ref{fig:harp})
to measure the momentum. All these tasks are performed by 
a sophisticated fitting, extrapolation and
matching package called RecPack~\cite{recpack}, 
which is based on the well known Kalman Filter technique~\cite{kalman}.

The interaction vertex in this analysis is well defined. The transverse 
coordinates $(x,y)$ are obtained by extrapolating the trajectory of the
incoming beam particle, measured with the MWPCs (with an error
of the order of 1~\mm), and the $z$
coordinate can be taken as that of the nominal plane of the target (which is
19.80~\mm thick). 

Consequently,
the momentum of a track can be determined by imposing the constraint that
it emanates from the vertex, that is, by connecting a 3D segment 
downstream the dipole magnet 
with a 3D point upstream the magnet. 
Tracks of this type are called  `VERTEX2 tracks', and
the estimator of the momentum obtained by connecting a 3D segment
with the vertex 3D point is denoted `$p_2$'. Specifically, this is done
by extrapolating the downstream 3D segment to the nominal plane of the
target, and imposing that the distance between the
transverse coordinates thus obtained
$(x_s,y_s)$  and the $(x,y)$ coordinates defined above
is less than 10~\cm (in practice one builds a $\chi^2$ which also
takes into account the measurement errors). Tracks which extrapolate
to distances larger than 10~\cm are not considered (in fact, the
inefficiency of the $p_2$ algorithm, a few percent comes 
almost exclusively from this source).

Alternatively, one can measure the momentum connecting
a 3D segment downstream 
of the dipole with a 3D segment in the NDC1 module. These
are called `VERTEX4 tracks', and the estimator of the momentum is
denoted `$p_4$'. The way a downstream segment is connected with a
NDC1 3D segment is described with some detail below. In essence, 
one requires a reasonable collinearity in the non-bending plane and
obtains the momentum from the curvature in the bending plane. 

The availability of two independent momentum estimators allows
the tracking efficiency to be measured from the data themselves. This
is possible, since, a) the reconstruction methods providing the estimators
$p_2$ and $p_4$ are independent
b) $p_2$ and $p_4$ have a Gaussian distribution around the true momentum
$p$ (The distribution is expected to be Gaussian in the
  variable $1/p$
  rather than $p$. With the relatively good resolution the difference
  is negligible.) 
This makes it possible to use one of the estimators ($p_2$) to measure
the yields while the other ($p_4$) is used to measure tracking efficiency. The
estimator $p_2$ is preferred to measure yields since it does not involve
the use of the NDC1 module, where tracking efficiency is lower than in
the downstream modules (see the discussion on tracking efficiency 
in Section~\ref{sec:tracking-eff}). 

\begin{figure}[tb]
\begin{center}
\hspace{0mm} \epsfig{file=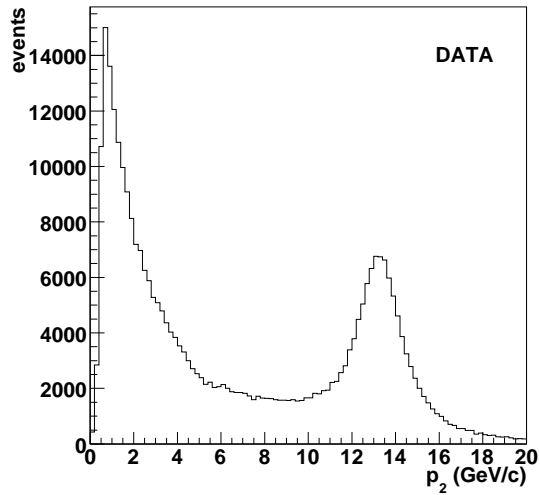,width=8cm}
\end{center}
\caption{Inclusive $p_2$ momentum distribution of reconstructed tracks 
(see text for definition). 
The peak from non-interacting beam particles is visible at 12.9~\GeVc.
}
\label{fig:p2}
\end{figure}

\begin{figure}[tbp]
\begin{center}
\epsfig{file=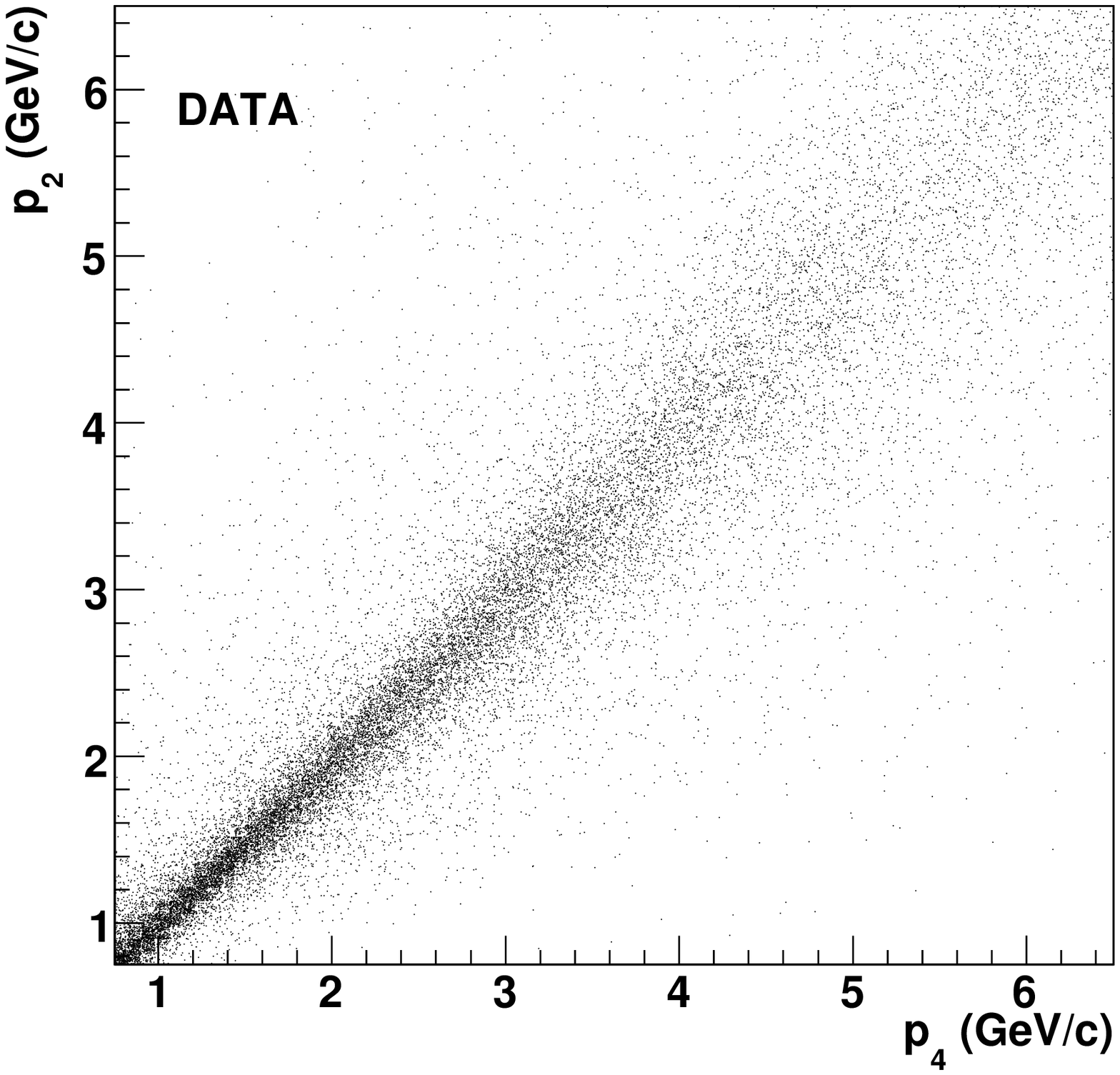,width=0.45\textwidth}
\epsfig{file=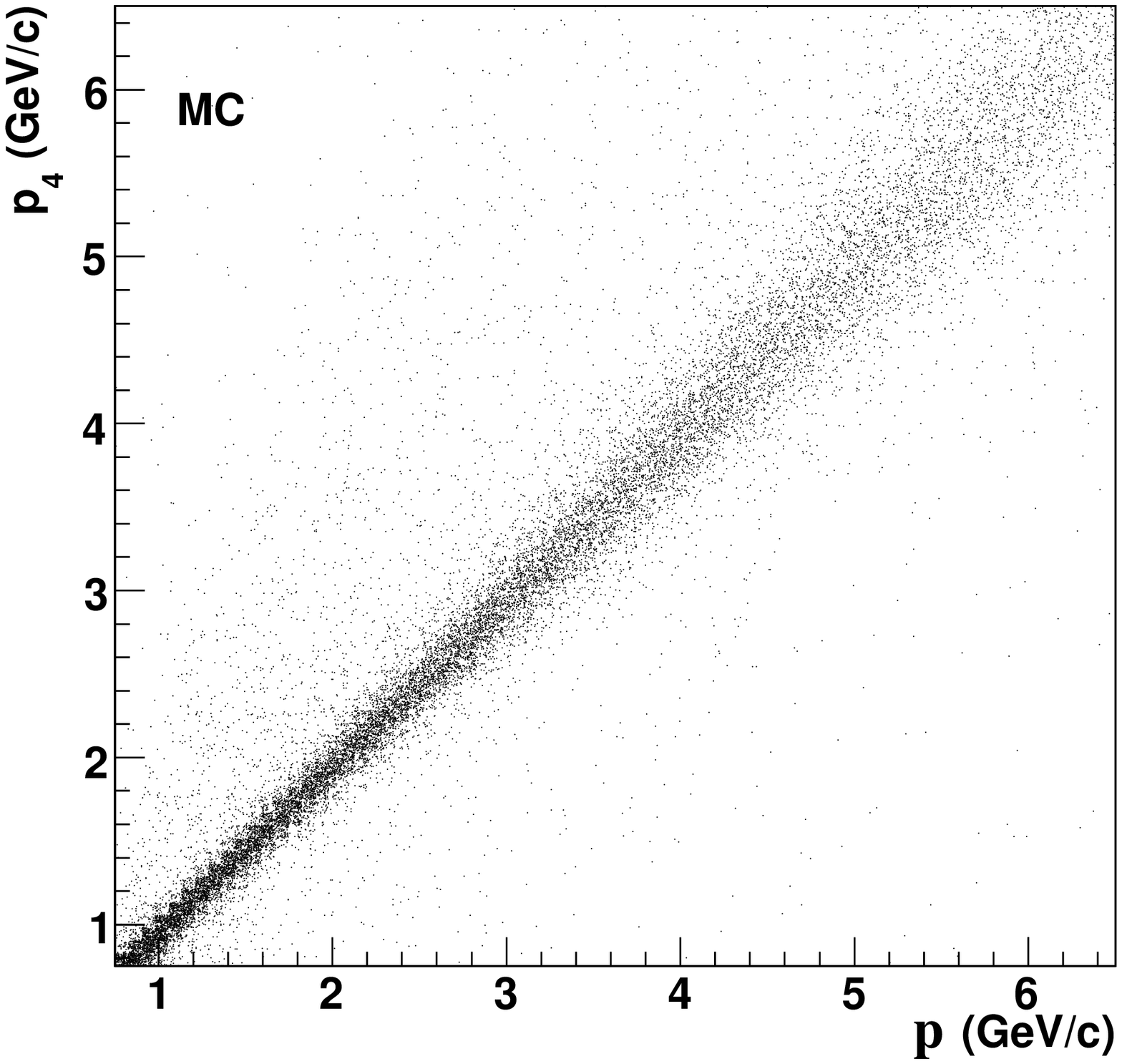,width=0.45\textwidth}
\end{center}
\caption{Left panel: The correlation between $p_2$ and $p_4$, showing that
both are estimators of the same quantity; right panel: the correlation
between $p_4$ and $p$ shows 
that $p_4$ is an unbiased estimator of the momentum $p$
within the momentum resolution and binning (from 500~\MeVc up to 1.5~\GeVc) 
used in the analysis.
}

\label{fig:p2_vs_p4}
\end{figure}

Figure~\ref{fig:p2}
shows the inclusive $p_2$ distribution measured for events
with an FTP trigger. 
The hole in the FTP largely suppresses the peak of beam protons at
12.9 \GeVc.
The remaining peak  corresponds
to events with an FTP trigger caused by elastically scattered protons,
protons with multiple scattering in the tail of the angular
distribution, and by protons accompanied by soft particles produced
upstream of the FTP.   
The linear correlation between $p_2$ and
$p_4$, shown in Fig.~\ref{fig:p2_vs_p4} (left panel) for simulated tracks, 
illustrates the fact that
both are estimators of the same quantity, while the correlation
between $p_4$ and $p$ (Fig.~\ref{fig:p2_vs_p4}, right panel) 
shows that both are unbiased estimators of $p$. 
The small non-linearities and disagreements between $p_2$ and $p_4$
and between $p_4$ and $p$ have negligible contribution
to the total systematic uncertainty.
 
\subsection{Momentum and angular resolution}
\label{subsec:resolution}
 
Following the previous discussion, the momentum measurement 
for VERTEX2 tracks is performed by extrapolating tracks
built downstream of the magnet to the vertex plane. The algorithm
performs a loop over allowed momenta. 
For each value of $p$ one
computes the extrapolated position $(x_{\mathrm{t}}, y_{\mathrm{t}})$
at the target reference $z_0$ coordinate and the matching
$\chi^2$ with the event $(x_0, y_0)$ coordinates. The momentum is
then calculated by minimizing this $\chi^2$. 
For VERTEX4 the algorithm is
similar, but in this case, a 3D segment downstream of the magnet is
matched to a 3D segment in NDC1 module. 
In both cases, an upper cut on the 
minimum matching $\chi^2$ decides whether the matching is accepted or not. 
This reduces the background from tertiary particles (not coming from the primary vertex) 
in the case of matching with the vertex (VERTEX2 tracks). For VERTEX4, 
this cut reduces the background from particles interacting in the 
region between NDC1 and NDC2. 

\begin{figure}[tbp]
  \begin{center}
  \epsfig{file=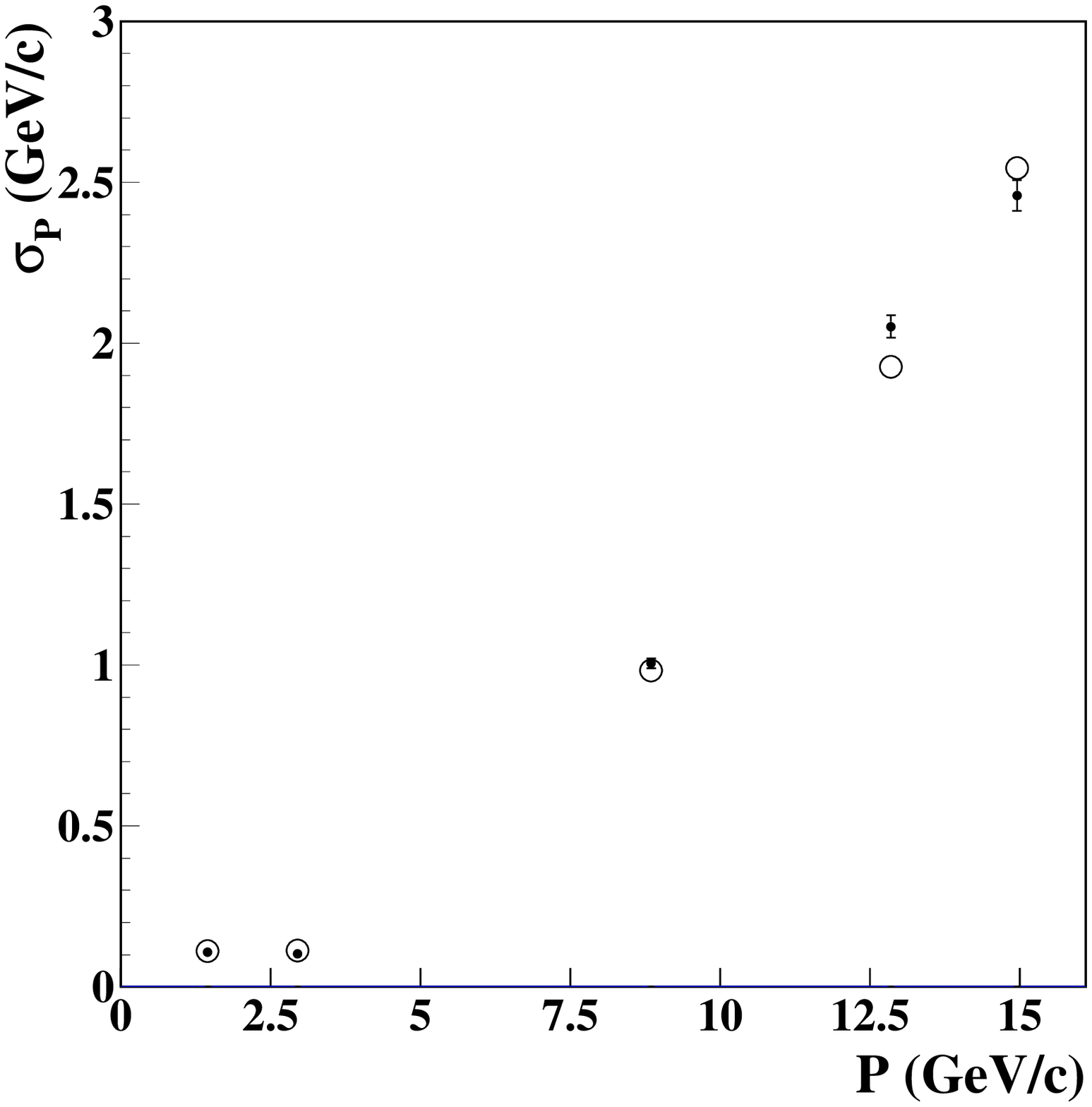,width=0.45\textwidth}
  \epsfig{file=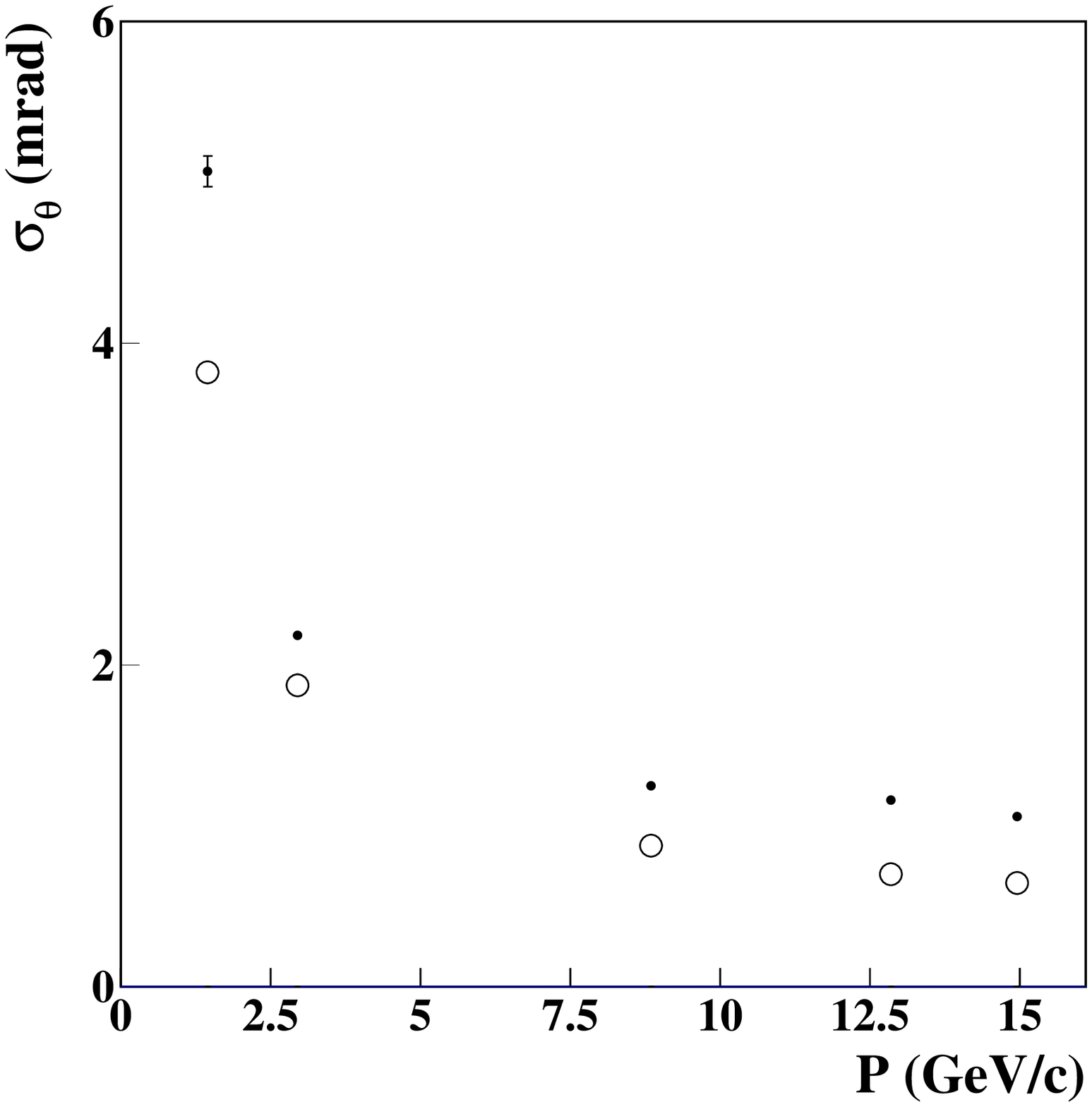,width=0.45\textwidth}
  \caption{
Left panel: momentum resolution ($p_2$) obtained from fits to 
data (points with error bars) taken using several well-defined discrete 
beam momenta
and no target. Also shown (open circles) is the corresponding
resolution found using the Monte Carlo. 
Right panel: angular resolution obtained from fits to 
data (points with error bars) taken using several well-defined discrete 
beam momenta
and no target. The open circles show again the corresponding
resolution found using the Monte Carlo.  In the region of interest,
the agreement between data and Monte Carlo is good for
momentum resolution with values much smaller than the
binning used in the analysis (from 500~\MeVc for $p <$ 4~\GeVc
up to 1.5~\GeVc at $p$ = 6.5~\GeVc). Similarly the difference
between measured and predicted angular resolution is
negligible compared to the 30~\mrad binning adopted in the
analysis (see text).
}
\label{fig:momres}
  \end{center}
\end{figure}

The momentum resolution as a function of the momentum 
is shown in Fig.~\ref{fig:momres} (left panel), 
for the case of $p_2$. The resolution
can be measured using beam particles of several
momenta. Also shown (open circles) is the corresponding
resolution found using the Monte Carlo. 

The momentum resolution does not improve below 3~\GeVc due
to details of the momentum reconstruction algorithm and also
because the particles traverse the material at
larger angles so that the multiple scattering term is no longer
a constant.
This feature is well reproduced by the simulation.
Figure~\ref{fig:momres} (right panel), shows the angular resolution. 
Both the momentum and angular resolutions are small compared with the size
of the bins used in this analysis (500~\MeVc momentum bins, up to 
4~\GeVc, 1000~\MeVc from 4 to 5~\GeVc, 1500~\MeVc from 5 to 6.5~\GeVc, and
30~\mrad angular bins). 
In the region of interest, the agreement between
data and Monte Carlo is good for the momentum resolution, while for
the angular resolution the difference is less than 1~\mrad, negligible
compared to the bin size. 
Thus effects due to the finite resolution are small, and it is safe to
apply a Monte Carlo based correction. 

The charge misidentification rate has been estimated by computing the 
fraction of protons that are reconstructed with negative charge. 
This is done by measuring the fraction of negative particles with
momenta above the pion CHE threshold that give
no signal in CHE. 
The upper limit of 0.5\% for the charge misidentification 
probability is found to be consistent with the known CHE
inefficiency.

\subsection{Definition of kinematical variables}
\label{subsec:kine}

The final cross section, being rotationally invariant around the 
beam axis, can be expressed in polar coordinates $(p, \theta)$, where $p$ is 
the true total momentum of the particle and $\theta$ is the true angle with respect to 
the beam axis (approximately equivalent to the $z$ axis). 
However,  given the rectangular geometry of the dipole and of the drift chambers, 
some of the corrections needed to compute the cross-section 
are most naturally expressed in terms of $(p, \theta_x, \theta_y)$,
where $\theta_x = \arctan (p_x/p_z)$ and  $\theta_y = \arctan (p_y/p_z)$. 
Thus the conversion from rectangular to polar coordinates is
carried out at a later stage of the analysis.

\section{Track reconstruction efficiency}
\label{sec:tracking-eff}

The track reconstruction efficiency,
$\varepsilon^{\mathrm{track}} (p,\theta _x, \theta _y)$, is defined as
the fraction of tracked 
particles (with position and momentum measured) 
$N^{\mathrm{track}}$ with respect to the total
number of particles $N^{\mathrm{parts}}$  reaching
the fiducial volume of the HARP
spectrometer as a function of the true momentum, $p$, and angles, 
$\theta_x, \theta_y $:
\begin{equation}
\varepsilon^{\mathrm{track}} (p,\theta_x,\theta_y)= 
\frac{N^{\mathrm{track}}(p,\theta_x,\theta_y)}{N^{\mathrm{parts}}(p,\theta_x,\theta_y)}
\ .
\label{eq:track_eff}
\end{equation} 

The track reconstruction efficiency can be computed using
the redundancy of the drift chambers taking advantage of
the multiple techniques used for the track reconstruction. 
The rest of this section
details the steps leading to this calculation.
The efficiency was calculated for positively charged particles only. 

\subsection{The use of the $p_4$ estimator to measure tracking efficiencies}

The calculation
of the cross-section requires the knowledge of tracking efficiency
and acceptance in terms of the true kinematical variables of the particle.
Strictly speaking, this is only possible if one uses the Monte Carlo to
compute these quantities. This would make the calculation
sensitive to the details of the Monte Carlo simulation of the spectrometer.

The existence of two independent estimators of the momentum allows 
the tracking efficiency to be measured in terms of $p_4$, taking
advantage of the fact that it
is Gaussian distributed around $p$ and therefore can be used to
approximate the latter.

Therefore, a sample of $p_4$ tracks is selected with well measured momentum
imposing the additional constraint that the tracks emanate from the
primary vertex.  
This is achieved by requiring that the distance of the track
extrapolation to the MWPC vertex is smaller than 10~\mm.
By construction, the vertex of VERTEX2 tracks clusters at a small radius
around the nominal vertex origin (defined by the MWPC resolution) which
is fully covered by these VERTEX4 tracks.

\subsection{Module efficiency}

Using the selected sample of VERTEX4 tracks, one can measure
the tracking efficiency and acceptance 
of individual NDC modules in terms of VERTEX4 kinematical
quantities.

\begin{figure}[tbp!]
\begin{center}
\includegraphics[width=1.5in]{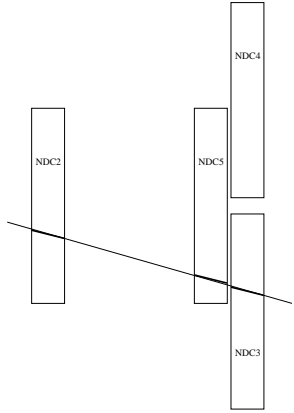}
\end{center}
\caption{Schematic layout of the downstream modules of the drift chambers (top view).}
\label{fig:ndcSchematic}
\end{figure}

The measurement of $p_2$ requires
a downstream segment which is then connected to the event vertex. In turn,
a downstream segment can be made of a segment in NDC2, or a segment in 
any of the modules downstream NDC2 (that is 
NDC3, NDC4 and NDC5, see Fig.~\ref{fig:ndcSchematic}). For the purpose of
the analysis one can treat {\it conceptually} those three NDC modules as 
a single 
module
which we call {\it back-plane}. Thus, a downstream segment is defined
as a NDC2 segment, a back-plane segment or a long segment which combines
both NDC2 and back-plane.
In all cases the measurement of $p_2$ requires
that the true particle has crossed NDC2. By definition the control sample
of VERTEX4 tracks verifies that the true particle crossed NDC2 (since the
measurement of $p_4$ requires a downstream segment connected to NDC1),
but not necessarily that a segment was reconstructed in NDC2 (since a good
VERTEX4 track can be built with a back-plane segment and a segment in NDC1). 
In practice, the required condition is that at least six hits are found inside 
the road defined by the extrapolation of the downstream track to NDC2. It was 
verified that this condition has a negligible effect on the efficiency
determination. 

The NDC2 efficiency $\varepsilon_2$ is defined as the number of
segments reconstructed in NDC2 (in terms of VERTEX4 kinematical quantities, 
$p, \theta_x, \theta_y$) divided by the number of tracks in the VERTEX4
control sample. This is equivalent to finding the number of segments
reconstructed in NDC2 divided by the number of particles reaching NDC2. Thus
$\varepsilon_2$ measures the `true' tracking
efficiency of the NDC2 module, 
unfolded
from other effects such
as acceptance, absorption or decay. 
If a particle decays
or is absorbed before reaching NDC2 it will not be included in the
control sample and therefore it will not be included 
in the calculation of $\varepsilon_2$.
 
\begin{figure}[tbp]
\begin{center}
\includegraphics[width=\textwidth]{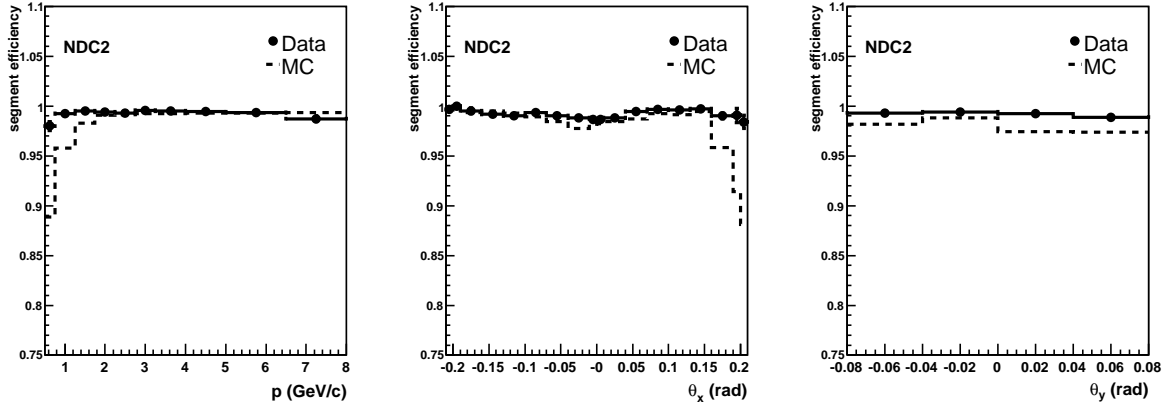}
\end{center}
\caption{Segment efficiency of NDC2 module (see text for definition),
as a function of $p$ (left panel), $\theta_x$ (centre panel)
and $\theta_y$ (right panel). The efficiency is computed as a
three-dimensional function of the above variables. The plots show
the individual projections. Points with error bars correspond to data,
the dashed line to Monte Carlo. The agreement between Monte Carlo and data
calculation is good except in the region of large $\theta_x$
and small $p$ (see text).
}
\label{fig:ndc2}
\end{figure}

Figure~\ref{fig:ndc2} shows $\varepsilon_2$ as a function of
$p, \theta_x, \theta_y$ (estimated from the VERTEX4 control sample). As expected
the distribution is flat in terms of all three variables. The dots represent
the calculation from the data themselves, and show that the NDC2 tracking
efficiency is essentially 100\%. The dashed line represents the Monte Carlo
calculation, which agrees with the data calculation except in the region of
low momentum and large, positive $\theta_x$.
The inefficiencies in these regions are correlated and are due to edge
effects which are not perfectly described in the Monte Carlo. 
This region is not used in the analysis.

\begin{figure}[tbp]
\begin{center}
\includegraphics[width=\textwidth]{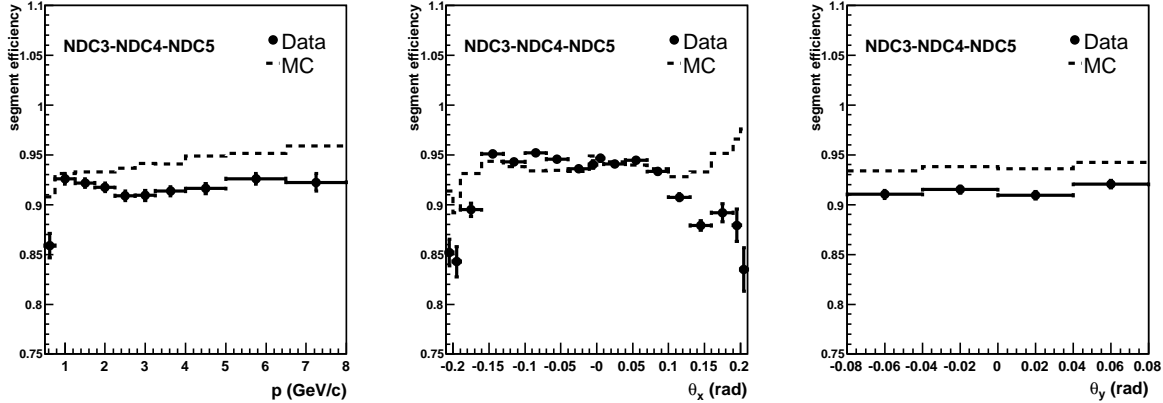}
\end{center}
\caption{Segment efficiency of the back-plane,
defined by modules NDC3, NDC4 and NDC5
shown as a function of $p$ (left panel), $\theta_x$ (center panel)
and $\theta_y$ (right panel). The efficiency is computed as a
three-dimensional function of the above variables. The plots show
the individual projections.  Points with error bars correspond to data,
the dashed line to Monte Carlo. }
\label{fig:ndcb}
\end{figure}

The back-plane efficiency, $\varepsilon_b$, is defined as the number
of segments reconstructed 
in the back-plane divided by the number of tracks in the VERTEX4
control sample.  This definition 
folds
tracking efficiency with acceptance and other effects such
as absorption or decay (for example, one could have a well 
reconstructed VERTEX4 track with a NDC2 segment and a NDC1 segment, decaying
or undergoing a nuclear interaction in NDC2).
The Monte Carlo tends to overestimate the efficiency by less than 5\% 
on average (see Fig.~\ref{fig:ndcb}).
At large positive $\theta_x$ the Monte Carlo predicts a rise in the
efficiency which is not seen in the data.
This is a region where large angle tracks are further deflected by the
magnet and traverse the drift chambers at large angles.
As will be shown below, owing to the redundancy of the chambers, the
overall downstream efficiency is well reproduced by the simulation.
 
The knowledge of the efficiency of NDC1 is not needed, since the yields
are computed in terms of $p_2$ which does not use it. However, one
could choose to measure the yields in terms of $p_4$ and use $p_2$
as an estimator of true momentum to measure tracking efficiency. In
that case, NDC1 would play the same role that NDC2 plays in the current
approach. 

Indeed, it is illustrative to compute $\varepsilon_1$ for 
VERTEX4 tracks in terms of VERTEX2 kinematical quantities, 
$p_2, \theta_x, \theta_y$). 
The efficiency $\varepsilon_1$ is defined 
as the number of segments reconstructed
in NDC1 divided by the number of tracks in a VERTEX2
control sample. This is equivalent to requiring the number of segments
reconstructed in NDC1 over the number of particles reaching NDC1. Thus
$\varepsilon_1$ measures the `true' tracking
efficiency of the NDC1 module, 
unfolded
from other effects such
as acceptance, absorption or decay. 
 
\begin{figure}[tbp]
\begin{center}
\includegraphics[width=\textwidth]{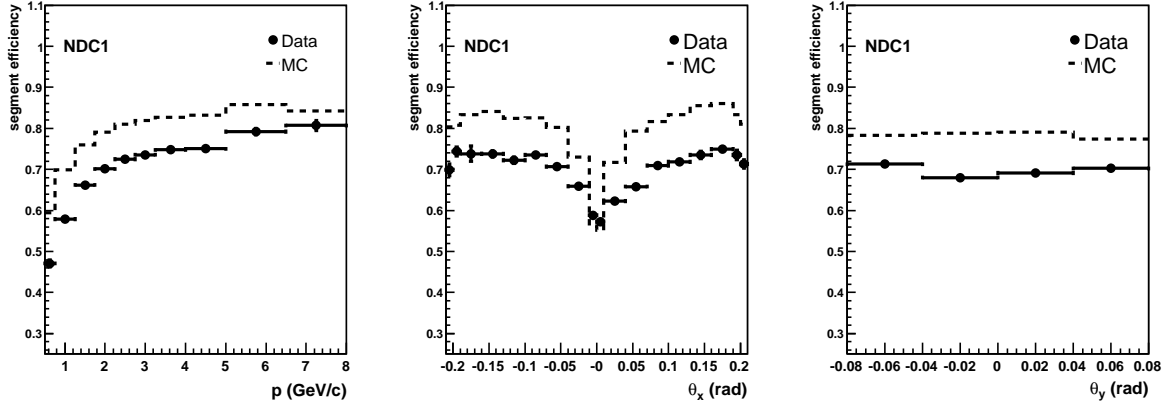}
\end{center}
\caption{Segment efficiency of NDC1 (see text for definition)
as a function of $p$ (left panel), $\theta_x$ (center panel)
and $\theta_y$ (right panel). The efficiency is computed as a
three-dimensional function of the above variables. The plots show
the individual projections.  Points with error bars correspond to data,
the dashed line to Monte Carlo.}
\label{fig:ndc1}
\end{figure}

Figure~\ref{fig:ndc1} shows $\varepsilon_1$ as a function of
$p, \theta_x, \theta_y$ (estimated from the VERTEX2 control sample). 
The distribution is relatively flat in $\theta_y$, while it
has a marked dependence on $\theta_x$, with a minimum at 
$\theta_x = 0$. The distribution of $\varepsilon_1$ as a function
of $\theta_x$ shows the inefficiency of NDC1 associated
with a saturation of the chambers due to the primary beam intensity. NDC1
is the only module affected by this saturation effect, which is
negligible downstream of the dipole magnet (as proved by inspection of 
$\varepsilon_2$). 
The Monte Carlo tends to overestimate the efficiency by 15\%.
The dip in the efficiency at $\theta_x$ close to the origin is induced
by the saturation effect of the beam.  This feature is simulated in the
Monte Carlo by artificialy lowering the efficiency of the drift
regions most traversed by undeflected beam particles.
Although one can measure the NDC1 tracking efficiency with
good precision using the data, the marked dependence on
$\theta_x$ (which translates also in a dependence on $p$) suggests as
a better strategy to measure the yields in terms of VERTEX2, which does not
use NDC1. 

\subsection{Downstream tracking efficiency}

The downstream tracking efficiency, 
$\varepsilon^{\mathrm{down}}$, is defined
as the number of tracks reconstructed downstream the dipole magnet
(those include NDC2 single segments, back-plane single segments and
NDC2-back-plane combined segments) divided by the number of 
particles reaching NDC2 (which defines the fiducial volume). 
Since $\varepsilon_2 $ and $\varepsilon_b $ are uncorrelated, this
quantity is easily computed 
from the individual segment efficiency described above, as:

\begin{equation}
\varepsilon^{\mathrm{down}} = \varepsilon_2 + \varepsilon_b - 
\varepsilon_2 \cdot \varepsilon_b \ .
\label{eq:downstreamEff}
\end{equation}   

\begin{figure}[tbp]
\begin{center}
\includegraphics[width=\textwidth]{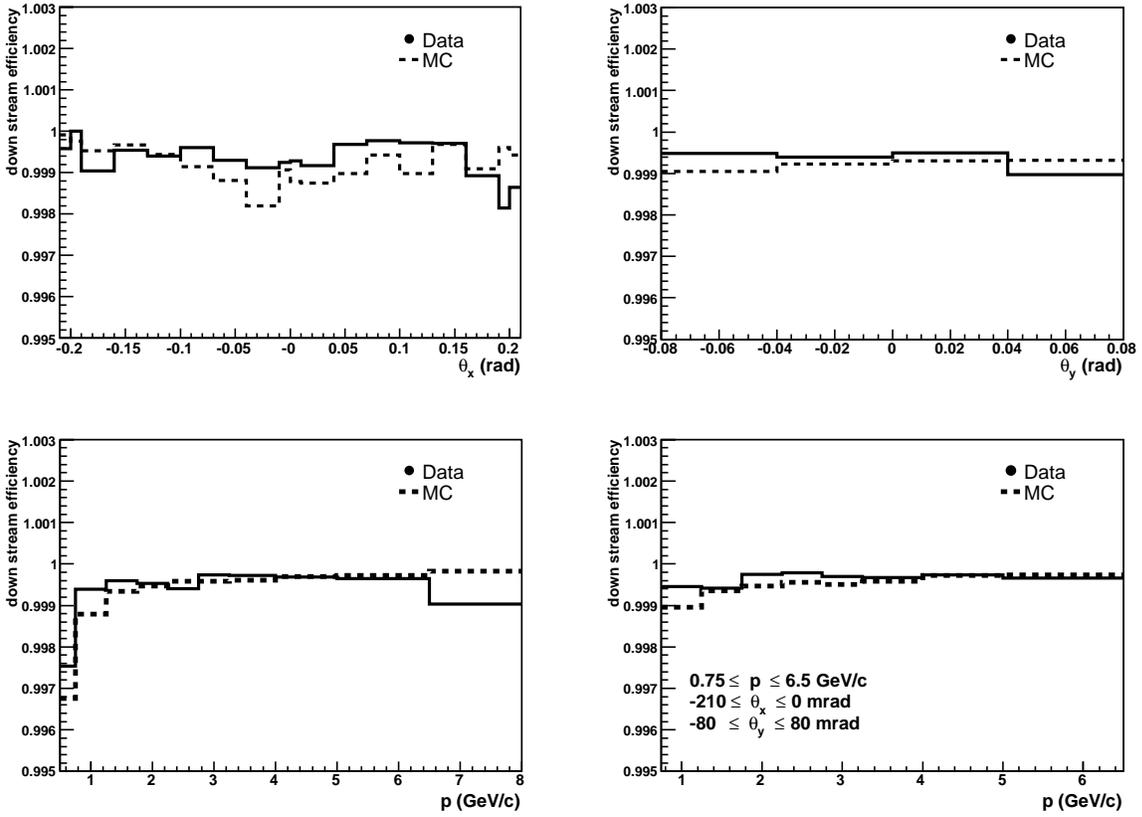}
\end{center}
\caption{Downstream tracking efficiency as a function 
of kinematic variables, $p$, $\theta_{x}$, 
and $\theta_{y}$, at production for positively 
charged particles emanating from the vertex. Upper-left panel: As a function
of $\theta_x$. Upper-right panel: As a function
of $\theta_y$. Lower-left panel: As a function of $p$. Lower-right panel: 
As a function of $p$ averaged over the $\theta_x$ and $\theta_y$ regions 
used in the present analysis only. The efficiency is flat in all variables and
close to 100\%. The solid histograms correspond to data,
the dashed line to Monte Carlo.}
\label{fig:downstream}
\end{figure}

Figure~\ref{fig:downstream} shows the
total downstream segment efficiency in terms of the kinematical quantities
(from VERTEX4) $p, \theta_x, \theta_y$. Due to the very high tracking
efficiency of the individual modules (the apparent drop of the
efficiency of the back-plane visible in Fig.~\ref{fig:ndcb}  
is largely due to decay and absorption) the downstream
tracking efficiency is almost 100\%, flat in all variables, within the 
angular acceptance considered.  The overall downstream tracking
efficiency is well reproduced by the Monte Carlo.

\subsection{Upstream tracking efficiency}

For the contribution of the vertex matching to the 
overall track reconstruction efficiency, 
one needs to compute the fraction of times that a good 
downstream track segment was correctly matched to the vertex point
resulting in an acceptable momentum.
This is also done using a VERTEX4 control sample, 
with the additional constraint that the particle
emanates from the target volume.
The latter condition is ensured by requiring 
$r_{\mathrm{V4}} \leq 30 \ \mm, \ \chi^2_{\mathrm{match}} \leq 10$,
where $r_{\mathrm{V4}}$ is the distance of the track extrapolation to
the MWPC vertex in the target reference plane, and 
$\chi^2_{\mathrm{match}}$ the goodness of the matching of the track
with the beam particle track extrapolated to the nominal target
position. 
The cut in $\chi^2_{\mathrm{match}}$ guarantees
good matching with the vertex.  
These tracks enter the
denominator of the efficiency calculation. The numerator is made of
the accepted VERTEX2 tracks within this sample:
\begin{equation}
\varepsilon^{\mathrm{vertex2}} = 
\frac{N^{\mathrm{vertex2}} (\exists p_2;
\exists p_4, r_{\mathrm{V4}} \leq 30 \ \mm,
  \chi^2_{\mathrm{match}} \leq 10)}
{N^{\mathrm{vertex4}} (\exists p_4, r_{\mathrm{V4}} \leq 30 \ \mm,
  \chi^2_{\mathrm{match}} \leq 10)} \ ,
\end{equation}

where the condition $\exists p_{4(2)}$ guarantees that 
the track momentum was estimated by the two reconstruction algorithms,
respectively.

\begin{figure}[tbp!]
\begin{center}
\includegraphics[width=\textwidth]{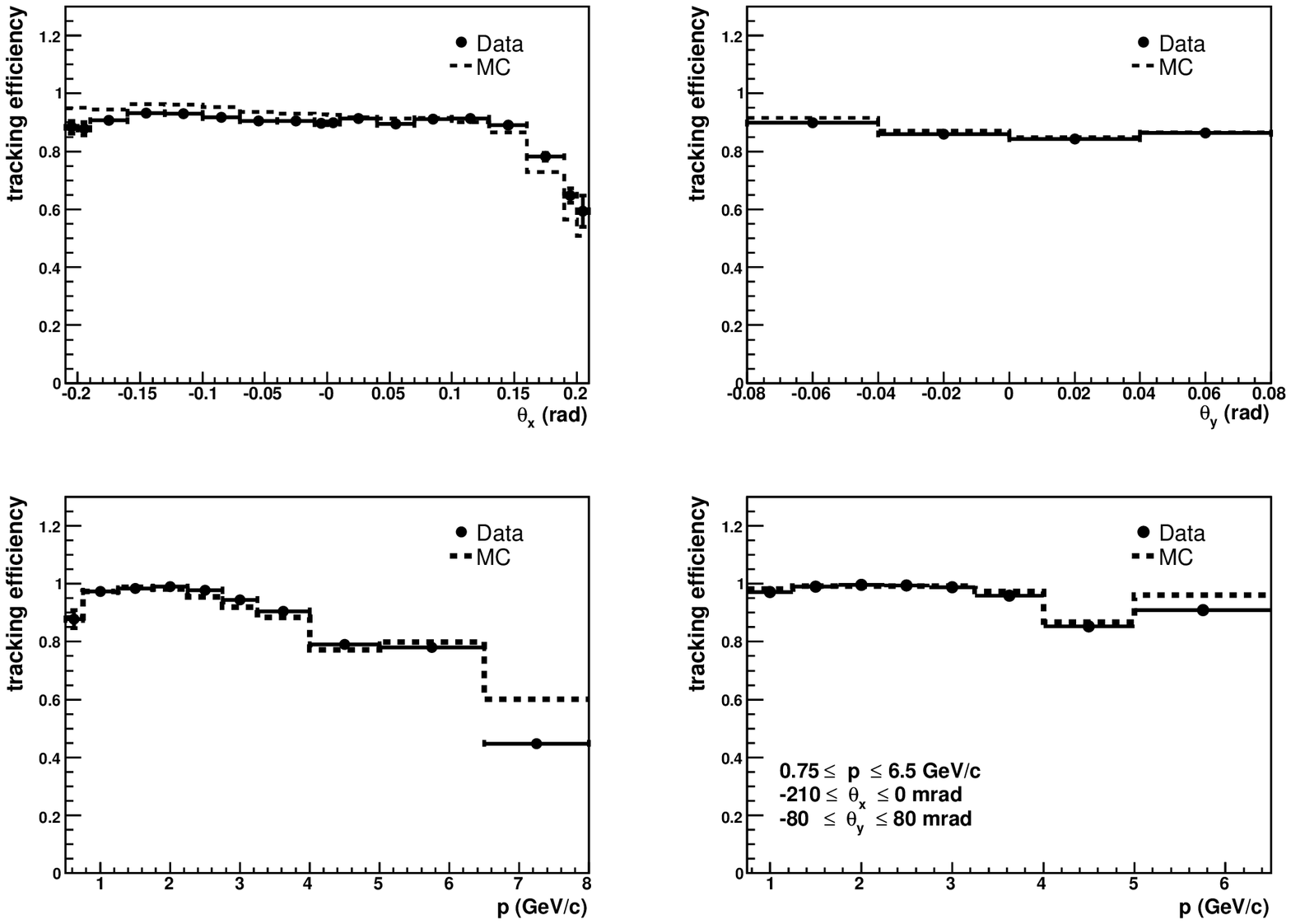}
\end{center}
\caption{Upstream tracking efficiency as a function of 
kinematic variables, $p$, $\theta_{x}$, and $\theta_{y}$, 
at production for positively charged particles emanating from the vertex.  
Upper-left panel: As a function
of $\theta_x$. Upper-right panel: As a function
of $\theta_y$.  Lower-left panel: As a function of $p$. Lower-right panel: 
As a function of $p$ averaged over the $\theta_x$ and $\theta_y$ regions 
used in the present analysis only. The efficiency is close to 100\% for negative
$\theta_x$ and momenta less than 4~\GeVc, and drops for high values
of $\theta_x$ (due to the dipole's acceptance) and for high momenta
(due to a weakness in the reconstruction algorithm, see text).
Points with error bars correspond to data,
the dashed line to Monte Carlo. The agreement is excellent, except in the
momentum bin $6.5 \ \GeVc \le p < 8.0 \ \GeVc$, where there is a 25\%
difference.  This bin is not used in the analysis (see text).}
\label{fig:total}
\end{figure}

The efficiency for reconstructing a
VERTEX2 track for a particle coming from the target 
is shown in Fig.~\ref{fig:total}. 
Figure~\ref{fig:total} (upper-left), shows that the efficiency 
as a function of $\theta_x$ is flat and close to 
95\%, up to 150~\mrad and drops above this value. This drop is due a
the momentum-dependent acceptance limitation imposed by the dipole magnet, as
clearly demonstrated by Fig.~\ref{fig:total} (upper-right), which shows a flat 
distribution in the non-bending plane $\theta_y$ and Fig.~\ref{fig:total} (lower-left),
which shows the efficiency as a function of $p$, integrated for all
$\theta_x$. The drop of the efficiency for large values of $\theta_x$ is fully
correlated with the drop at low $p$. This can be seen in 
Fig.~\ref{fig:total} (lower-right), where the reconstruction efficiency as a function
of $p$ for negative $\theta_x$ (particles fully contained in the dipole
acceptance) is shown.  The efficiency is flat and close to 100\%
up to 4~\GeVc, and drops for the last two bins. 
The efficiency as a function of $\theta_x$ ($\theta_y$) for
momenta less than 4~\GeVc  is close to 100\%, indicating that the loss
of efficiency is due to the efficiency drop at $p > 4 \ \GeVc$.
The efficiency of the momentum reconstruction algorithm is lower near
the edge of the acceptance because it requires a number of
trajectories around the best fit to be inside the aperture. 
This requirement reflects in a drop of overall efficiency
for $\theta_x > 150 \ \mrad$.

The drop in reconstruction efficiency at high momenta is due to the  
lack of optimization, at high momentum, 
of the reconstruction algorithm.
Its effect on the calculation of the cross-section is very small
(in practice it translates into a slightly larger error in the higher momentum bins, 
which are dominated by statistical errors). 
Except for the highest-momentum bin, the data are well described by
the Monte Carlo.

\subsection{Total reconstruction efficiency}

The total tracking efficiency,
$\varepsilon^{\mathrm{track}}$,
can be expressed as the product of two factors.
One factor represents the downstream (of the dipole magnet) 
tracking efficiency and the other represents the efficiency 
for matching a downstream
segment to a vertex (for tracks originating inside the target volume): 
\begin{equation}
\varepsilon^{\mathrm{track}} = \frac{N^{\mathrm{down}}}{N^{\mathrm{parts}}} \cdot 
\frac{N^{\mathrm{vertex2}}}{N^{\mathrm{down}}} =
\varepsilon^{\mathrm{down}} \cdot \varepsilon^{\mathrm{vertex2}} \ ,
\label{eq:totaleff}
\end{equation}

where $N^{\mathrm{vertex2}}$ is the number of VERTEX2 tracks, which
corresponds to $N^{\mathrm{track}}$ in Eq.~\ref{eq:track_eff}. 
In the momentum range of interest for this analysis, a good
time-of-flight measurement is essential for particle
identification. 
Therefore, one also requires a good TOFW hit matched
to the track. In addition to selecting the scintillator slab hit by 
the particle, this matching consists of a 
cut in the matching $\chi^2$ of the track with the TOFW hit coordinate 
that is measured by the time difference of the signals at the two 
sides of the scintillators.

The TOFW hit matching efficiency was computed using the same track
sample as in the previous section:
\begin{equation}
\varepsilon^{\mathrm{ToF}} = \frac{N^{\mathrm{ToF}} 
(\exists \mathrm{ToF};
\exists p_4, r_{\mathrm{V4}} \leq 30 \ \mm,
  \chi^2_{\mathrm{match}} \leq 10)}
{N^{\mathrm{vertex4}} (\exists p_4, r_{\mathrm{V4}} \leq 30 \ \mm,
  \chi^2_{\mathrm{match}} \leq 10)} \ ,
\end{equation}
where the condition $\exists \mathrm{ToF}$ guarantees that a TOFW hit was
associated with the track, and the total reconstruction efficiency is
found from:
\begin{equation}
\varepsilon^{\mathrm{recon}} = \varepsilon^{\mathrm{track}} \cdot
\varepsilon^{\mathrm{ToF}} \ .
\end{equation}

\begin{figure}[tbp!]
\begin{center}
\includegraphics[width=\textwidth]{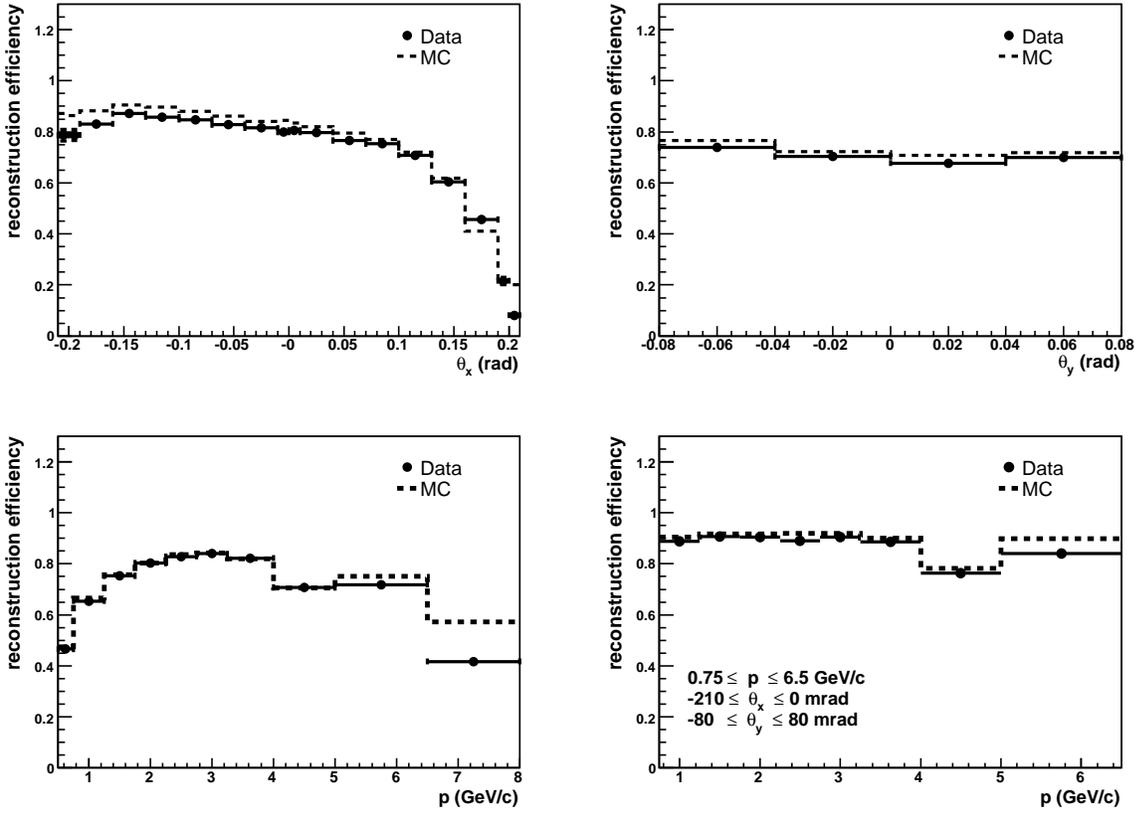}
\end{center}
\caption{Total reconstruction efficiency as a function of 
kinematic variables, $p$, $\theta_{x}$, and $\theta_{y}$, 
at production for positively charged particles emanating from the vertex.  
Upper-left panel: as a function
of $\theta_x$. Upper-right panel: as a function
of $\theta_y$.  Lower-left panel: as a function of $p$. Lower-right panel: 
as a function of $p$ averaged over the $\theta_x$ and $\theta_y$ regions 
used in the present analysis only. The efficiency is close to 90\% for negative
$\theta_x$ and momenta less than 4~\GeVc, and drops for high values
of $\theta_x$ (due to the TOFW acceptance) and for high momenta
(due to a weakness in the reconstruction algorithm, see text).
Points with error bars correspond to data,
the dashed line to Monte Carlo. The agreement is excellent, except in the
bin with highest momentum, where the difference is 6\%.
}
\label{fig:totalRecon}
\end{figure}

The total reconstruction efficiency is shown 
in Fig.~\ref{fig:totalRecon}. 
The inclusion of the TOF wall
enhances the momentum-dependent acceptance cut (due to the fact that
the TOF wall has a smaller geometrical acceptance than the NDC back-plane).
The total reconstruction efficiency as a function of $\theta_x$ has
a slight slope and drops for positive $\theta_x$ above 100~\mrad. 
The total reconstruction efficiency as a function of $p$ for negative
$\theta_x$ is flat (and about 90\%) for momenta below 4~\GeVc.  The
drop to 90\% is due to the inefficiency in matching tracks to a TOFW hit. 
The agreement between
data and Monte Carlo for the total reconstruction efficiency is 
excellent, except in the
last momentum bin, where there is a 25\% difference, reflecting the
same effect already observed in the upstream efficiency.

\section{Particle identification}
\label{sec:pid}

A set of efficient PID algorithms to select pions and reject other
particles is required for the current analysis.
A Monte Carlo prediction of the differential yields of the various particle types 
shows that the pion production cross-section is small 
above 6.5~\GeVc, which is set as the upper limit of
this analysis. The electron distribution peaks at low energy, while the
proton background increases with momentum. The kaon yield is expected to
be only a small fraction of the pion yield. 
In the momentum and angular range covered by the present measurements
the proton yield is of a similar order of magnitude as the pion yield.

The PID strategy  is 
based on the expectation of the yields of different particle types predicted 
by the Monte Carlo, and also on the momentum regions covered by the available 
PID detectors. The time-of-flight measurement with the combination of 
BTOF and TOFW systems (referred to as the TOFW measurement in what follows) 
allows pion--kaon and pion--proton 
separation to be performed up to 3~\GeVc and beyond 5~\GeVc respectively. 
The Cherenkov is used for hadron-electron separation below 2.5~\GeVc 
and pion--proton/kaon separation above 2.5~\GeVc in conjunction with the TOFW.  
The ECAL is used only to separate hadrons  
from electrons below 2.5~\GeVc to study the Cherenkov performance. 

As mentioned above, the electron(positron) background is concentrated
at low momentum ($p < 2.5 \ \GeVc$). 
It can be suppressed to negligible level with an upper limit on the CHE signal, 
given the fact that electrons are the only particles giving signal in the Cherenkov below
the pion Cherenkov light emission threshold, which 
is equal to 2.6~\GeVc
for the gas mixture used in HARP.
In practice, any particle that has a momentum below 2.5~\GeVc
and a signal in the CHE exceeding 15 photo-electrons is called an electron. 
In the following we will refer to this cut as the e-veto cut. The remaining 
electron background after the e-veto cut is negligible as studied in
Ref.~\cite{ref:pidPaper}. 

Having applied the e-veto cut to reject electrons
and keeping in mind that there is a small fraction of
kaons, one builds PID estimators for protons 
and pions by combining the information from 
TOFW and CHE using likelihood techniques (Sec.~\ref{sec:pid_prob}). 
Then, a cut on these PID estimators is applied to select pions or protons. The selected 
samples (raw pion and proton samples)
will contain a small fraction of kaons, which can be 
estimated from the data, as described in Ref.~\cite{ref:pidPaper}.
This background is subtracted from the dominant yields of pions and protons. 

The quantities that enter the cross-section calculation are 
the raw pion and proton yields and the PID efficiencies and purities (PID
corrections) obtained  
by the application of the e-veto cut and cuts in the PID estimators. 
The PID corrections include the e-veto efficiencies, the kaon
subtraction corrections and the pion-proton efficiency matrix  
($M^{id}$, described in Sec.~\ref{sec:two-analyses}). A precise knowledge of 
these quantities (as a function of momentum and angle) 
requires the understanding of the responses of different PID 
detectors to the particle types considered. This is studied in detail
using the data, taking advantage of the redundancy between the PID 
detectors. 

These steps are explained briefly in the following sections. More details 
are given in Ref.~\cite{ref:pidPaper}.

\subsection{Response of the PID detectors}

The TOFW--CHE Probability Density Function (PDF), 
$P(\beta, N_{\mathrm{phe}} | i, p, \theta)$, 
describes the probability that a particle of type $i$ (pion or proton) 
with momentum $p$ and polar angle $\theta$ results in simultaneous measurements 
$\beta$ in TOFW and $N_{\mathrm{phe}}$ in CHE. The latter comes directly from
the calibrated CHE signal, while $\beta$ is the particle velocity, computed as 
$\beta = l_{\mathrm{tof}}/(t_{\mathrm{tof}} \cdot c)$, 
where $l_{\mathrm{tof}}$ is the track length measured from the nominal vertex 
position to the TOFW hit position, $t_{\mathrm{tof}}$ is the measured time-of-flight and 
$c$ is the speed of light. 
Assuming that the PID from both detectors are independent\footnote{
The use of the reconstructed momentum instead of the 
true momentum in the probability density 
function $P(\beta, N_{\mathrm{phe}} | i, p, \theta)$ introduces a small 
correlation between TOFW and CHE. However, 
the good momentum resolution of the forward spectrometer ($\sigma_p/p<10\%$) makes this 
correlation very small. At the level of 
precision required by this analysis this effect can be neglected.
} 
the TOFW--CHE PDFs can be factorized
in independent TOFW and CHE PDFs, such that
$P(\beta, N_{\mathrm{phe}} | i, p, \theta)=P(\beta | i, p, \theta) 
 \cdot P(N_{\mathrm{phe}}| i, p, \theta)$.

\subsubsection{TOFW response}
\label{subsubsec:tofw}

\begin{figure}[tbp]
  \begin{center}
  \epsfig{file=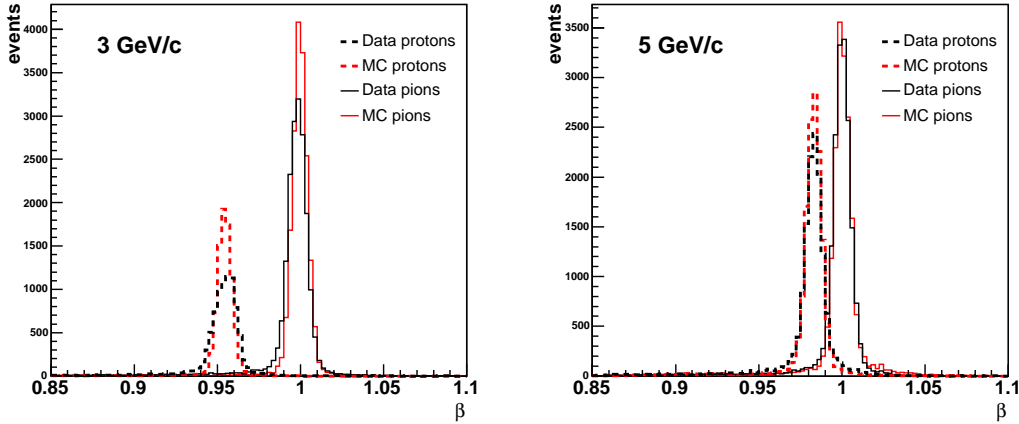,width=0.9\textwidth}
  \caption{\label{fig:beta_emptyTarg} The distribution of the particle
  velocity, $\beta$, for pions and protons of 3 and 5~\GeVc, for both
  data (corresponding to a sample of beam particles, selected as
  pions or protons by the beam instrumentation) and its corresponding
  Monte Carlo simulation. At 3~\GeVc the separation between the two
  populations is $\approx 5 \sigma$, and the separation is still
  $\approx 2.2 \sigma$ at 5~\GeVc.  The proton and pion peaks in the
  Monte Carlo have been separately normalized to the area of the
  corresponding peaks in the data.
}  
  \end{center}
\end{figure}

The use of the particle velocity, $\beta$, to characterize the TOFW response 
has several advantages. Its
distribution is nearly Gaussian--the agreement between data and Monte Carlo
is sufficient for the current analysis--and
it discriminates very effectively
between pions and protons up to momenta around 5~\GeVc
(at this energy the separation between the average values of
the proton and pion Gaussians is around $2.2 \sigma$).
These points are illustrated in Fig.~\ref{fig:beta_emptyTarg}. 

\begin{figure}[htbp]
  \begin{center}
  \epsfig{file=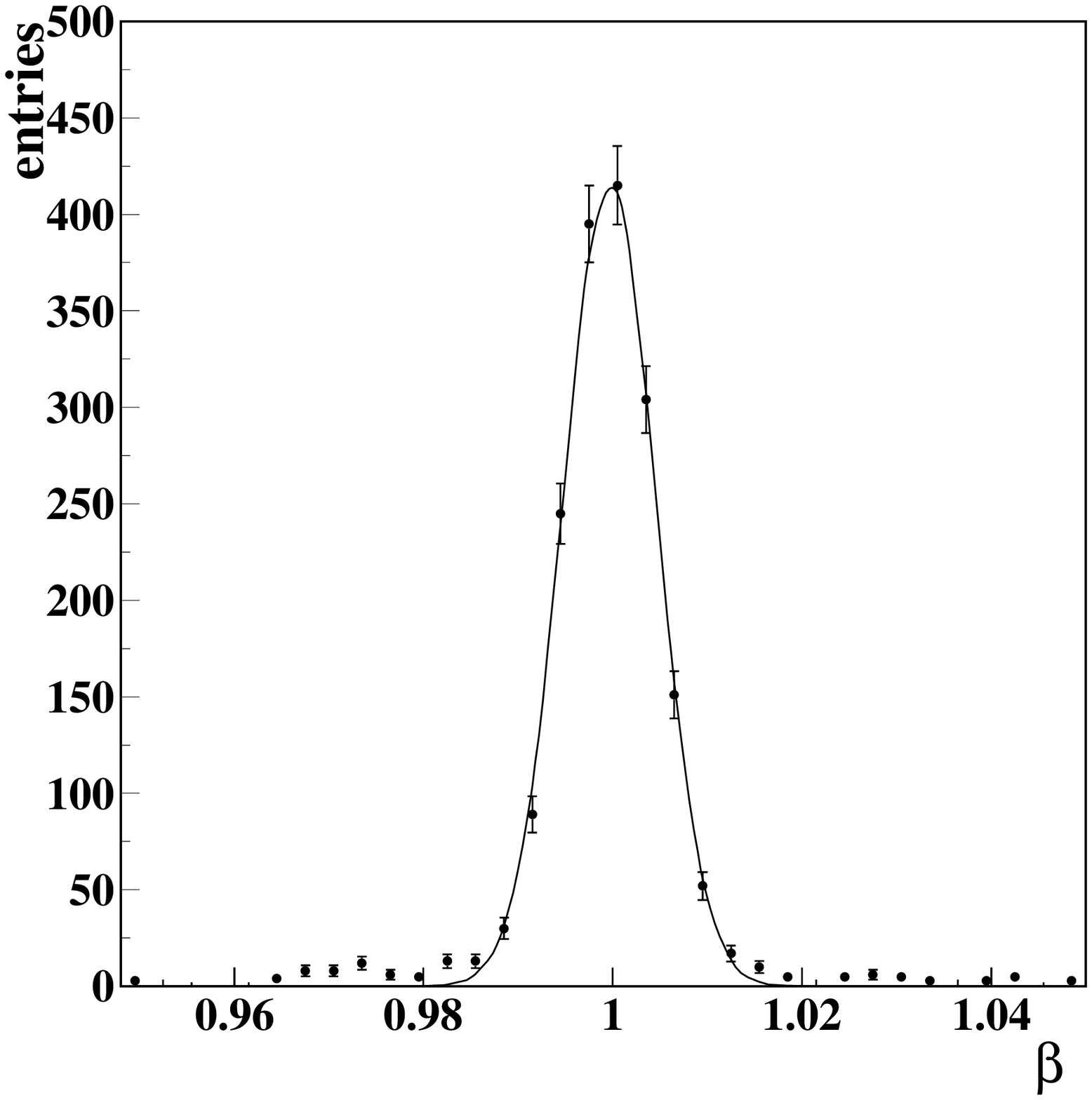,width=0.3\textwidth} 
  \epsfig{file=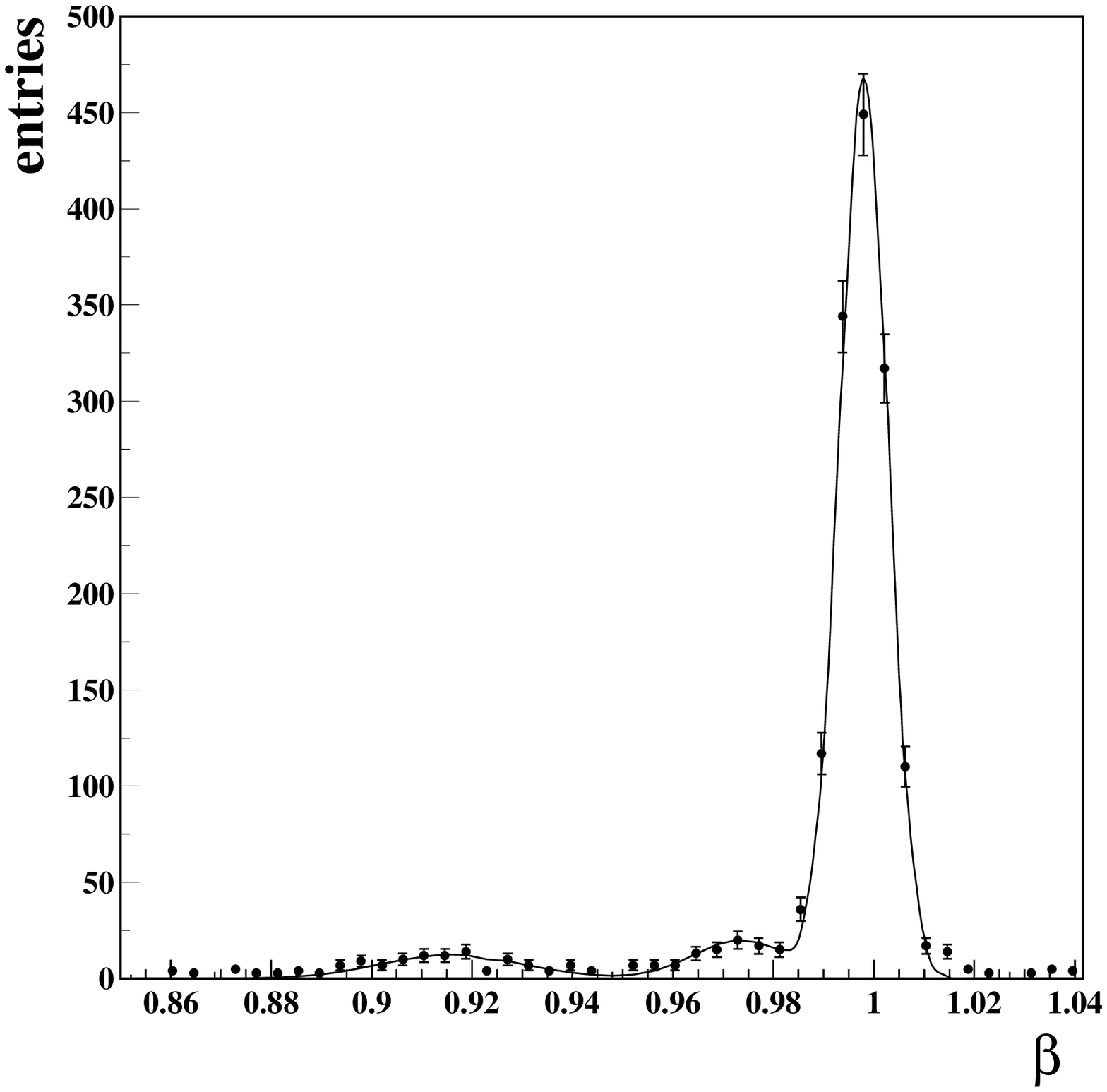,width=0.3\textwidth}
  \epsfig{file=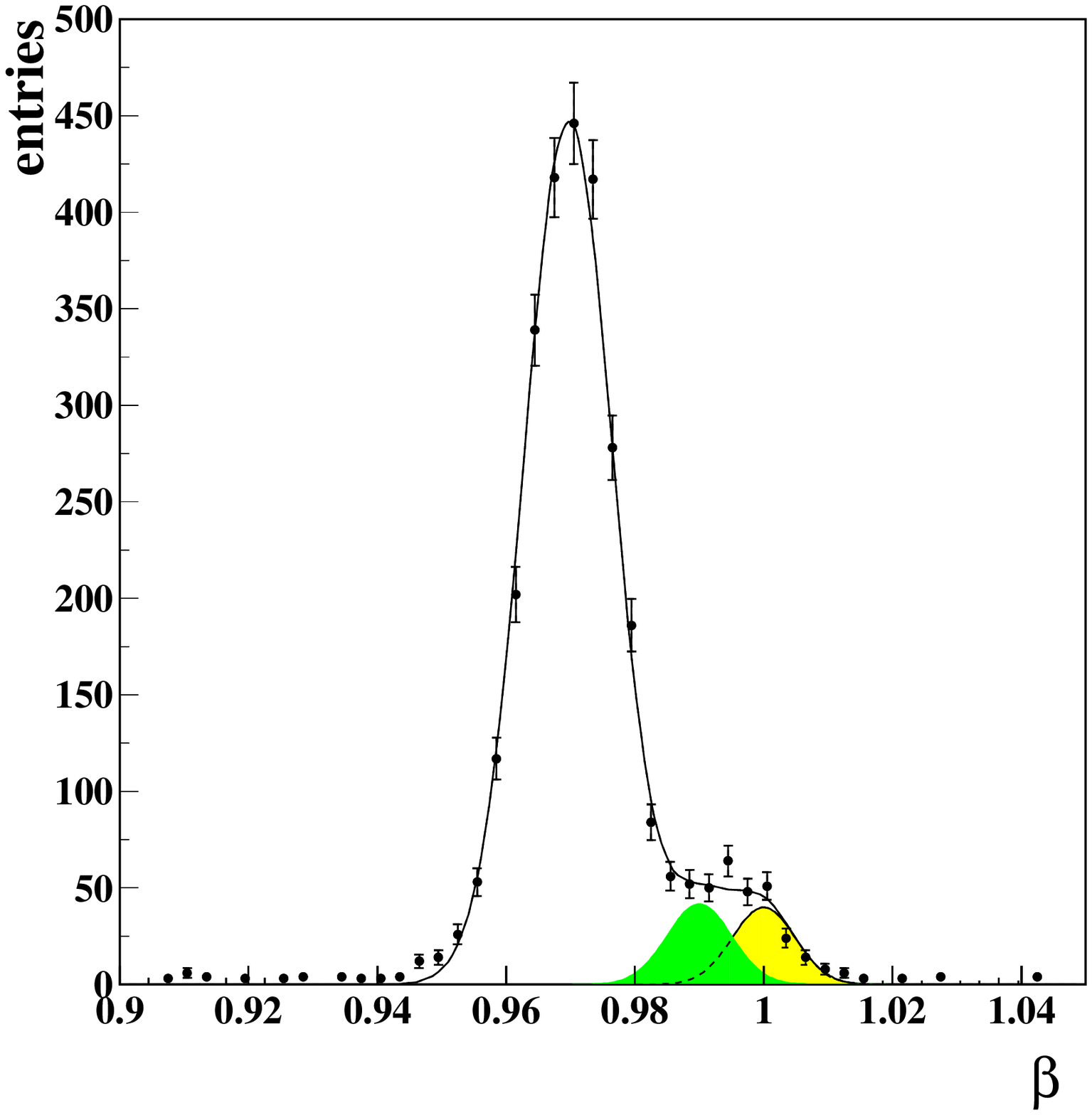,width=0.3\textwidth} 
  \caption{\label{fig:beta_pion_proton_rec_mc}
    Inclusive $\beta$ distribution for pions, kaons and protons passing the e-veto cut. 
    Left panel: negative particles (essentially pions) 
    with reconstructed momentum between $3.25$ and $4$~\GeVc.
    Central panel: positive particles with reconstructed momentum between $1.75$ and $2.25$~\GeVc.
    Right panel:  positive particles (mostly protons) 
    with reconstructed momentum between $3.25$ and $4$~\GeVc.
    In the right panel, the kaon and pion Gaussians are also shown (shaded areas).
    }
  \end{center}
\end{figure}

To build the TOFW PDFs one should
know the $\beta$ distribution of pions
and protons as a function of the particle momentum and angle. 
In order to maximize the efficiency of the selection algorithm and to avoid any possible 
bias in the PID corrections related with a data--MC disagreement, those 
distributions have been measured from the data. This requires the
ability to select pure and unbiased samples of pions and protons from the
data. Samples of pions with negligible contamination from other species can be obtained 
selecting particles of negative charge passing the e-veto cut 
(Fig.~\ref{fig:beta_pion_proton_rec_mc}, left panel). 
At low momentum, the proton parameters can 
be obtained by simply fitting to a Gaussian the proton part of the 
$\beta$ distribution, since this is well separated from pions 
(Fig.~\ref{fig:beta_pion_proton_rec_mc}, central panel). At large 
momentum ($>2.5$~\GeVc), the proton and pion distributions overlap significantly. 
In this case, pions are rejected to the $1\%$ level by a cut on the CHE signal 
(Sec.~\ref{subsubsec:che}). 
The resulting sample contains a majority of protons and residual contaminations  
from pions and kaons. In this momentum range the proton parameters can 
be obtained by fitting the inclusive $\beta$ distribution to a triple Gaussian 
with fixed pion and kaon shapes\footnote{The pion shape is obtained from the data, 
while the kaon shape is calculated using the MC information.}, 
as shown in the right panel of Fig.~\ref{fig:beta_pion_proton_rec_mc}. 
A full description of this technique is given in Ref.~\cite{ref:pidPaper}.

\begin{figure}[tbp]
  \begin{center}
  \epsfig{file=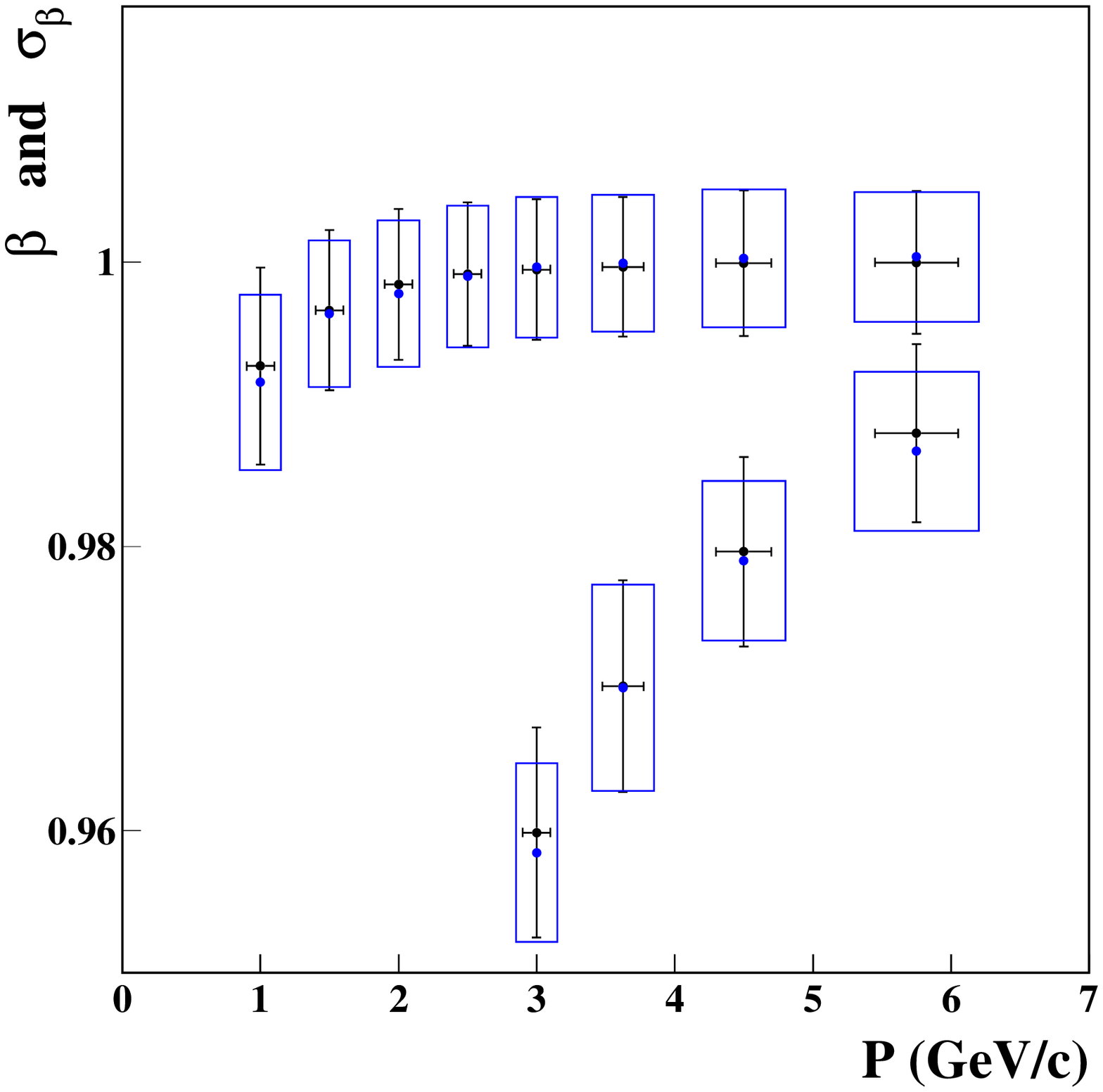,width=0.45\textwidth} 
  \epsfig{file=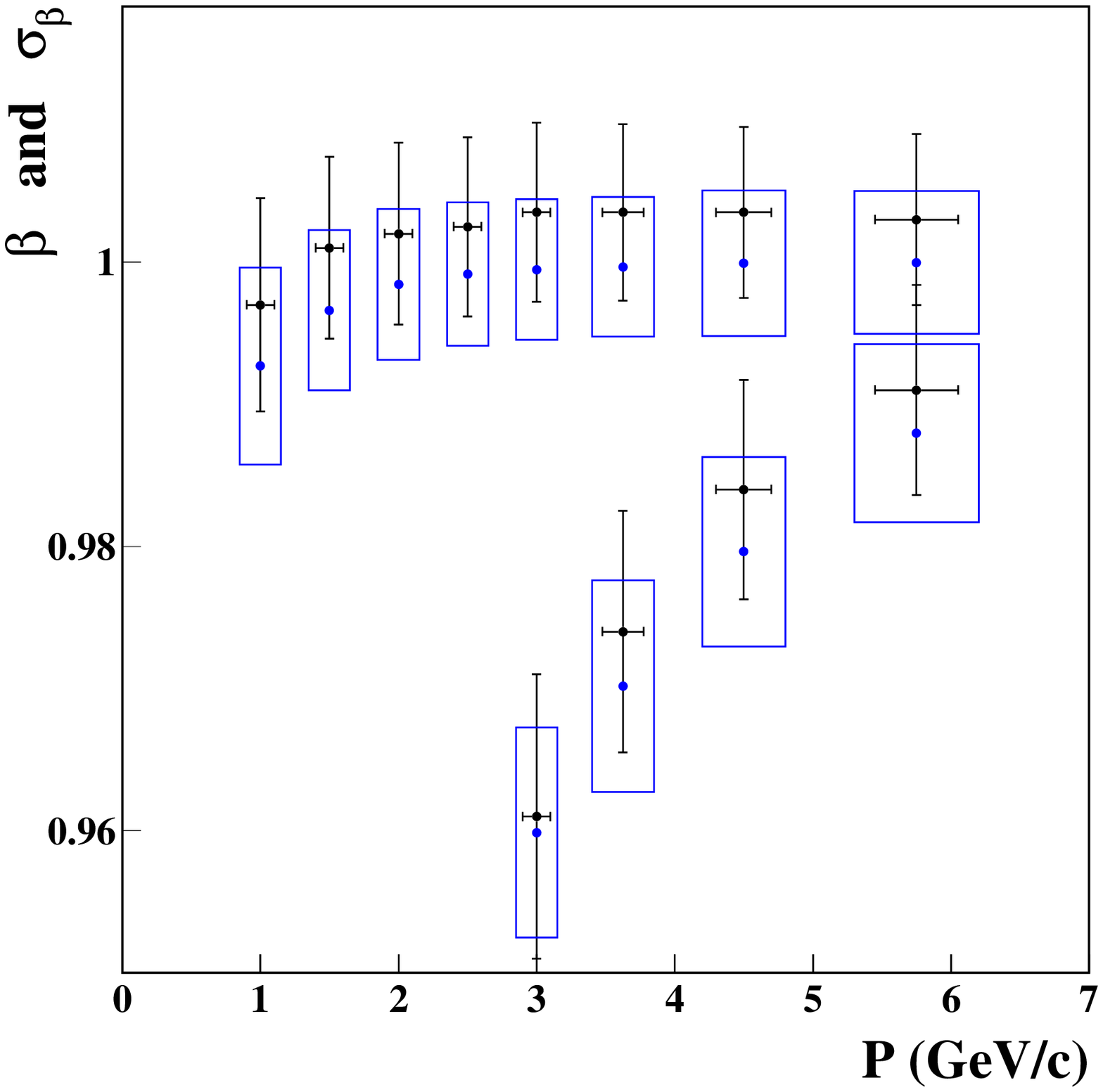,width=0.45\textwidth}
  \caption{\label{fig:beta_true_rec}
    Mean value of $\beta$ (points) and its standard deviation 
    (error bars and rectangles) 
    as a function of momentum for protons and pions. 
    On the left, MC events only, obtained 
    using only reconstructed quantities to 
    select pure samples (points with error bars)
    and using the MC information
    about the true particle type (rectangles). 
    On the right, points with error bars correspond to
    similarly selected events from real data, 
    and the rectangles correspond to equivalent MC events.
    The centres of the rectangles are indicated with points.  
    } 
  \end{center}
\end{figure}

Figure~\ref{fig:beta_true_rec} shows the mean value and the standard deviation
of $\beta$ for pions and protons. The left panel of
Fig.~\ref{fig:beta_true_rec} 
shows the comparison between the values obtained with the method explained above 
and the ones obtained using the MC information about the true particle type.   
In both cases, the same sample of Monte Carlo data has been used.  
The observed good agreement confirms that the proposed technique
does not bias the pion and proton parameters. 
The result of the application of this method to the data
is shown in the right panel of Fig.~\ref{fig:beta_true_rec}. This plot also 
shows the comparison between data and Monte Carlo when using the same method for the 
selection of pure particle samples.  
We observe a global shift of about 0.003 in $\beta$ which can be attributed 
either to a TOFW misalignment of 3~\cm along the $z$ direction or to a time offset 
of 0.1~\ns between the TOFW and the BTOF system. 
In order to account for this difference the analysis uses the
PDFs based on the 
$\beta$ distribution measured in the data directly when treating 'data-events'
and the PDFs based on 
the Monte Carlo distribution when treating 'Monte Carlo events'.
The TOFW PDFs are parametrized as the sum of 
a dominant Gaussian function and a term 
accounting for non-Gaussian outliers, normalized to the observed effect.

\subsubsection{Cherenkov response}
\label{subsubsec:che}

\begin{figure}[tbp!]
\begin{center}
\includegraphics[width=0.49\textwidth]{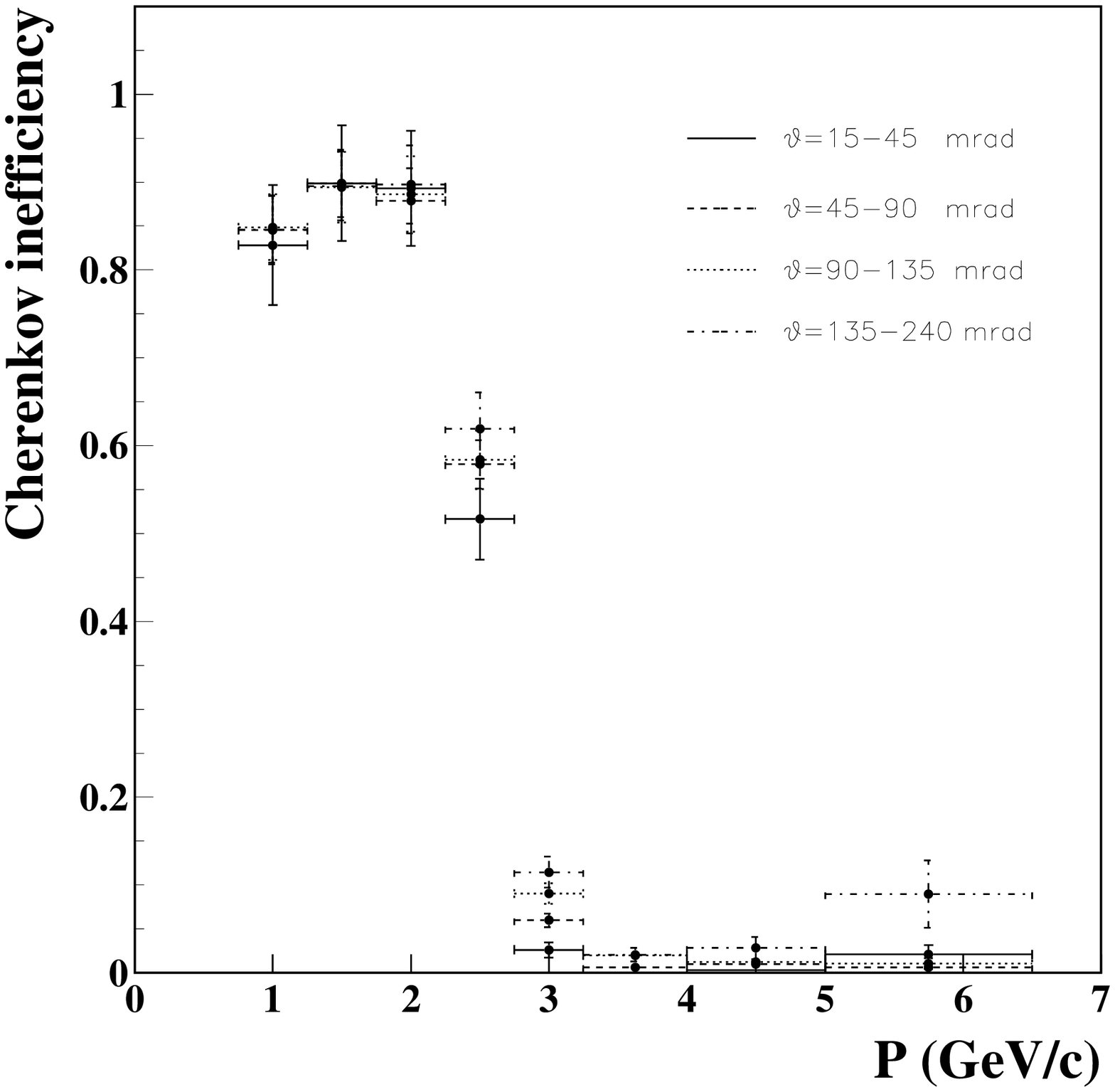}
\includegraphics[width=0.49\textwidth]{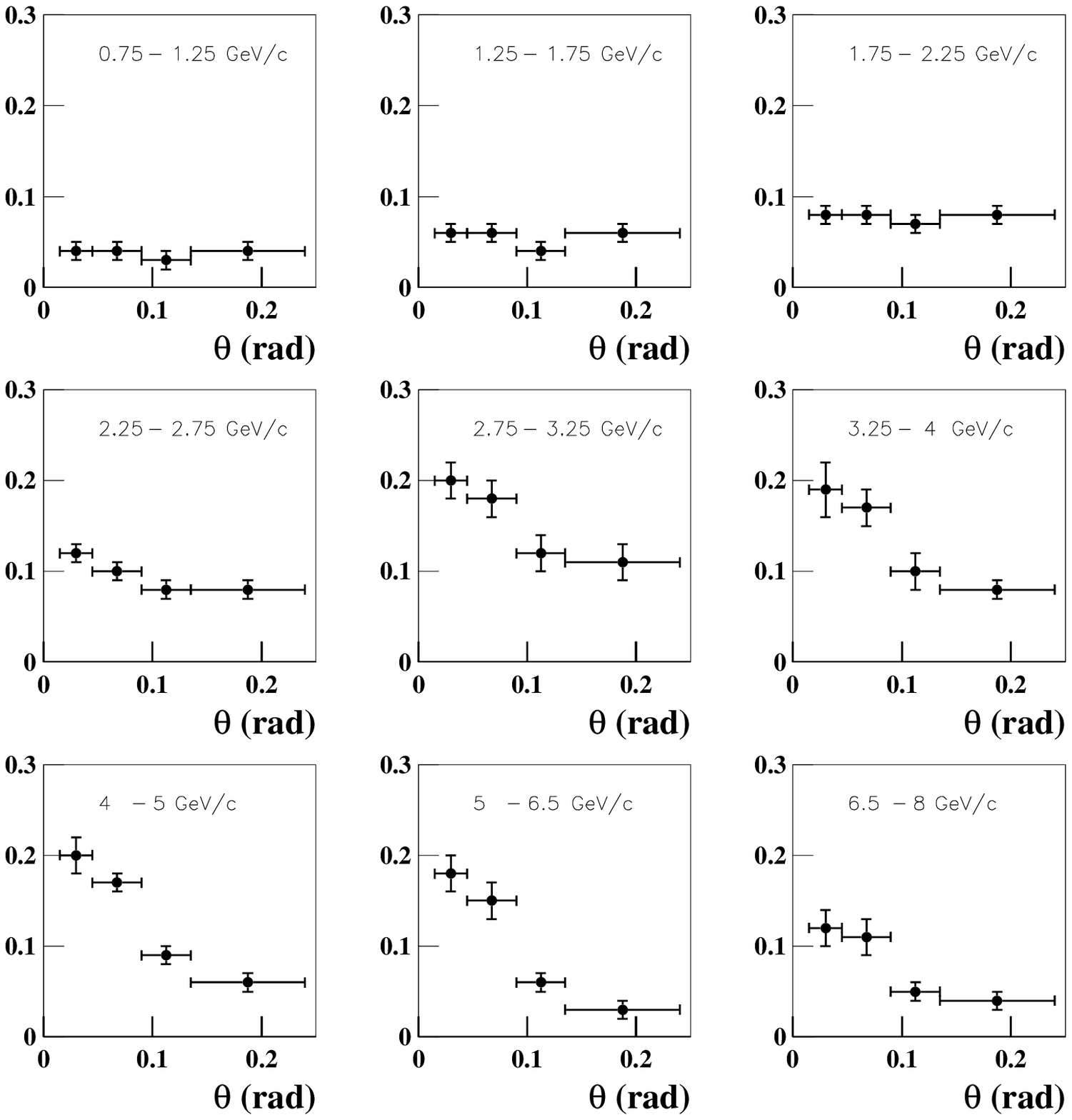}
\end{center}
\caption{Left panel: CHE pion inefficiency as a function of the
momentum for different angular regions. Right panel: CHE proton
efficiency as a function of the angle for different momentum regions. 
}

\label{fig:chePDF}
\end{figure}

The Cherenkov detector is used digitally in this analysis: a signal is
accepted if the number of photoelectrons is larger than 2.
To obtain the CHE PDFs for a given momentum and angular bin, 
the fraction of true pions (protons) with negative (positive) signal
in the CHE is measured. 
The number of true particles of a given type is obtained with a technique similar 
to the one used for the TOFW detector, in this case by applying a strict cut
to the TOFW measurement (see Ref.~\cite{ref:pidPaper} for the details).  
Fig.~\ref{fig:chePDF} (left panel) shows the CHE inefficiency 
for particles of negative charge (essentially pions) as a function of 
the reconstructed momentum and angle. The asymptotic 
inefficiency for pions, {\em i.e.} above a momentum of 3.5~\GeVc, is
estimated to be $(1.0 \pm 0.5)\%$.        

In the momentum range studied no signal is expected in the
CHE for protons. However, in a fraction of events, the
reconstruction algorithm wrongly associates the CHE hit from a pion or
an electron to the proton and consequently, 
a fraction of 
protons has a non-negligible amount of associated photoelectrons. 
This is a potential source of background (as well as of
pion inefficiency), particularly important at high momentum, where the TOFW 
is not applicable. The efficiency of the CHE for protons has 
been measured as a function of momentum and angle. The results are shown in  
Fig.~\ref{fig:chePDF} (right panel). This non-zero 
efficiency is fully taken into account in this analysis as explained 
in Ref.~\cite{ref:pidPaper}. 
The CHE PDFs are given in Fig.~\ref{fig:chePDF}.

\subsection{The pion-proton PID estimators}
\label{sec:pid_prob}

The assignment of a particle type $i=\pi,{\mathrm{p}}$ to a reconstructed track 
is based on a cut $P_i>P_{\mathrm{cut}}$ in a PID estimator $P_i$, the so-called combined 
PID probability, which is built by using Bayes' theorem:
\begin{equation}
P_i = P(i | \beta, N_{\mathrm{phe}}, p, \theta) = 
\frac
    { P(\beta, N_{\mathrm{phe}} | i, p, \theta)            \cdot P(i, p, \theta) } 
    { P(\beta,N_{\mathrm{phe}} | \pi, p, \theta)           \cdot P(\pi, p, \theta)  
   +  P(\beta,N_{\mathrm{phe}} | {\mathrm{p}}, p, \theta)  \cdot P(\mathrm{p}, p, \theta)} \ ,
\label{eq:bayes_1}
\end{equation}
where $P(\beta, N_{\mathrm{phe}} | i, p, \theta)$ are the TOFW--CHE PDFs described above, and 
$P(i, p, \theta)$ is the prior, describing the a priori probability
that a particle passing the event and  
track selection criteria (Secs.~\ref{subsec:event_sel} and \ref{subsec:track_sel}) 
is of type $i$ and has momentum $p$ and polar angle $\theta$.  

Several simplifications have been made to the general formula~(\ref{eq:bayes_1}).  
They result in a slightly less efficient particle type selection, 
which implies a larger error on the PID corrections. 
However, the additional error is negligible when added in quadrature 
to the dominant non-PID errors.  
As a first approximation, the PDFs that enter the PID estimator were
averaged over all angles.  
As second approximation, equally probable priors were used, so that
they cancel. In this case only the information from  
the current track is used to build the PID estimator. 
The PID estimator built with no priors 
does not have a full probabilistic meaning and cannot be used directly to estimate 
the particle yields. Instead, the raw pion and proton yields must be corrected by the 
efficiencies and purities obtained by the application of the cut $P_i>P_{\mathrm{cut}}$, 
as will be described in Sec.~\ref{sub:extract}. Finally, the TOFW--CHE PDFs can be 
factorized in independent TOFW and CHE PDFs, as explained before. 
The final PID estimator is then represented by the formula:

\begin{equation}
P_i = P(i | \beta, N_{\mathrm{phe}}, p) = 
\frac
    { P(\beta | i, p) \cdot P(N_{\mathrm{phe}}| i, p)}
    { P(\beta | \pi, p) \cdot P(N_{\mathrm{phe}}| \pi, p) + P(\beta |
      {\mathrm{p}}, p) \cdot P(N_{\mathrm{phe}}| {\mathrm{p}}, p)} 
    \ .
\end{equation}

\section{Calculation of the cross-section}
\label{sec:xsec}

The double-differential cross-section for the production of a particle of 
type $\alpha$ can be expressed in the laboratory system as:

\begin{equation}
{\frac{{d^2 \sigma_{\alpha}}}{{dp_i d\theta_j }}} =
\frac{1}{{N_{\mathrm{pot}} }}\frac{A}{{N_A \rho t}}
M_{ij\alpha i'j' \alpha'}^{-1} \cdot
{N_{i'j'}^{\alpha'}} 
\ ,
\label{eq:cross}
\end{equation}

where $\frac{{d^2 \sigma_{\alpha}}}{{dp_i d\theta_j }}$
is expressed in bins of true momentum ($p_i$), angle ($\theta_j$) and
particle type ($\alpha$),  
and the terms on the right-hand side of the equation are:
\begin{itemize}
\item 
  $N_{i'j'}^{\alpha'}$ 
  is the number of particles of observed type $\alpha'$ in bins of reconstructed
  momentum ($p_{i'}$) and  angle ($\theta_{j'}$). These particles must satisfy the event, track and PID 
  selection criteria, explained below. This is the so called `raw yield'. 
\item 
  $ M_{ij\alpha i'j' \alpha'}^{-1}$ 
   is a correction matrix which corrects for finite efficiency and resolution of the detector. It 
   unfolds the true variables $ij\alpha$ from the reconstructed variables $i'j'\alpha'$ and corrects 
   the observed number of particles to take into account effects such as 
   reconstruction efficiency, acceptance, absorption, pion decay, tertiary production,
   PID efficiency and PID misidentification rate. 
\item
  $\frac{A}{{N_A \rho t}}$ 
  is the inverse of the number of target nuclei per unit area
  ($A$ is the atomic mass,
  $N_A$ is the Avogadro number, $\rho$ and $t$ are the target density and thickness).
\item 
  $N_{\mathrm{pot}}$ is the number of incident protons on target.
\end{itemize}

The summation over reconstructed indices $i'j'\alpha'$ is implied in
the equation.
It should be noted that the experimental procedure bins the result
initially in terms of the angular variable $\theta$, while the
final result will be expressed in terms of the solid angle $\Omega$. 
Since the background from misidentified protons in the pion sample is
not negligible, the pion and proton 
raw yields ($N_{i'j'}^{\alpha'}$, for
$\alpha'=\pi, \mathrm{p}$) 
have to be measured
simultaneously.

For practical reasons, the background due to interactions of the primary
proton outside the target (called `Empty target background') 
has been taken out of the correction matrix $M^{-1}$. Instead, a
subtraction term 
is introduced in Eq.~\ref{eq:cross}: 
\begin{equation}
\frac{{d^2 \sigma_{\alpha}}}{{dp_i d\theta_j }}  = 
%
\frac{1}{{N_{\mathrm{pot}} }}\frac{A}{{N_A \rho t}}
%
M_{ij\alpha i'j' \alpha'}^{-1} \cdot  \left[ 
N_{i'j'}^{\alpha'}(\mathrm{T}) - 
N_{i'j'}^{\alpha'}(\mathrm{E})
\right] \ ,
\label{eq:cross-complete}
\end{equation}
where (T) refers to the data taken with the aluminium target and (E) 
refers to the data taken with no target (Empty target). 

The event, track and particle identification selection criteria will
be described first, then
the method used to obtain the cross-section and
each of the corrections will be described in more detail.

\subsection{Event Selection}
\label{subsec:event_sel}

\begin{figure}[tb]
\begin{center}
\epsfig{file=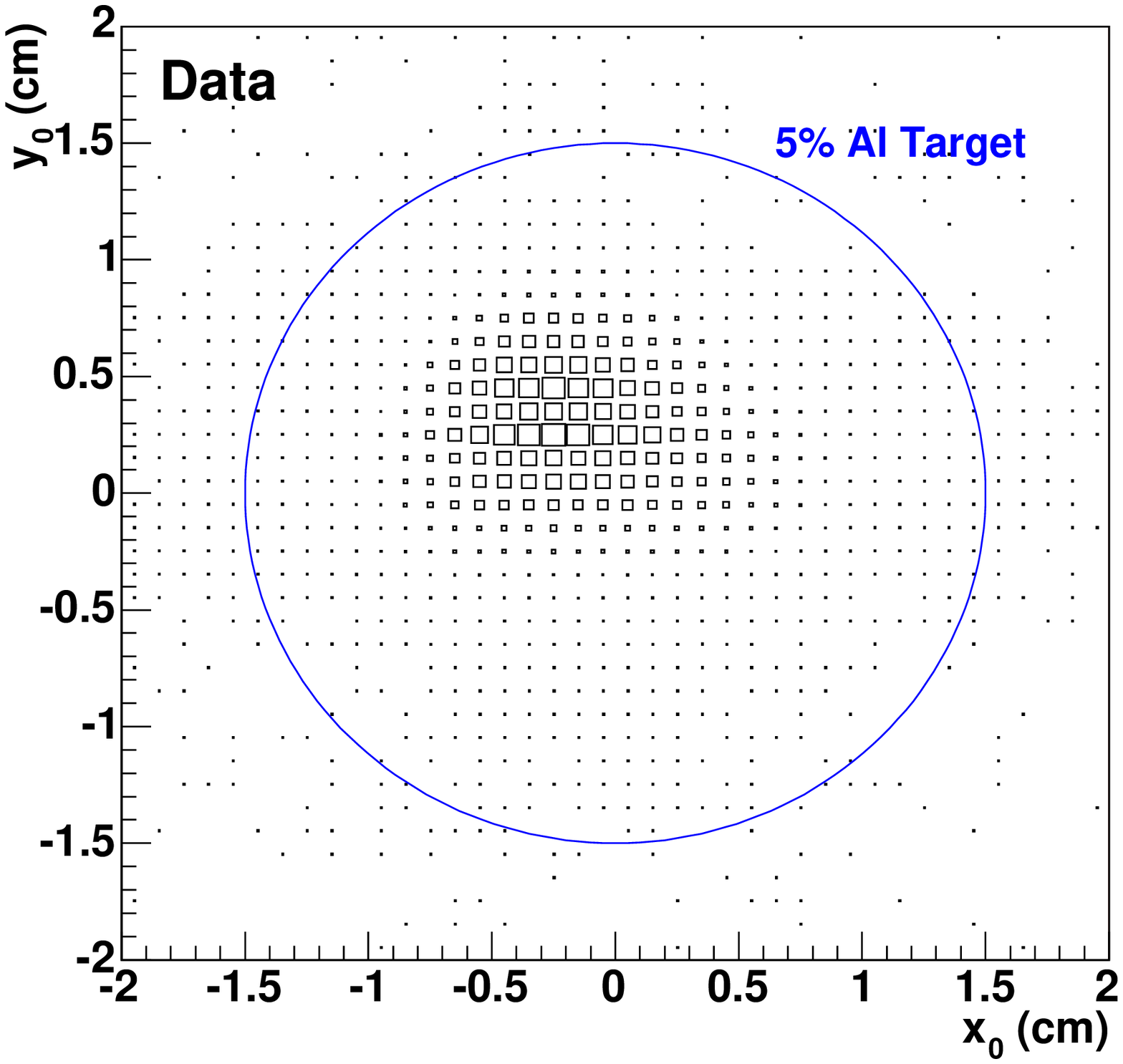,width=0.41\textwidth}
\epsfig{file=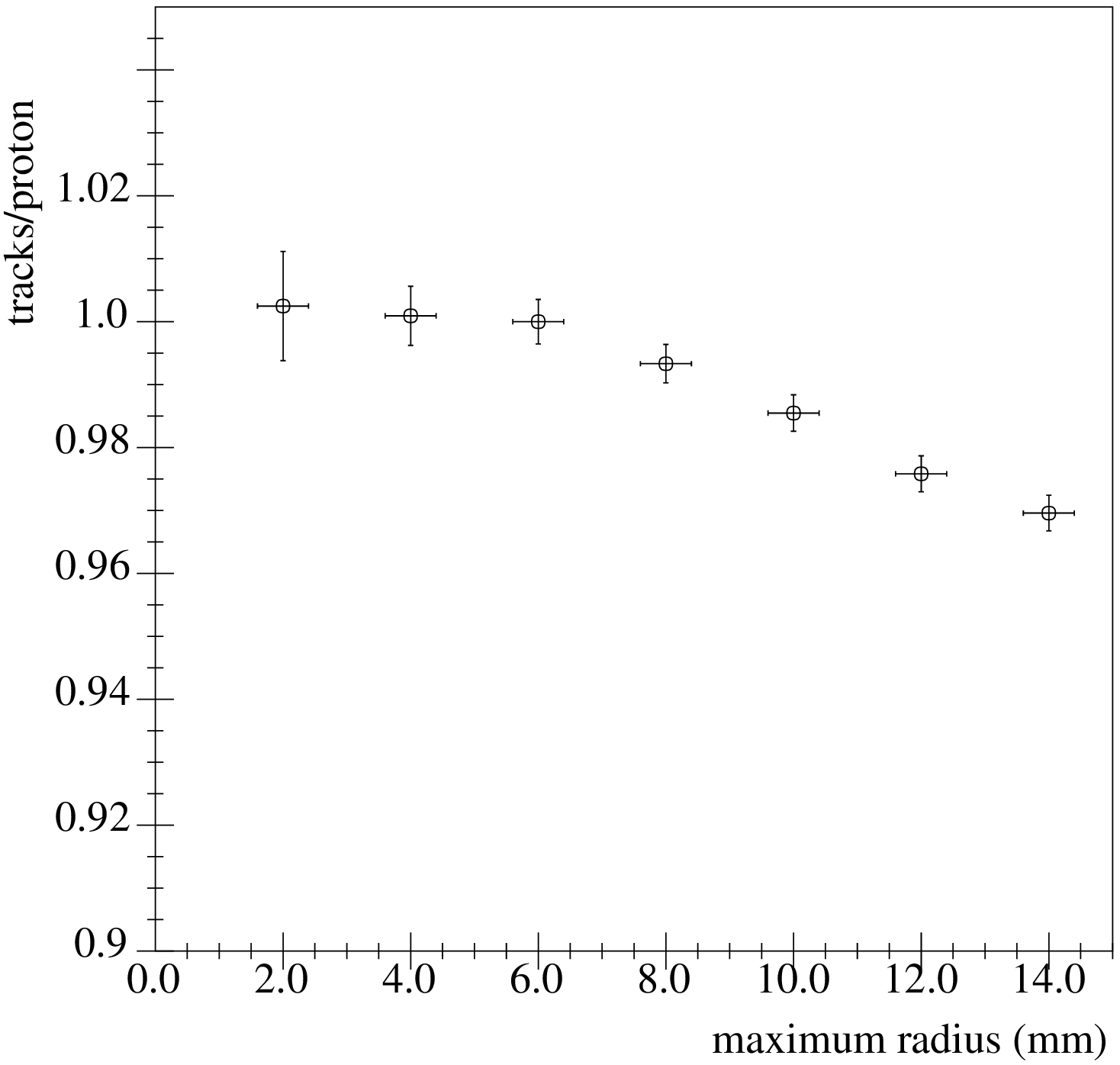,width=0.37\textwidth}
\caption{Left panel: reconstructed position $(x_0, y_0)$ of beam
  particles at the reference $z_0$ plane of the target.
  The circle gives the position and size of the target.
  Right panel: observed number of tracks per incident proton as a
  function of the maximum accepted measured radius of incidence of the
  beam track.  The ratio is normalized to unity at 6~mm.
}
\label{fig:mwpc}
\end{center}
\end{figure}

In the 12.9~\GeVc beam protons are selected by vetoing particles which
give a signal in any of the beam Cherenkov detectors. 
Only particles which give a good timing signal in all
three beam timing detectors, leave a single track in the MWPCs,
and are not seen in the halo detectors are accepted.
A good timing measurement is defined as a set of three hits, one in
each of the timing detectors, with their relative time difference
consistent with a beam particle.
The distribution of the position of beam particles 
extrapolated to the target is shown
in Fig.~\ref{fig:mwpc} (left panel).  
The size of the target is indicated by a circle.  Only particles
extrapolated within a radius of 10~\mm are accepted.
By evaluating the number of tracks reconstructed in the spectrometer
as a function of the extrapolated impact point of the MWPC track to
the target, it was determined that ($1.5 \pm 0.5$)\% of the proton
tracks selected according to these criteria miss the target,
as shown in Fig.~\ref{fig:mwpc} (right panel).
A correction for this loss has been applied.
The MWPC track was required to have a measured direction within
5~\mrad of the nominal beam direction to further reduce halo particles.
The purity of this proton sample is estimated to be better than 99.5\%.

Prior to those cuts, the beam particle was required to satisfy the
trigger conditions described in Section~\ref{sec:beamtrigger}.
Applying the above selection cuts to the 12.9~\GeVc aluminium
5\%~$\lambda_{\mathrm{I}}$ target data set and the 12.9~\GeVc empty
target data set results in the
total statistics listed in Table~\ref{tb:al5events}.
The total number of protons on target ($N_{pot}$ in Eq.~\ref{eq:cross}) 
listed in the table
is exactly the number to be used in the overall  
normalization of the cross-section results, and is known to better
than 1\%.  The total number of protons on target is counted using
prescaled `beam' triggers 
that were continuously recorded at the time of data taking.
The trigger condition for the prescaled beam triggers only involved a
simple coincidence of scintillators in the beam line with no
requirement of an interaction in the target.
Using subsamples of the triggers the prescale
factor was checked to confirm it had its preset value 1/64.
Because the selection criteria for beam protons used in event analysis
and prescaled beam proton events are the same, the efficiencies for
these cuts cancel, and the total normalization can be known 
without additional systematic uncertainty.

Events to be used in the analysis must also contain one or more hits
in the forward trigger plane (FTP).  
 
\begin{table}[tbp!] 
\begin{center}
\begin{tabular}{ l  c  c} \hline
\bf{Data Set} & \bf{Al 5\% 12.9 \bfGeVc} & \bf{12.9 \bfGeVc Empty Target}\\ \hline
    Protons on target                & 17,954,688 & 4,769,408 \\
    Total events processed           & 4,710,609 & 771,330\\ 
    Events with accepted beam proton & 3,404,372 & 547,838 \\ 
    Prescaled triggers with accepted beam proton & 280,542 & 74,522 \\ 
    FTP triggers                     & 2,087,732 & 225,639 \\
    FTP trigger rate = ( FTP triggers / pot ) & 0.116 & 0.047 \\
    Total good tracks                & 209,929  & 11,704\\ 
\end{tabular}
\caption{Total number of events in the 12.9~\GeVc aluminium
  5\%~$\lambda_{\mathrm{I}}$ target 
  and empty target data sets, and the number of protons on target as
  calculated from the prescaled trigger count.} 
\label{tb:al5events}
\end{center}
\end{table}

\subsection{Track Selection}
\label{subsec:track_sel}

The recorded events have been processed according to 
the track selection criteria listed below:  

\begin{itemize}
\item The VERTEX2 track momentum is measured (see Section~\ref{subsec:resolution}). 
\item A track segment in NDC2 or in the back-plane is used in track reconstruction.
\item Number of hits in a road around the track in NDC1 $\geq$ 4 (this is
  applied to reduce non-target interaction backgrounds).
\item The average $\chi^2$ for hits with respect to the track in NDC1 $\leq$ 30, 
\item Number of hits in the road around the track in NDC2 $\geq$ 6
  (this is applied to reduce background of tracks not coming from the
  target).
\item The track has a matched TOFW hit.
\end{itemize}

 The result of applying these cuts to the entire 12.9~\GeVc
aluminium 5\%~$\lambda_{\mathrm{I}}$ and empty target data sets is  
listed in Table~\ref{tb:al5events}.

In addition, geometrical cuts are applied.
As described in Sec.~\ref{sec:tracking-eff} for positive $\theta_x$
the efficiency is momentum dependent.
This region is avoided in the analysis by defining the fiducial volume
as $-210 \le \theta_x < 0$~\mrad (thus, only particles in the negative half of the
bending plane of the dipole are accepted) and $-80< \theta_y < 80$~\mrad. 
The
restricted acceptance in $\theta_y$ is imposed to avoid edge effects of the
dipole, possible fringe effects in the magnetic field, etc. 
Since the behaviour of the spectrometer is calibrated with 
beam particles (at $\theta_y = 0$) the analysis restricts $\theta_y$ to a
rather small region around the horizontal mid-plane of the
spectrometer. 
In order to avoid a correction for the acceptance of the FTP-trigger
and to avoid background from beam protons, cross-sections are given for
$\theta > 30 \ \mrad$.

\subsection{PID selection}

\begin{figure}[tb]
  \begin{center}
  \epsfig{file=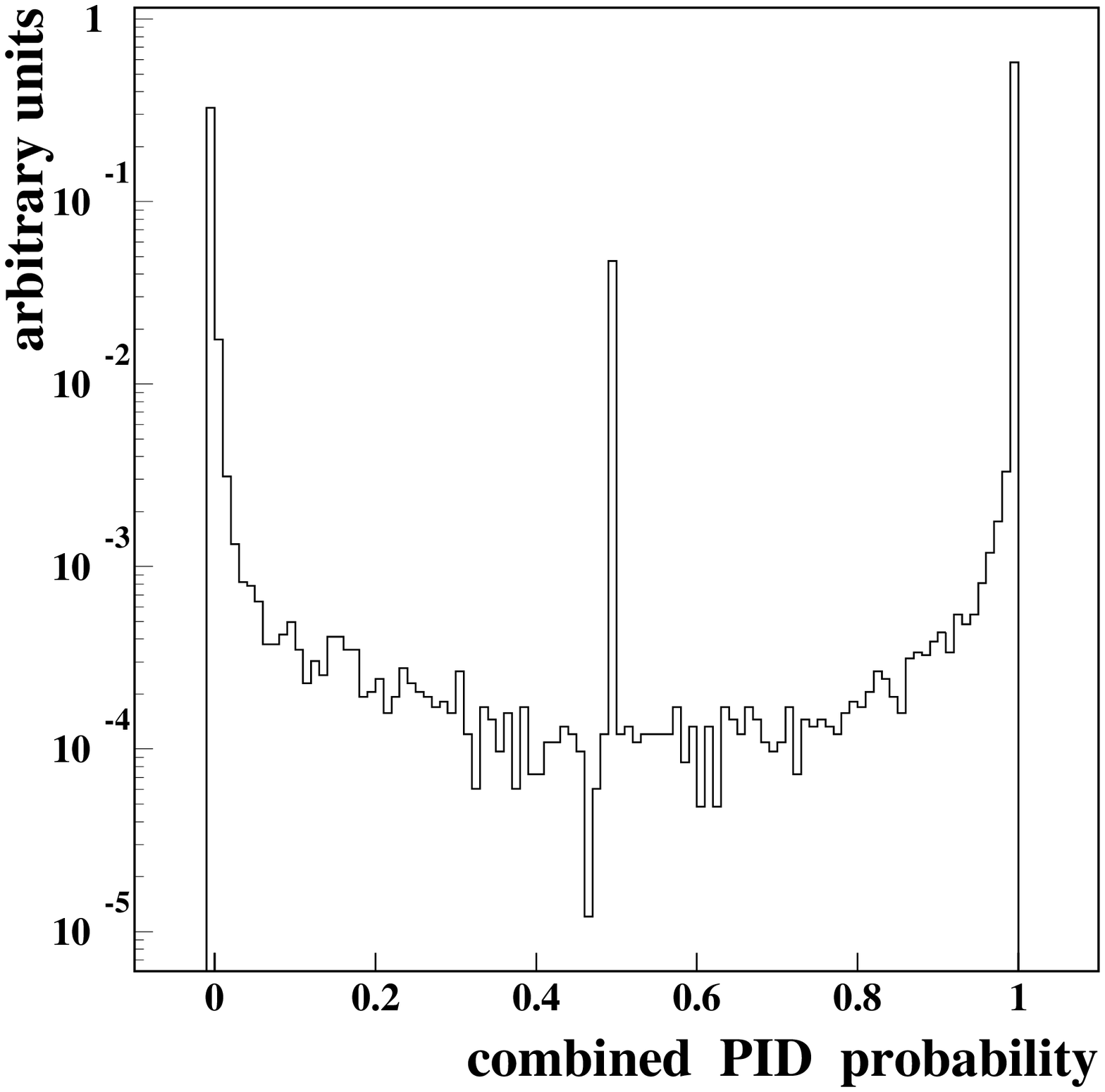,width=0.4\textwidth}
  \epsfig{file=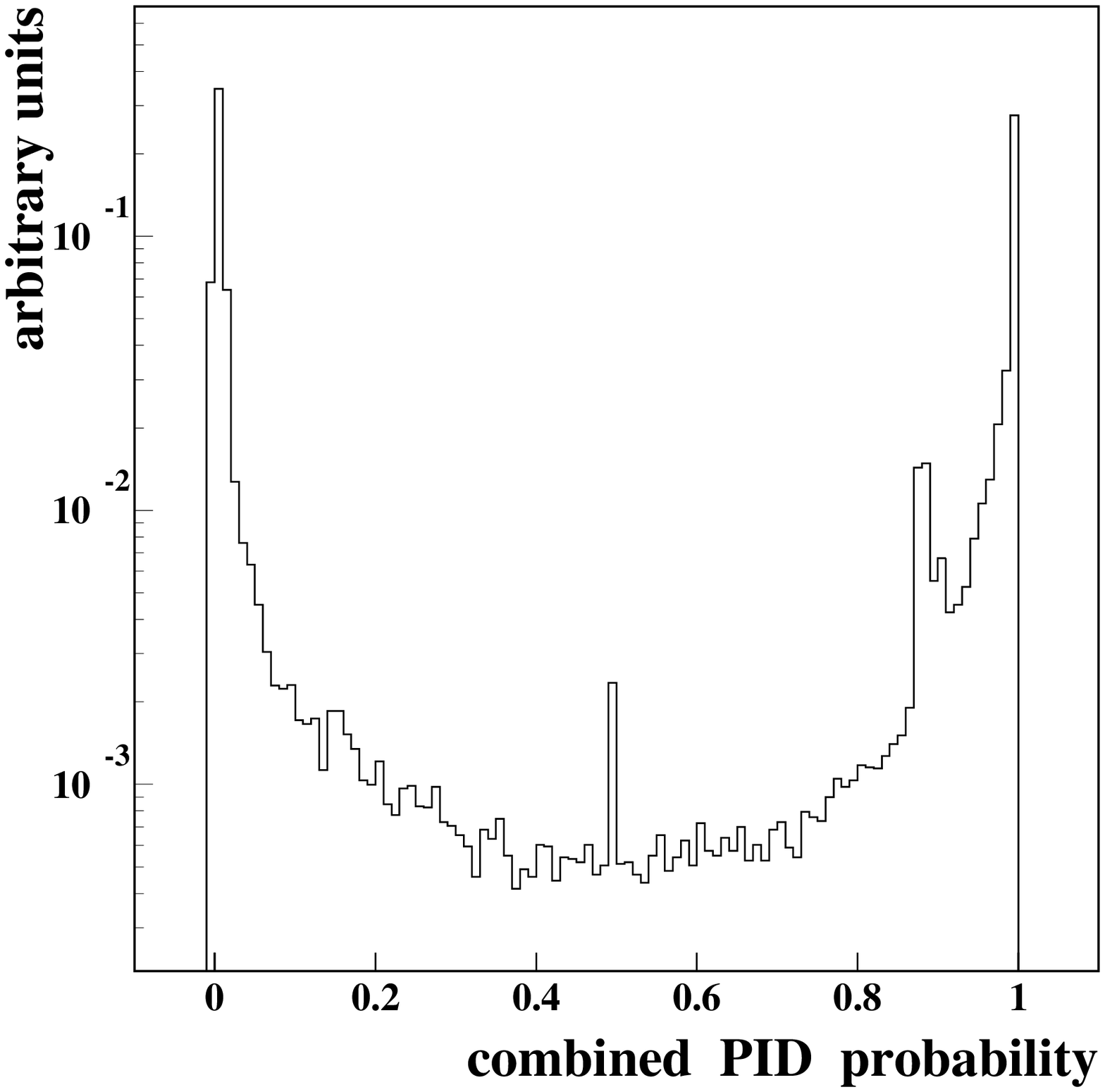,width=0.4\textwidth}
  \caption{\label{fig:pion_prob}  
  Combined pion probability (logarithmic scale) for positive particles
  passing the e-veto cut.  
  The left panel shows positive particles below CHE threshold. The right panel 
shows positive particles above CHE threshold. 
  The meaning of the different spikes is explained in the text. 
	}
  \end{center}
\end{figure}

Particle identification criteria (described in 
Sec.~\ref{sec:pid}) are applied to the tracks passing the event and track 
selection criteria.  First the e-veto cut is applied to reject
electrons and then a cut in the PID estimator is applied to
distinguish between pions and protons.  
Figure~\ref{fig:pion_prob} shows the combined pion probability 
(PID estimator for pions) for positive particles passing the e-veto cut. 
A large 
population of particles in the low probability region is attributed
to a contribution from protons. 
The small peaks at 0.5 correspond 
to particles which leave no useful information   
in either the TOFW or the Cherenkov. 
The peak near 0.9 in the right panel 
corresponds to particles which leave no useful information 
in the TOFW (presumably being non-Gaussian outliers) but which give   
a positive signal in the CHE.

It is found that the optimal cut to select pions with high efficiency and purity 
is $P_{\pi} > 0.6$. 
The cut is set at a value of probability
where the track population is low, and thus the result is not sensitive to small
changes in the exact value. 
Protons are selected by
the condition $P_{\mathrm{\pi}} < 0.4$ (equivalent to $P_{\mathrm{p}}
> 0.6$). 

\subsection{The Atlantic and UFO analyses}
\label{sec:two-analyses} 

Two complementary analyses have been performed with the aim of  
checking internal consistency, and 
checking for possible biases in the respective procedures.
The first, called Atlantic\footnote{
`Atlantic' for Analysis of Tracks at Low ANgle with Tof Id and
Cherenkov id}, simplifies the problem of unfolding by decomposing the 
correction matrix of Eq.~(\ref{eq:cross}) into distinct independent
contributions, which are computed mostly using the data themselves. 
The second analysis, called UFO (from UnFOlding), performs 
a simultaneous unfolding of $p$, $\theta$ and PID, with a correction
matrix $M^{-1}$ computed mainly using the Monte Carlo.

The UFO procedure uses an iterative Bayesian technique, described in
Ref.~\cite{dagostini},
in order to unfold the measured distribution.
The central assumption of the method is that the 
probability density function in the physical parameters (`physical
distribution') can be approximated by a histogram with bins
of sufficiently small width.
A population in the physical distribution of events in a
given cell $ij\alpha$ generates a distribution in the measured variables,
$M_{ij\alpha i'j'\alpha'}$, 
where the indices $ij\alpha$ indicate the binning in the physical angular, 
momentum and PID variables, respectively, and  $i'j'\alpha'$ the
binning in the measured variables. 
Thus the observed distribution in the measurements can be
represented by a linear superposition of such populations.
The task of the unfolding procedure consists then in finding the
number of events in the physical bins for which the predicted
superposition in the measurement space gives the best description of
the data.

In order to predict the population of the migration matrix element 
$M_{ij\alpha i'j'\alpha'}$, the resolution, efficiency
and acceptance of the detector are obtained from the Monte Carlo.
This is a reasonable approach, since the Monte Carlo
simulation describes most of these quantities correctly
(see 
Section~\ref{sec:tracking-eff}). Where some deviations
from the control samples measured from the data are found, 
the data are used to introduce (small) corrections to the Monte Carlo.

Although some corrections are common to both approaches,
large differences between the results of these two analyses 
would indicate inconsistencies in the simplifications adopted by Atlantic
for unfolding, the hypothesis of correct Monte Carlo description of the detector
on which UFO is based, or both. As it turns out, the analyses
are consistent within the overall systematic error, reinforcing  
our confidence in the correctness of the results
presented here. For clarity, in the rest of this paper 
only the Atlantic analysis will be discussed.

\section{The Atlantic analysis}
\label{sec:atlantic}

As discussed in section~\ref{subsec:resolution}, 
both the momentum and angular
resolution are small compared with the binning of the cross-section. Migration
effects are, therefore, small. In particular, angular migration 
can be neglected.
In addition, kinematic migration is almost decoupled from pion--proton
PID migration. As explained in Sec.~\ref{sec:pid} electron and kaon ID has
been decoupled from the dominant pion--proton ID so that
electron and kaon correction factors are diagonal in the PID variables. 
With the above considerations the correction matrix $M^{-1}$ can be
written as:
\begin{equation}
M_{ij\alpha i'j' \alpha'}^{-1} =
%
(M^{id}_{ij;\alpha \alpha'})^{-1} \cdot
%
\varepsilon_{ij\alpha'}^{-1} \cdot
%
(M^\theta_{jj'})^{-1} \cdot 
%
(M^p_{ii'})^{-1}  \ ,
\label{eq:cross-matrix}
\end{equation}
where again reconstructed indices are indicated with a prime. 
The corrections are applied in the order from right to left as they
appear in the equation. 
The symbols in Eq.~(\ref{eq:cross-matrix}) have the following meaning: 
\begin{list}{\mbox{}}{}
\item
$\varepsilon_{ij\alpha'}^{-1}$
is the collection 
of factors applying the corrections that are diagonal in the PID indices: 
reconstruction efficiency, acceptance, physical loss of particles (absorption,
decay), background from tertiary interactions, e-veto  efficiency 
and kaon subtraction;
\item
$(M^p_{ii'})^{-1}$
is the simplified unfolding matrix correcting for the momentum
smearing which only depends of the indices $i$ and $i'$ representing
the true and reconstructed momentum bins, respectively;
\item
$(M^\theta_{jj'})^{-1}$
is the identity matrix, representing the assumption that the smearing
effect in the angular measurement is negligible; and 
\item
$(M^{id}_{ij;\alpha \alpha'})^{-1}$
is the matrix which corrects for pion--proton PID inefficiency and migration, which
is diagonal in  $i,i'$ and $j,j'$, but built of two--by--two
sub-matrices, each different and non-diagonal in the PID 
variables $\alpha,\alpha'$.
\end{list}
The diagonal efficiency correction 
\begin{equation}
\varepsilon_{ij\alpha'}^{-1} =
w_{ij}^{\mathrm{recon}} \cdot
w_{ij}^{\mathrm{acc}} \cdot
w_{ij\alpha'}^{\mathrm{absorption}} \cdot
w_{ij\alpha'}^{\mathrm{tertiaries}} \cdot
\eta_{ij\alpha'}^{\mathrm{K}} \cdot
\eta_{ij\alpha'}^{\mathrm{e}} 
\label{eq:cross-efficiency}
\end{equation}
is composed of the following factors:
\begin{list}{\mbox{}}{}
\item
$w_{ij}^{\mathrm{recon}}$
the correction for the overall reconstruction efficiency;
\item
$w_{ij}^{\mathrm{acc}}$
the correction for the acceptance;
\item
$w_{ij\alpha'}^{\mathrm{absorption}}$
the correction for the loss of particles due to absorption and decay; 
\item
$w_{ij\alpha'}^{\mathrm{tertiaries}}$
the correction for the background of tertiary particles generated by
the secondaries produced in the target;
\item
$\eta_{ij\alpha'}^{\mathrm{K}}$
is the factor correcting for the kaon background; and
\item
$\eta_{ij\alpha'}^{\mathrm{e}}$
is the factor correcting for the effects of the electron veto. 
\end{list}

The first two corrections are the same for pions and protons while
the latter four also depend on the particle type. 
It is worth noting that the efficiency correction is expressed in terms 
of the true momentum and angle, and in terms of the reconstructed 
particle type ($\alpha'$). This is because these corrections are applied 
before PID unfolding, as explained below. 

As advanced in Sec.~\ref{subsec:kine}, some of the above corrections 
are computed as a function of $p$, $\theta_x$ and $\theta_y$, while some 
others are directly expressed in the final variables 
$(p, \theta)$. In the first case, the transformation to polar 
coordinates ($p$, $\theta$) is done integrating over all $\theta_x$ 
and $\theta_y$ resulting in a given $\theta$ bin. 
In particular, the four first corrections of Eq.~(\ref{eq:cross-efficiency}), 
denoted by $w$, 
are computed as a function of $(p, \theta_x, \theta_y)$.    

Each of the above corrections will be described in the 
sections below.

\subsection{Reconstruction and acceptance corrections}

The correction for the total reconstruction efficiency, requiring a momentum measured 
and a matched TOFW hit (computed in Sec.~\ref{sec:tracking-eff}), is introduced 
as a weight 
$w_{ij}^{\mathrm{recon}} = [\varepsilon^{\mathrm{recon}}(p,\theta_x, \theta_y)]^{-1}$. 
 
\begin{figure}[tbp!]
\begin{center}
\includegraphics[height=2.0in]{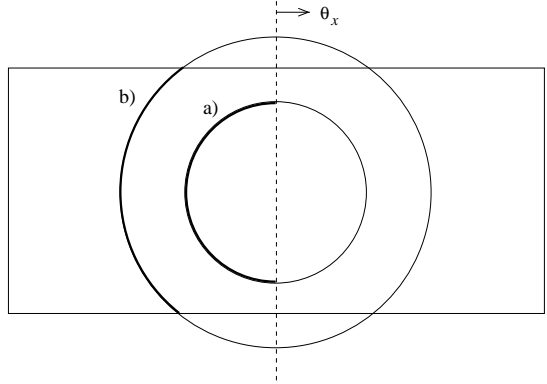}
\end{center}
\caption{Sketch of the forward detector showing the two 
kinds of acceptance corrections, 
a) for $\theta < \theta_{y}^{\mathrm{cut}}$ 
where $\varepsilon^{\mathrm{acc}}$ = 0.5, and b) 
for $\theta > \theta_{y}^{\mathrm{cut}}$ 
where $\varepsilon^{\mathrm{acc}}$ is given by Eq.~(\ref{eq:acc}).}
\label{fig:acc}
\end{figure}

It is necessary to correct for the restricted definition of fiducial volume.
Inside the $\theta_y$ acceptance
(that is, below the vertical cutoff at $\theta_{y} = \pm 80 \ \mrad$)
the correction is a simple factor of 2 due to the fact that tracks
with $\theta_x > 0$ are not used.
For values of $\theta$ above the $\theta_{y}$ cutoff,
the correction is:
\begin{equation}
\label{eq:acc}
\varepsilon^{\mathrm{acc}}(p,\theta_x, \theta_y) = \frac{1}{\pi} \cdot 
\arcsin \left( \frac{\tan(\theta_{y}^{\mathrm{cut}})}{\tan(\theta)} \right)
\ ,
\end{equation} 
describing the part of the circle which is inside the acceptance.
Figure~\ref{fig:acc} shows a sketch depicting the two forms 
of geometrical acceptance and the origins 
of the correction factors listed above. 
The acceptance correction is then applied as a weight defined in
Eq.~(\ref{eq:cross-efficiency}),
$w_{ij}^{\mathrm{acc}} = [\varepsilon^{\mathrm{acc}}(p,\theta_x, \theta_y)]^{-1}$.
The above corrections are independent of the particle type.

\subsection{Corrections for absorption, decay, and secondary interactions}

\begin{figure}[htbp]
\begin{center}
\includegraphics[width=\textwidth]{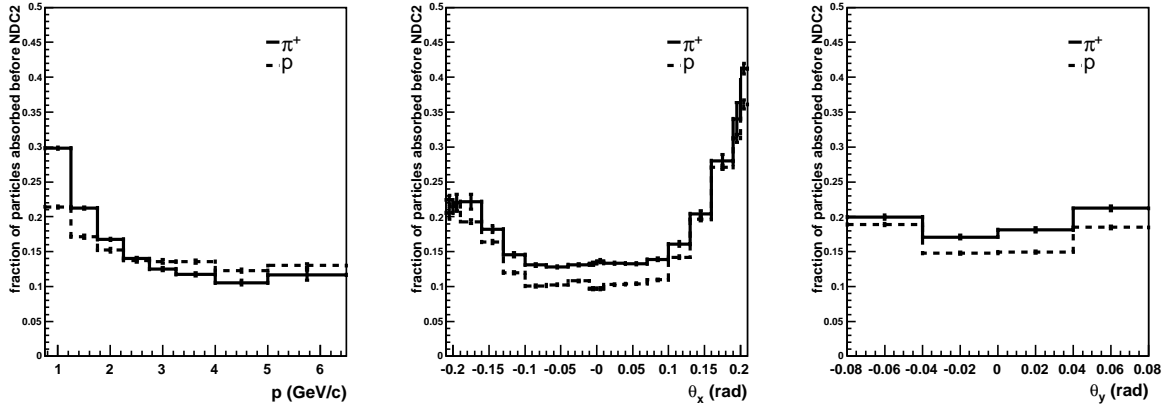}
\end{center}
\caption{Absorption and decay rate as a function of 
kinematic variables, $p$, $\theta_{x}$, and $\theta_{y}$, 
at production for positively charged particles emanating from the vertex.
Solid lines are pions, dashed lines are protons.   
Left panel: As a function of $p$. Central panel: as a function
of $\theta_x$. Right panel: As a function
of $\theta_y$.  The  acceptance effects for positive
$\theta_x$  due to particles hitting the dipole walls are clearly
visible. 
This effect is momentum dependent, as seen in the left panel. }
\label{fig:abs}
\end{figure}

An additional correction to consider is the absorption and decay
of secondary pions (protons) in the materials of the detector components
upstream of the magnet 
which prevent them from reaching the downstream region of the
detector. These missing particles are not considered by the
reconstruction efficiency described above.  
In some cases, {\em e.g.} pion decay, tertiaries are reconstructed as
part of the original track.
A correction is applied to take these cases into account.
The overall absorption and decay rate 
is determined using the Monte Carlo, as a three-dimensional function
$\varepsilon_{\alpha}^{\mathrm{absorption}}(p,\theta_x,\theta_y)$, whose
projections are shown in Fig.~\ref{fig:abs}.   
The overall effect is between 10\% and 30\% depending on $p$ and
$\theta_x$. This effect is verified to be correct within 10\% of its
magnitude using beam particles.
The `absorption' correction is then applied as a weight introduced in
Eq.~(\ref{eq:cross-efficiency}),
$w_{ij\alpha}^{\mathrm{absorption}}= [1-\varepsilon_{\alpha}^{\mathrm{absorption}}(p,\theta_x,\theta_y)]^{-1}$  
for both particle types separately.

\begin{figure}[htbp]
\begin{center}
\includegraphics[width=\textwidth]{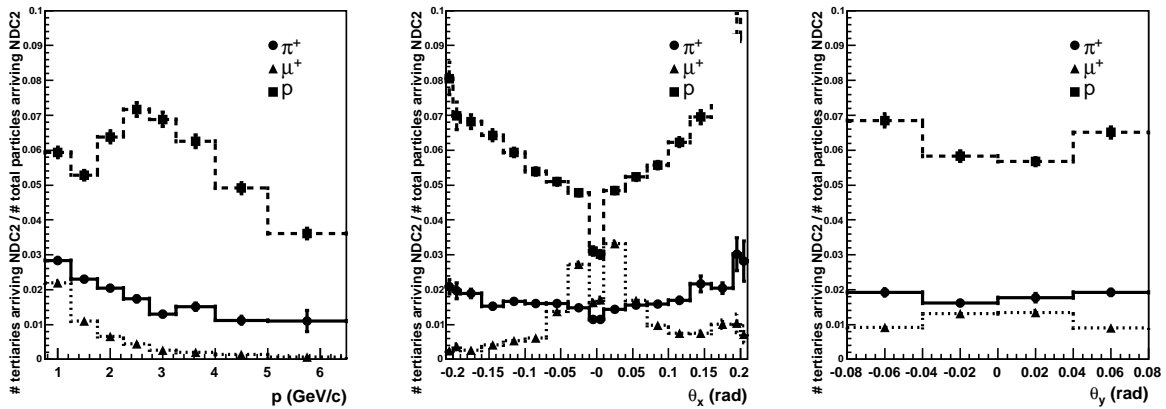}
\end{center}
\caption{Tertiary particle rate as a function of 
kinematic variables, $p$, $\theta_{x}$, and $\theta_{y}$, 
at production for positively charged particles emanating from the vertex.
Circles are pions, squares are protons, triangles are muons 
falsely identified as pions.   
Left panel: As a function of $p$. Central panel: as a function
of $\theta_x$. Right panel: As a function
of $\theta_y$.  }
\label{fig:ter}
\end{figure}

A correction of opposite sign to the one above stems from positively
charged particles which are not produced in the primary interaction
between the incident proton and the target nucleus. These 
{\em tertiary particles} (`tertiaries')
can come directly from the target (nuclear re-interactions, which is a
small effect, since the target is only 5\% $\lambda_{\mathrm{I}}$ in
thickness) or from the 
region outside the target area.
Another different background is due
to particles produced by the interactions of the primary proton with
material outside the target; this is corrected by taking data with
empty target settings, and is described in
Section~\ref{sec:empty-target}. 
The correction for tertiaries is
also computed with Monte Carlo, as a three-dimensional function
$\varepsilon_{\alpha}^{\mathrm{tertiaries}}(p,\theta_x,\theta_y)$, shown in 
Fig.~\ref{fig:ter}.   
The overall effect is between 2--3\% (for pions) and 7\% 
(for protons). In addition
one needs to correct for pion decay resulting in muons which tend
to be collinear with the original pion. 
The `tertiary' correction is then applied as a weight introduced in
Eq.~(\ref{eq:cross-efficiency}),
$w_{ij\alpha}^{\mathrm{tertiaries}} =  1-\varepsilon_{\alpha}^{\mathrm{tertiaries}}(p,\theta_x,\theta_y)$.  
The correction is model-dependent, and has been assigned a systematic
uncertainty of 100\%.

The above corrections are computed separately for true pions and protons. However,  
for practical reasons, they are applied before PID unfolding assuming 
that they correspond to reconstructed pions and protons (hence the
index $\alpha'$ in Eq.~\ref{eq:cross-efficiency}). The bias introduced  
by this approximation is negligible since pion--proton mixing 
is very small ( $<5\%$), as demonstrated in Fig.~\ref{fig:pid_eff_matrix_prob0.6_data}, 
and these corrections are either small ($<7\%$ for tertiaries) or similar for both 
particle types (the absorption is similar while the decay of pions
introduces a relatively small correction). Thus, the maximum bias 
would be of the order of $0.4\%$.

\subsection{Empty target subtraction}
\label{sec:empty-target}

Several additional background sources need to be corrected for.
The `empty target' background is defined as the particles accepted by
the selection criteria which are generated by interactions of the
primary protons outside the target.  The effect of this background is
measured experimentally to a good approximation by taking data without
placing the target in its holder.

\begin{figure}[htbp]
\begin{center}
\includegraphics[width=0.85\textwidth,width=0.5\textheight]{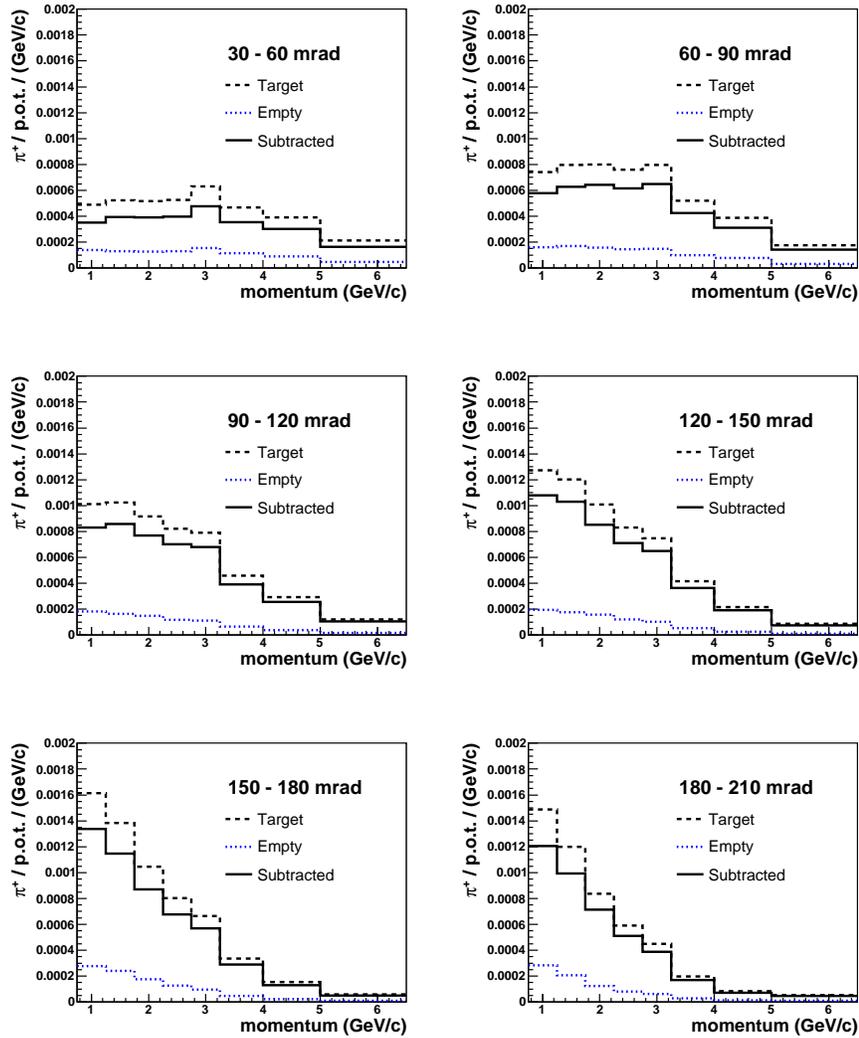}
\end{center}
\caption{
Positive pion yields, defined as pions per p.o.t (protons on target) as a function of
momentum.
Empty target yields are subtracted from target 
yields to remove backgrounds.  The subtraction ranges 
from negligible to approximately 20\%.}
\label{fig:empty}
\end{figure}

The corrections described in the previous section are applied 
to both 5\%~$\lambda_{\mathrm{I}}$ target and empty target data sets.  
The empty target yield $N_{i'j'}^{\alpha'}(\mathrm{E})$ undergoes
similar corrections to the yields measured with the target in place. 
The corrected empty target yields are then 
subtracted bin--by--bin from the corrected yields measured with target to remove
this background.
The relative normalization of the data with target and the empty
target data is calculated using the number of protons on target
accepted in the prescaled beam trigger.
The overall subtraction is approximately 20\% 
as shown in Fig.~\ref{fig:empty}.
The approximation used in this approach is to assume that the target
itself does not influence the primary proton beam.  
To first order, this assumption introduces an error of 5\% on the
subtraction, given by the interaction length of the target.

\subsection{Corrections for electron-veto and kaon background}

\begin{figure}[htbp]
  \begin{center}
  \epsfig{file=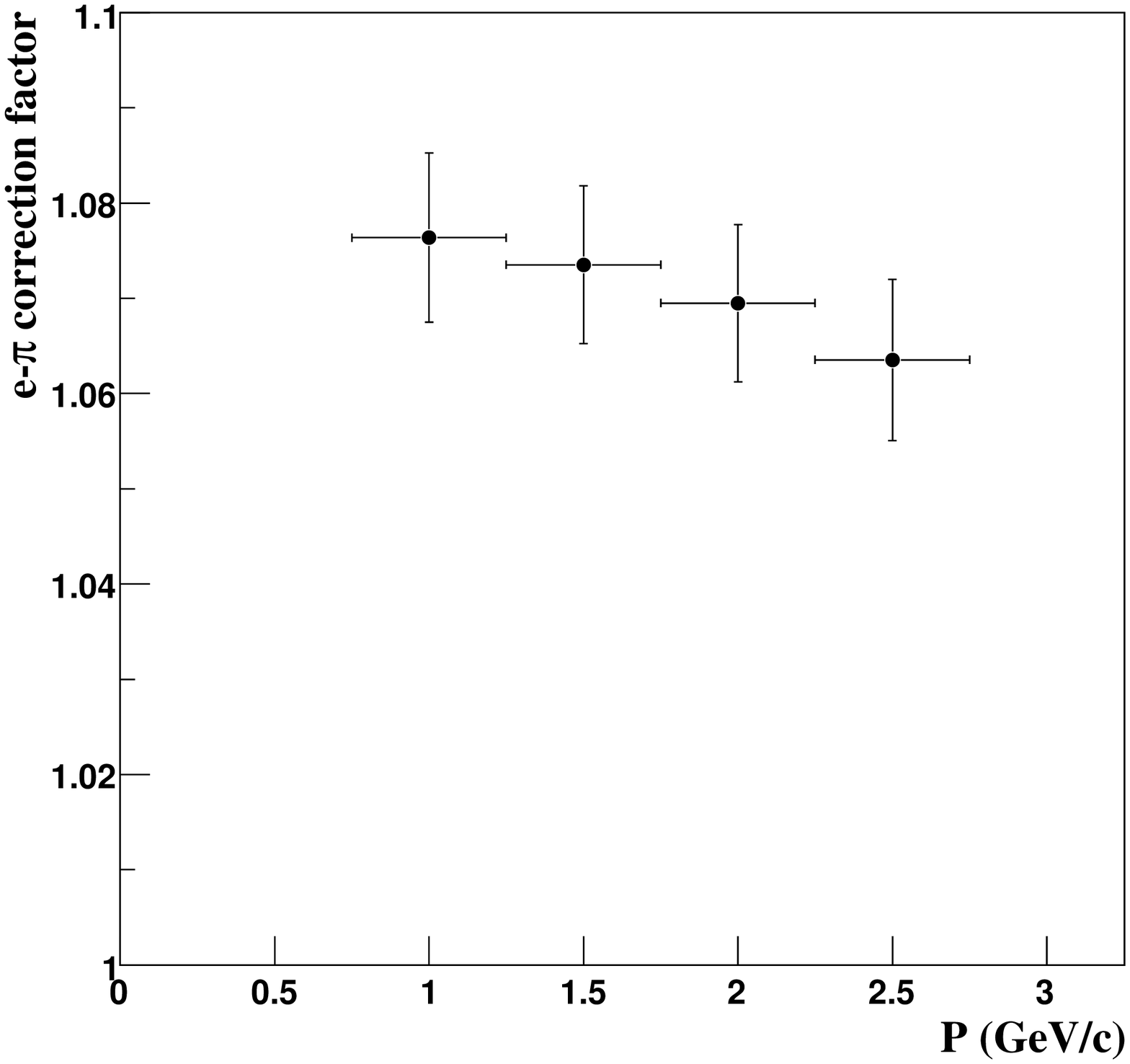,width=0.39\textwidth,height=0.238\textheight} 
  \epsfig{file=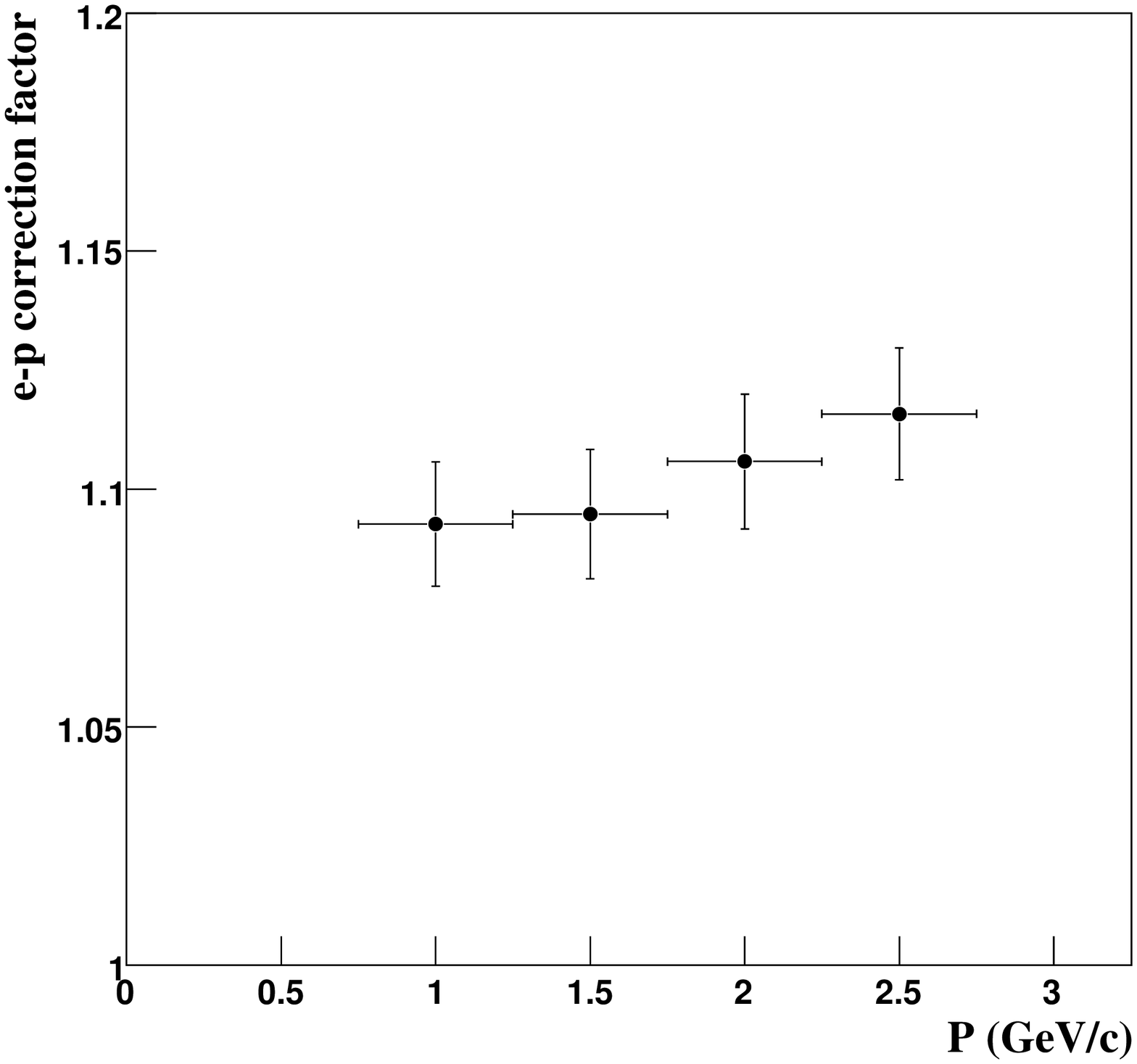,width=0.39\textwidth,height=0.238\textheight} 
  \epsfig{file=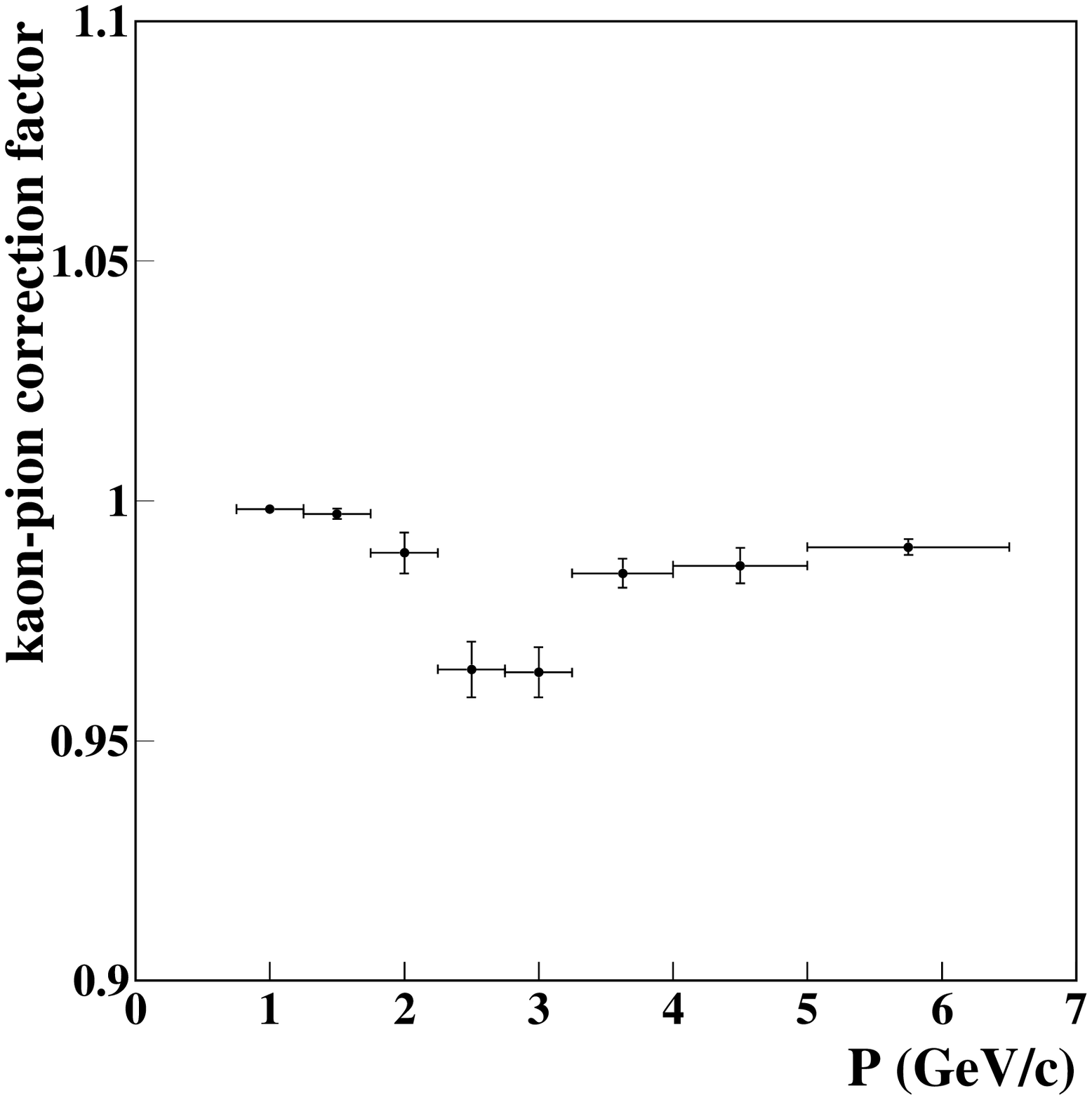,width=0.4\textwidth,height=0.233\textheight} 
  \epsfig{file=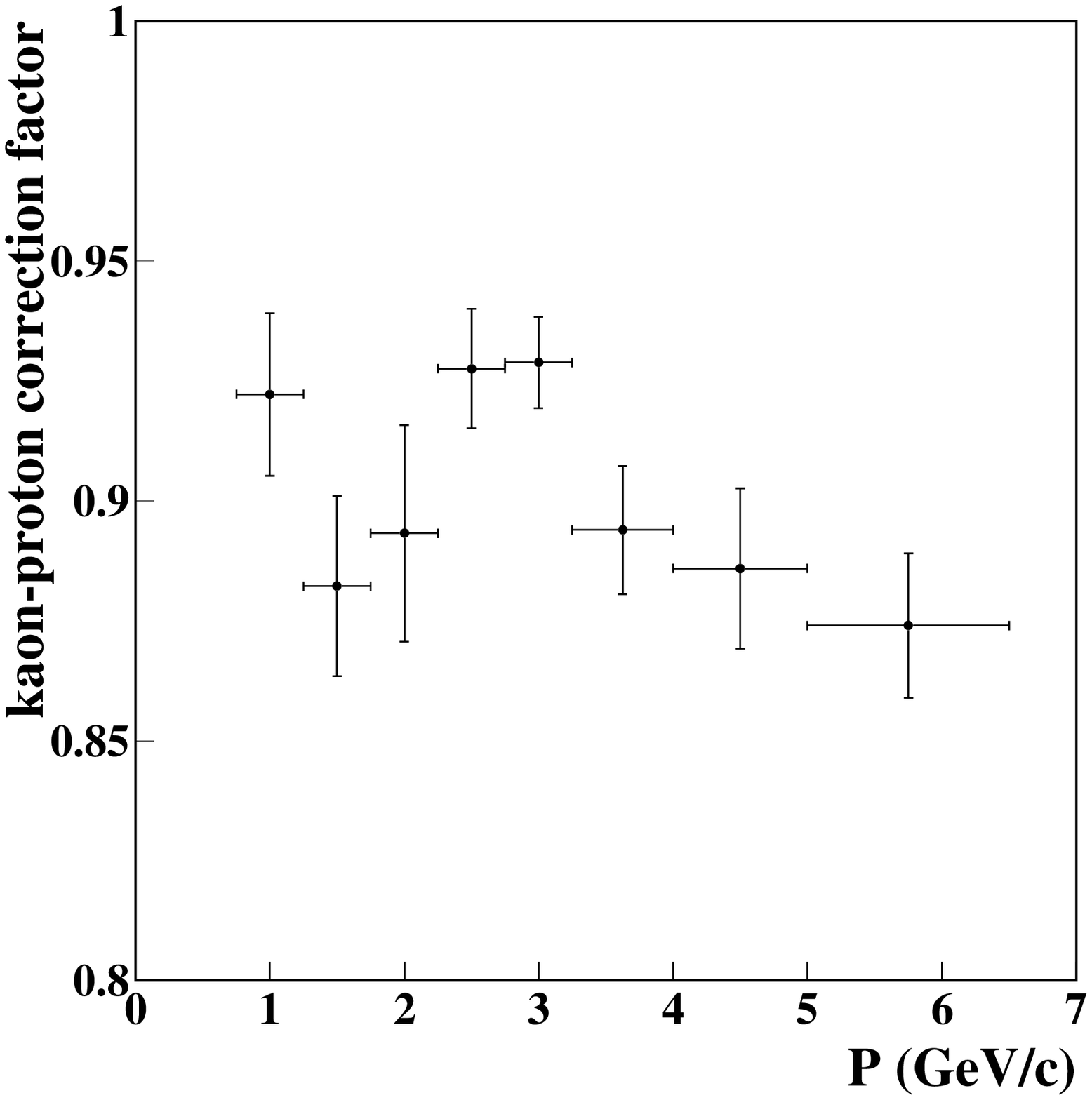,width=0.4\textwidth,height=0.233\textheight} 
  \caption{\label{fig:e_corr}
    On the top, pion (left) and proton (right) e-veto correction factors 
    averaged over angles. On the bottom the kaon-pion (left) and kaon-proton (right)  
    correction factors in real data for $\theta=90~\mrad$. 
     } 
  \end{center}
\end{figure}

As discussed, the electron and kaon hypotheses are not considered by 
the PID selection algorithm. 
Kaons produced in the target and identified as either pions or protons
have to be subtracted.
Electrons are rejected by applying the CHE veto described above.
This veto introduces a loss of efficiency for pions and protons.
Multiplicative corrections are applied to the raw yields of pions and
protons to compensate for these two effects as shown in
Fig.~\ref{fig:e_corr}.   

The electron veto, described in Section~\ref{sec:pid}, has some effect 
on pion and proton efficiency (around 7--10\% of the pions and protons
also give a signal in CHE below pion threshold, due to spatially
associated electrons, {\em e.g.} the emission of hard $\delta$-rays)
which has been measured using both the Monte Carlo and the 
data (pure 
electron and hadron control samples were selected for that purpose using
ECAL~\cite{ref:pidPaper}). 
The weights $\eta_{ij\alpha'}^{\mathrm{e}}$
(Eq.~(\ref{eq:cross-matrix})) are the inverse of the 
efficiencies for pions and protons to survive the electron veto
requirement, respectively.

Kaons are subtracted from the raw yields of pions and protons.
The kaon yield has being estimated from the data by fitting the 
inclusive TOFW-$\beta$ distribution to the sum of three Gaussians 
(corresponding to protons, kaons and pions). 
An example is shown in Fig.~\ref{fig:beta_pion_proton_rec_mc}. The kaon--to--pion and 
kaon--to--proton migration rates have also been determined 
from the data using the same technique as for pion--proton migration. 
The subtraction can be expressed as a correction factor
$\eta_{ij\alpha'}^{\mathrm{K}}$, introduced in 
Eq.~(\ref{eq:cross-matrix}). The correction factors for pions 
and protons are shown in the bottom panels of Fig.~\ref{fig:e_corr}. 
The correction for pions is only relevant 
($\sim 3\%$) in the region of transition between 
TOFW and CHE ($\sim 3 \, \GeVc$), while the correction for 
protons is of the order of $10\%$ in the entire phase space. 

A detailed description of the methods used to determine the 
electron and kaon correction factors can be found in~\cite{ref:pidPaper}.

\subsection{Unfolding the momentum dependence and the pion--proton yields}
\label{sub:extract}

A yield of tracks with reconstructed momentum $p_2$ falling in a bin $i'$ can be expressed 
as a superposition of tracks with true momentum in a bin $i$ ($p_{i}$).  
The coefficients of this expansion are the elements of the momentum migration matrix 
of Eq.~(\ref{eq:cross-matrix}), $M^p_{ii'}$, that is  
$n(p_2)_{i'} = M^p_{ii'} \cdot {n(p)_i}$. 
To perform the momentum unfolding each observed track with measured quantities  
($p_2$, $\theta_x$, $\theta_y$) populates several bins in a histogram 
of true variables ($p$, $\theta$) with weights $M^p_{ii'}$. This is 
mathematically equivalent to the matrix inversion of Eq.~(\ref{eq:cross-matrix}).

\begin{figure}[tbp]
  \begin{center}
    \epsfig{file=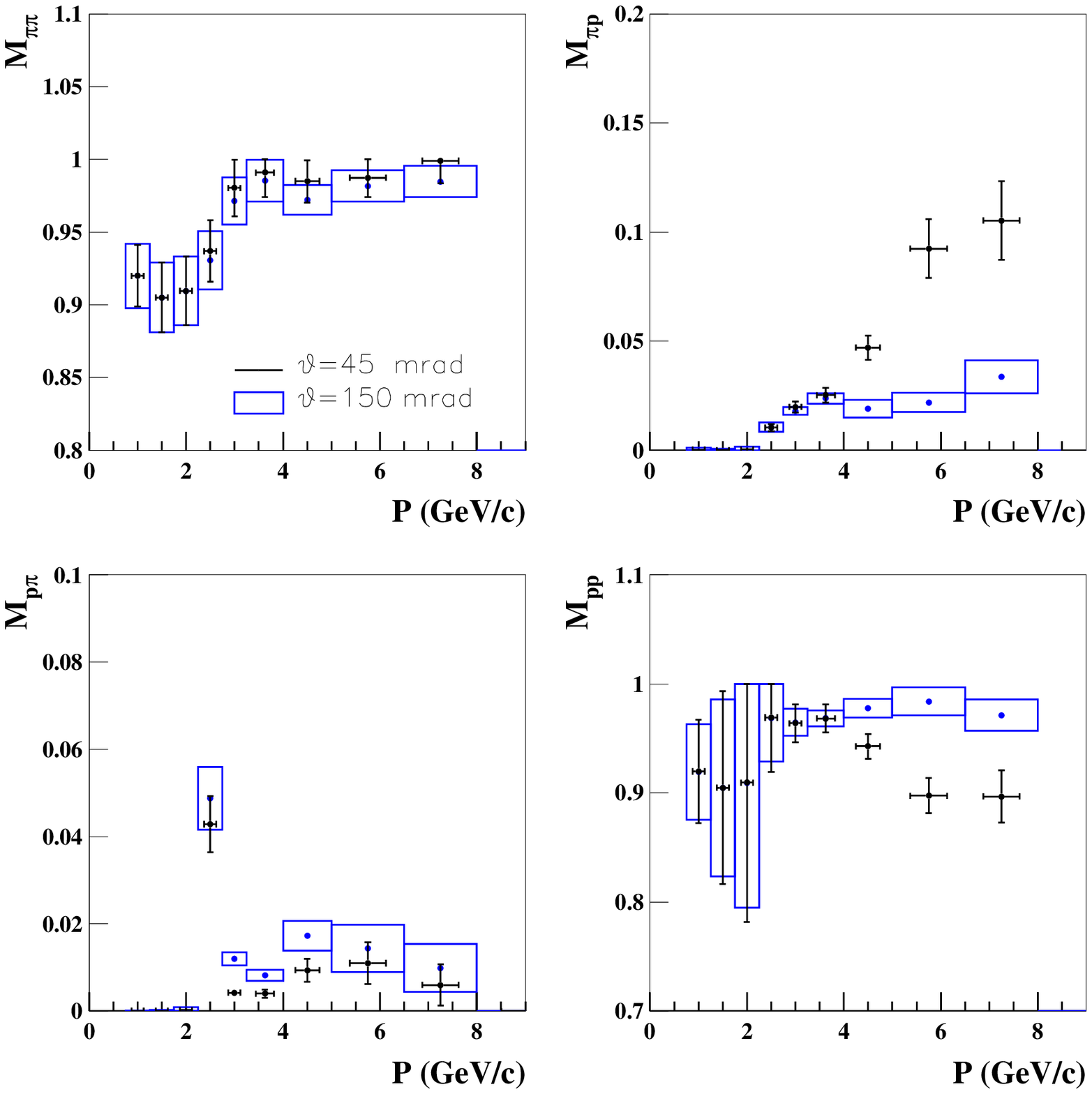,width=12cm} 
    \caption{\label{fig:pid_eff_matrix_prob0.6_data}
    The elements of the PID efficiency matrix in the data, as a
    function of momentum, for two different angles. 
The probability cut is placed at 0.6. }
  \end{center}
\end{figure}

\newcommand{\mel}{\ensuremath{M}\xspace}
The raw yield of identified pions and protons once corrected for 
all terms that are diagonal in the PID variables,
$n_{ij}^{\alpha'}=\varepsilon_{ij\alpha'}^{-1} \cdot 
(M^p_{ii'})^{-1}\cdot (M^{\theta}_{jj'})^{-1} \cdot  
[N_{i'j'}^{\alpha'}(T)- N_{i'j'}^{\alpha'}(E)]$, 
is related
to the true pion and proton yield, $n_{ij}^{\alpha}$,  
by the PID migration matrix (also called PID efficiency matrix), 
$M^{id}_{ij;\alpha \alpha'}$, 
introduced in Sec.~\ref{sec:two-analyses}.
In each bin of $p$ and $\theta$ the matrix $M^{id}$ is defined by:
\begin{equation}
\left( {\begin{array}{c}
  n^{\pi'} \\
  n^{\mathrm{p}'}   \\
\end{array}} \right)
 = \left( {\begin{array}{*{20}c}
   {\mel _{\pi \pi } } & {\mel _{\pi {\mathrm{p}}} }  \\
   {\mel _{{\mathrm{p}}\pi } } & {\mel _{\mathrm{pp}} }  \\
\end{array}} \right) \cdot
\left( {\begin{array}{c}
  n^\pi \\
  n^{\mathrm{p}}   \\
\end{array}} \right) \ ,
\label{eq:pid_eff}
\end{equation}
where the elements of the matrix, in rows, 
are the fractions of observed pions that
are true pions ($\mel_{\pi\pi}$), observed pions that are true protons
($\mel_{\pi {\mathrm{p}}}$), observed protons that are true pions
($\mel_{{\mathrm{p}}\pi}$), and observed protons that are true protons
($\mel_{\mathrm{pp}}$). 
In this experiment this matrix can be computed using the
redundancy in the data, as described in~\cite{ref:pidPaper}. 
Figure~\ref{fig:pid_eff_matrix_prob0.6_data}  shows 
the elements of $M^{id}$ (and the corresponding errors), as a function of 
momentum,  for different angular intervals.

Then, the true yields can be computed by solving the system of
linear equations given by Eq.~(\ref{eq:pid_eff}). 
The covariance matrix of the true yield vector, $n^\alpha$, is computed by
error propagation, taking into account the covariance
matrices of $M^{id}$ and  of the observed yield vector, $n^{\alpha'}$:
\[
C[n^\alpha, n^\beta] = n^{\gamma'} \cdot C[(M^{id}_{\alpha \gamma'})^{-1}, (M^{id}_{\beta\delta'})^{-1}]
      \cdot n^{\delta'} + 
       (M^{id}_{\alpha\gamma'})^{-1} \cdot C[n^{\gamma'}, n^{\delta'}] \cdot (M^{id}_{\beta\delta'})^{-1} \ ,
\]

where all indices run over the pion and proton hypotheses.

\section{Results}
\label{sec:results}

\begin{figure}[tbp!]
\begin{center}
\includegraphics[width=12cm]{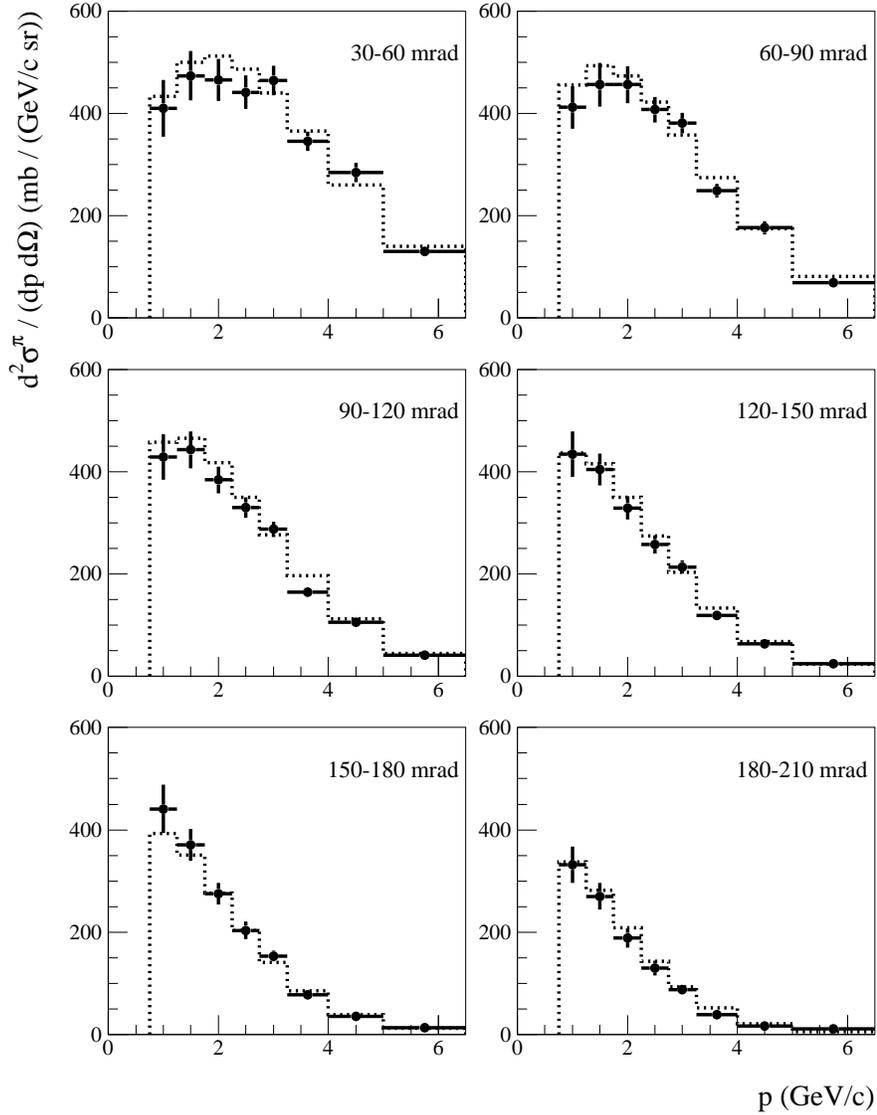}
\caption{\label{fig:harpswfit_2d}
 Measurement of the double-differential $\pi^+$ production
 cross-section in the laboratory system 
 $d^2\sigma/(dpd\Omega)$ for incoming protons of 12.9~\GeVc on an
 aluminium target as a function of pion momentum $p$, in bins
 of pion polar angle $\theta$. The data points are the measurements,
 the histogram represents the Sanford-Wang parametrization fitted to
 the data. }
\end{center}
\end{figure}

Figure \ref{fig:harpswfit_2d} and Table \ref{tab:xsec_results} show
the measurement of the double-differential cross-section for positive
pion production in the laboratory system as a function of
the momentum and the polar angle. Only diagonal errors are shown in the
plots and table (a full discussion of the error evaluation is given
below). Also shown in Figure \ref{fig:harpswfit_2d} 
is a fit to a Sanford-Wang parametrization, which will also be discussed
in this section.

\begin{table}[tbp!]
\centerline{
\begin{tabular}{|c|c|c|c|rcr|} \hline
$\theta_{\hbox{\small min}}$ &
$\theta_{\hbox{\small max}}$ &
$p_{\hbox{\small min}}$ &
$p_{\hbox{\small max}}$ &
\multicolumn{3}{c|}{$d^2\sigma^{\pi^+}/(dpd\Omega)$} 
\\
(mrad) & (mrad) & (\GeVc) & (\GeVc) &
\multicolumn{3}{c|}{(mb/(\GeVc sr))}
\\ \hline
 30 &  60 & 0.75 & 1.25 & $410$ & $\pm$ & $56$ \\
    &     & 1.25 & 1.75 & $473$ & $\pm$ & $49$ \\
    &     & 1.75 & 2.25 & $465$ & $\pm$ & $41$ \\
    &     & 2.25 & 2.75 & $441$ & $\pm$ & $33$ \\
    &     & 2.75 & 3.25 & $464$ & $\pm$ & $29$ \\
    &     & 3.25 & 4.00 & $346$ & $\pm$ & $18$ \\
    &     & 4.00 & 5.00 & $284$ & $\pm$ & $18$ \\
    &     & 5.00 & 6.50 & $129.7$ & $\pm$ & $ 8.1$ \\ \hline
 60 &  90 & 0.75 & 1.25 & $412$ & $\pm$ & $42$ \\
    &     & 1.25 & 1.75 & $456$ & $\pm$ & $42$ \\
    &     & 1.75 & 2.25 & $456$ & $\pm$ & $36$ \\
    &     & 2.25 & 2.75 & $407$ & $\pm$ & $24$ \\
    &     & 2.75 & 3.25 & $381$ & $\pm$ & $19$ \\
    &     & 3.25 & 4.00 & $249$ & $\pm$ & $13$ \\
    &     & 4.00 & 5.00 & $176$ & $\pm$ & $13$ \\
    &     & 5.00 & 6.50 & $ 68.9$ & $\pm$ & $ 6.3$ \\ \hline
 90 & 120 & 0.75 & 1.25 & $429$ & $\pm$ & $45$ \\
    &     & 1.25 & 1.75 & $442$ & $\pm$ & $36$ \\
    &     & 1.75 & 2.25 & $384$ & $\pm$ & $26$ \\
    &     & 2.25 & 2.75 & $330$ & $\pm$ & $20$ \\
    &     & 2.75 & 3.25 & $287$ & $\pm$ & $15$ \\
    &     & 3.25 & 4.00 & $164.7$ & $\pm$ & $ 9.8$ \\
    &     & 4.00 & 5.00 & $105.4$ & $\pm$ & $ 8.1$ \\
    &     & 5.00 & 6.50 & $ 41.4$ & $\pm$ & $ 4.3$ \\ \hline
120 & 150 & 0.75 & 1.25 & $434$ & $\pm$ & $44$ \\
    &     & 1.25 & 1.75 & $404$ & $\pm$ & $31$ \\
    &     & 1.75 & 2.25 & $329$ & $\pm$ & $23$ \\
    &     & 2.25 & 2.75 & $258$ & $\pm$ & $18$ \\
    &     & 2.75 & 3.25 & $213$ & $\pm$ & $13$ \\
    &     & 3.25 & 4.00 & $119.1$ & $\pm$ & $ 7.9$ \\
    &     & 4.00 & 5.00 & $ 62.8$ & $\pm$ & $ 5.2$ \\
    &     & 5.00 & 6.50 & $ 24.2$ & $\pm$ & $ 3.4$ \\ \hline
150 & 180 & 0.75 & 1.25 & $441$ & $\pm$ & $47$ \\
    &     & 1.25 & 1.75 & $371$ & $\pm$ & $31$ \\
    &     & 1.75 & 2.25 & $275$ & $\pm$ & $21$ \\
    &     & 2.25 & 2.75 & $203$ & $\pm$ & $17$ \\
    &     & 2.75 & 3.25 & $153$ & $\pm$ & $10$ \\
    &     & 3.25 & 4.00 & $ 77.5$ & $\pm$ & $ 7.1$ \\
    &     & 4.00 & 5.00 & $ 35.5$ & $\pm$ & $ 4.5$ \\
    &     & 5.00 & 6.50 & $ 13.3$ & $\pm$ & $ 1.7$ \\ \hline
180 & 210 & 0.75 & 1.25 & $332$ & $\pm$ & $35$ \\
    &     & 1.25 & 1.75 & $270$ & $\pm$ & $26$ \\
    &     & 1.75 & 2.25 & $189$ & $\pm$ & $19$ \\
    &     & 2.25 & 2.75 & $130$ & $\pm$ & $14$ \\
    &     & 2.75 & 3.25 & $ 87.8$ & $\pm$ & $ 7.1$ \\
    &     & 3.25 & 4.00 & $ 38.3$ & $\pm$ & $ 3.4$ \\
    &     & 4.00 & 5.00 & $ 16.6$ & $\pm$ & $ 1.7$ \\
    &     & 5.00 & 6.50 & $ 10.4$ & $\pm$ & $ 3.2$ \\ \hline
\end{tabular}
}
\caption{\label{tab:xsec_results}
HARP results for the double-differential $\pi^+$ production
 cross-section in the laboratory system,
 $d^2\sigma^{\pi^+}/(dpd\Omega)$. Each row refers to a 
 different $(p_{\hbox{\small min}} \le p<p_{\hbox{\small max}},
 \theta_{\hbox{\small min}} \le \theta<\theta_{\hbox{\small max}})$ bin,
 where $p$ and $\theta$ are the pion momentum and polar angle, respectively.
 The central values quoted are the ones obtained via the Atlantic
 analysis discussed in the text. The square-root of the diagonal elements
 of the covariance matrix are also given.
}
\end{table}

\subsection{Error estimates}
\label{subsec:errors}

A detailed error analysis has been performed to evaluate the accuracy
of the pion cross-section measurement.
 The main errors entering in this measurement are listed below.

First, 
the statistical uncertainties
 associated with the track yields measured from the aluminium target
 setting and from the empty target setting (needed for subtraction, as
 explained above) have been included in the pion production cross-section 
 uncertainty estimates.

Second, several uncertainties associated with the corrections needed
 to convert the measured track yields to true track yields have been
 evaluated. The track reconstruction efficiency correction is
based on the combination of thin target aluminium and beryllium data
sets. The main error associated with this computation is given by the
size of the statistical sample.  The correction to the pion and proton
 yields due to absorption or decay is computed via a Monte Carlo
 simulation. An uncertainty of 10\% for both proton and pion yields has been
 assumed for this correction, in addition to the uncertainty due to the finite size
 of the simulated data sample used to estimate this correction.
 Similarly, simulated data (and their
associated uncertainties) were used to estimate the correction
 for the contamination in
the sample due to tertiary particles that are not produced in the
target, but rather by the decay of secondaries, or by the interaction
of secondaries in the spectrometer material. An uncertainty of 100\%
 has been assumed for this subtraction, for both proton and pion
 yields. Furthermore, an uncertainty has been assigned to the
 empty target subtraction, in order to account for
 the effect of the target itself which attenuates the proton beam.

Third, uncertainties associated with the particle
 identification of tracks, and with the corrections needed to convert
 yields of tracks identified as pions to true pion yields, have been included.
Among the several error sources
 associated with the pion--proton PID selection, the dominant one is due
 to the uncertainty in the (small) fraction
 of pions and protons with an associated anomalous TOFW
 $\beta$ measurement, that is a $\beta$ measurement which exhibits a
 non-Gaussian behaviour. 
Estimates of the uncertainty in the kaon contamination and in the
correction for the electron veto have been obtained from an analysis
of the data as explained in Section~\ref{sec:pid}.
The robustness of the pion PID selection
and its associated correction has been evaluated by performing the analysis
with tighter and looser PID probability cuts with
respect to their nominal values, while correcting for the PID efficiency
and migration corresponding to the probability cuts.

Fourth, we have included uncertainties associated with the momentum
reconstruction performance of the spectrometer, and with the
corrections needed to convert the measured momenta to `true'
momenta. 
Concerning the momentum, biases
and resolution effects are taken into account using both real and
simulated data. 
It was found that momentum biases do not
exceed the 5\% level from a study of beam particles at different
momenta and from a comparison between the reconstructed momenta and
the momenta inferred from $\beta$ measurements 
with the TOFW and the threshold curves in the Cherenkov.

Finally, an overall normalization uncertainty of 4\% has been
estimated. The dominant sources for this uncertainty are the
targeting efficiency uncertainty, which is deduced from the measurement of
transverse beam spot size on target, as well as the reconstruction and
PID uncertainties that are fully correlated across different
$(p,\theta )$ pion bins, and which are not included in the above evaluation. On the
other hand, the aluminium target thickness and density were carefully
measured, and the effect on the overall cross-section
normalization due to these uncertainties is negligible.

\subsection{Results of the Error Evaluation}
\label{subsec:results-errors}
The impact of the error sources discussed in the previous section on
 the final cross-section measurement has been evaluated, either by
 analytic error propagation, or by Monte Carlo techniques. Correlation effects
 among different particle types, and among different $(p,\theta )$ bins,
 have also been taken into account. 

  The cross-section uncertainty level is quantified 
 by adopting two different conventions. The rationale is that both
 the errors on the `point--to--point', double-differential cross-section,
 and the error on the cross-section integrated over the entire pion phase space
 measured, might be of interest.

 First, the dimensionless quantity $\delta_{\hbox{\small {{diff}}}}$
 is  defined, expressing the typical error on the double-differential
 cross-section, as follows:
\begin{equation}
\delta_{\hbox{\small {diff}}}\equiv 
\frac{\sum_i (\delta [\Delta^2\sigma^{\pi}/(\Delta p\Delta\Omega)])_i}{
\sum_i (\Delta^2\sigma^{\pi}/(\Delta p\Delta\Omega))_i} \ ,
\label{eq:deltadiff}
\end{equation}
where $i$ labels a given pion $(p,\theta)$ bin,
 $(\Delta^2\sigma^{\pi}/(dp\cdot d\Omega))_i$ is the central value for the
  double-differential cross-section measurement in that bin,
 and $(\delta [\Delta^2\sigma^{\pi}/(dp\cdot d\Omega)])_i$ is the error
 associated with this measurement. 
\begin{table}[tb]
\centerline{
\begin{tabular}{ c c c c } \hline
{\bf Error Category} & {\bf Error Source}       & $\delta_{\hbox{\small diff}}$ (\%) & $\delta_{\hbox{\small int}}$ (\%) \\ \hline
Statistical    & Al target statistics              & 1.6                      &    0.3                               \\
               & Empty target subtraction (stat.)  & 1.3                      &    0.2                     \\ 
               & {\bf Sub-total}                   & {\bf 2.1}                &    {\bf 0.4}               \\ \hline
Track yield corrections & Reconstruction efficiency         & 0.8                     &    0.4          \\
                        & Pion, proton absorption           & 2.4                     &    2.6   \\
                        & Tertiary subtraction              & 3.2                     &    2.9         \\ 
                        & Empty target subtraction (syst.)  & 1.2                     &    1.1         \\
                        & {\bf Sub-total}                   & {\bf 4.5}               &    {\bf 4.1}      \\ \hline
Particle identification & PID Probability cut               & 0.2                                &    0.2     \\ 
                        & Kaon subtraction                  & 0.3                                &    0.1     \\
                        & Electron veto                     & 2.1                                &    0.5     \\
                        & Pion, proton ID correction        & 2.5                                &    0.4     \\ 
                        & {\bf Sub-total}                   & {\bf 3.5}                          &    {\bf 0.7}     \\ \hline
Momentum reconstruction & Momentum scale                    & 3.0                                &    0.3   \\
                        & Momentum resolution               & 0.6                                &    0.6   \\
                        & {\bf Sub-total}                   & {\bf 3.2}                          &    {\bf 0.7}   \\ \hline
Overall normalization   & {\bf Sub-total}                   & {\bf 4.0}                          &    {\bf 4.0}   \\ \hline
All                     & {\bf Total}                       & {\bf 8.2}                          & {\bf 5.8} 
\end{tabular}
}
\caption{\label{tab:errorsummary_delta}
Summary of the uncertainties affecting the double-differential
 cross-section ($\delta_{\hbox{\small {{diff}}}}$) and
 integrated cross-section ($\delta_{\hbox{\small {{int}}}}$) measurements.
 See text for details.
}
\end{table}

\begin{figure}[tbp!]
\centerline{
\includegraphics[width=0.8\textwidth]{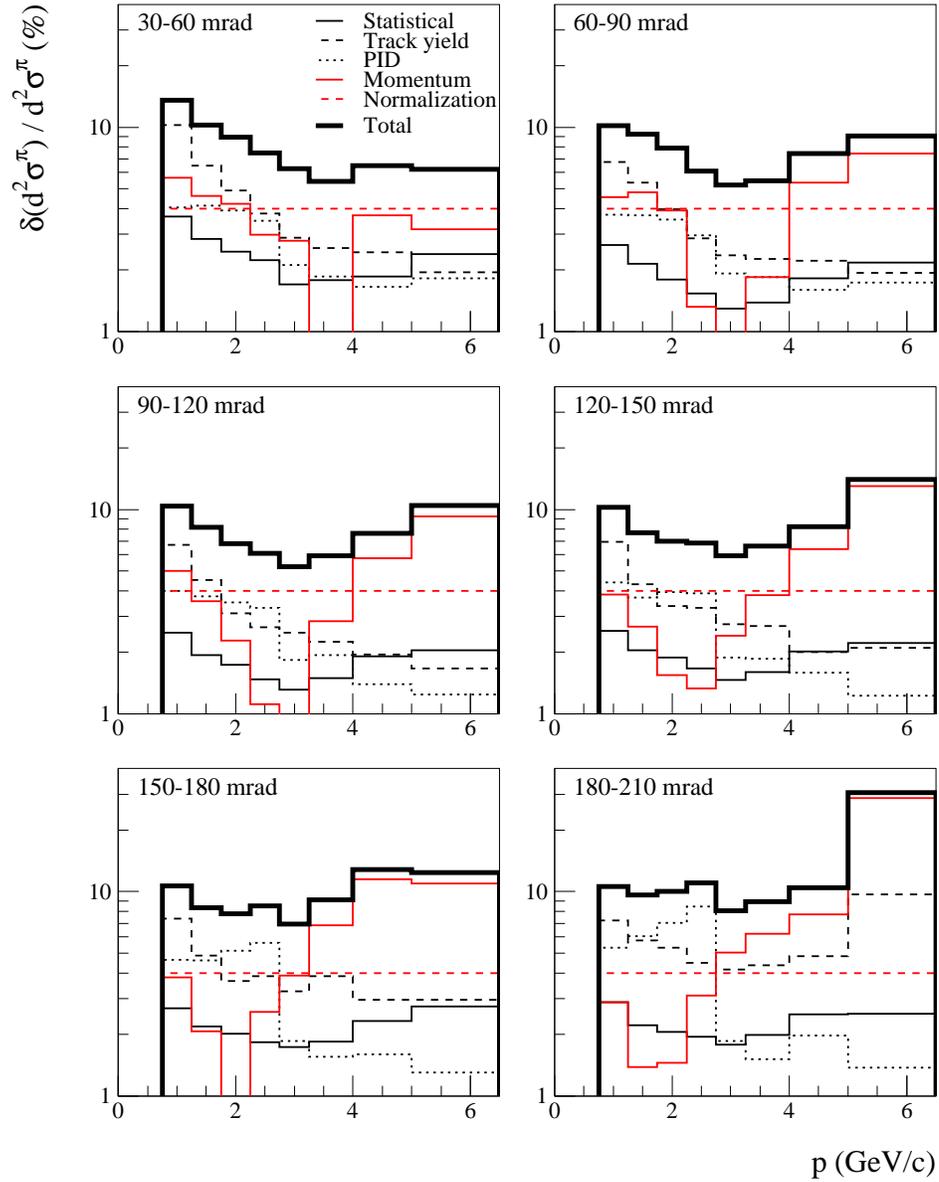}
}
\caption{\label{fig:errorsummary_rel}
Estimate of the fractional errors on the double-differential pion production
 cross-section measured as a function of pion
 momentum $p$ and polar angle. The errors shown are in percent.
 The contributions from the error categories in
 Tab.~\ref{tab:errorsummary_delta} (thin histograms) as well as the
 total error (thick solid) are shown. 
}
\end{figure}

 The individual and cumulative effect of the error sources discussed
 above on the $\delta_{\hbox{\small {{diff}}}}$ quantity are shown in
 Table~\ref{tab:errorsummary_delta}.
 The typical error on the double-differential cross-section is about
 8.2\%. 
The dominant error contributions to $\delta_{\hbox{\small {{diff}}}}$ arise
 from overall normalization (4\%), subtraction of tertiary tracks
 (3.2\%), and momentum scale
 (3.0\%). More details on the relative double-differential
 cross-section  uncertainties are shown in Fig.~\ref{fig:errorsummary_rel}
 for all measured $(p,\theta )$ bins. In Figure \ref{fig:errorsummary_rel}
 and in Tab.~\ref{tab:errorsummary_delta}, the individual cross-section
 uncertainties are grouped into five categories: statistical, track yield
 corrections, particle identification, momentum reconstruction, and overall
 normalization uncertainties. Uncertainties associated with the track yield
 corrections discussed above dominate the cross-section uncertainties in the
 low momentum region, while the dominant errors in the high momentum
 region are due to the momentum reconstruction and to the overall
 normalization.  

 Second, we define the dimensionless quantity
 $\delta_{\hbox{\small {{int}}}}$,
 expressing the fractional error on the integrated pion cross-section,
 $\sigma^{\pi}(0.75 \ \GeVc \le p<6.5 \ \GeVc$, $30 \ \mrad \le \theta<210 \ \mrad)$,
 as follows:
\begin{equation}
\delta_{\hbox{\small {{int}}}}\equiv 
\frac{
\sqrt{\sum_{i,j}(\Delta p\Delta\Omega)_i
 C_{ij}
 (\Delta p\Delta\Omega)_j
}}{
\sum_i (\Delta^2\sigma^{\pi})_i} \ ,
\label{eq:deltaint}
\end{equation}
 where $(\Delta^2\sigma^{\pi})_i$ is the double-differential
 cross-section in bin $i$, $(\Delta^2\sigma^{\pi}/(\Delta p\Delta\Omega))_i$,
 multiplied by its corresponding phase space element
 $(\Delta p\Delta\Omega)_i$. Here, $C_{ij}$ is the covariance matrix
 of the double-differential cross-section obtained by summing
 thirteen matrices from the error sources listed in
 Table~\ref{tab:errorsummary_delta}, and whose square root of the
 diagonal elements, $\sqrt{C_{ii}}$, corresponds to the error
 $(\delta (\Delta^2\sigma^{\pi}/(\Delta p\Delta\Omega)))_i$ appearing in
 Eq.~\ref{eq:deltadiff}. This covariance matrix is used to
 compare the two independent analyses of the same cross-section
 measurement, and to obtain the best-fit values, errors, and
 correlations for the coefficients entering into the Sanford-Wang
 formula used to parametrize the HARP measurements. The correlation
 coefficients among distinct $(p,\theta)$ bins in
 $C_{ij}$ vary between $-0.19$ and $+0.95$.

 The contributions to $\delta_{\hbox{\small {{int}}}}$ from all the
 error sources considered, as well as the total error estimate
 on the integrated cross-section, are also given
 in Table~\ref{tab:errorsummary_delta}. As expected, (mostly) correlated errors
 such as the one from the normalization or tertiary subtraction remain
 (almost) as large
 as they were for the point-to-point error. On the other hand, the
 contribution of the momentum scale uncertainty is negligible here,
 since its effect tends to be anti-correlated among 
 different phase space bins. In addition to the normalization and tertiary
 subtraction,
 other uncertainty sources which have some
 impact on the integrated cross-section include the pion absorption
 correction and the empty target subtraction systematic uncertainty.
 Overall,
 the total uncertainty on the pion production cross-section measured
 over the entire phase space ($0.75 \le p<6.5$~\GeVc, $30 \le \theta<210$~mrad)
 is estimated to be about 6\%.

 In the following section, the cross-section results
 are also expressed in a parametrized form. 


\subsection{Sanford-Wang parametrization}
\label{subsec:sw-param}

The $\pi^+$ production data was fitted with a Sanford-Wang
parametrization~\cite{ref:sanford-wang}, which has the functional form: 
\begin{equation}
\label{eq:swformula}
\frac{d^2\sigma (\hbox{p+A}\rightarrow \pi^++X)}{dpd\Omega}(p,\theta) =
 c_{1} p^{c_{2}}(1-\frac{p}{p_{\hbox{\footnotesize beam}}})
 \exp [-c_{3}\frac{p^{c_{4}}}{p_{\hbox{\footnotesize beam}}^{c_{5}}}-c_{6}
 \theta (p-c_{7}
 p_{\hbox{\footnotesize {beam}}} \cos^{c_{8}}\theta )] \ ,
\end{equation} 
 where $X$ denotes any system of other particles in the final state,
 $p_{\hbox{\footnotesize {beam}}}$ is
 the proton beam momentum in \GeVc, $p$ and $\theta$ are the $\pi^+$
 momentum and angle in units of \GeVc and radians, respectively, 
 $d^2\sigma/(dpd\Omega)$ is expressed in units of mb/(\GeVc\ sr),
 $d\Omega\equiv 2\pi\ d(\cos\theta )$, 
 and the parameters $c_1,\ldots ,c_8$
 are obtained from fits to $\pi^+$ production data. 

 The parameter $c_1$ is an overall normalization factor,
 the four parameters $c_2,c_3,c_4,c_5$ can be interpreted as
 describing the momentum distribution of the secondary
 pions, and the three parameters $c_6,c_7,c_8$ as describing the
 angular distribution for fixed secondary and proton beam momenta,
 $p$ and $p_{\hbox{\footnotesize {beam}}}$. This formula is purely
 empirical. In the $\chi^2$ minimization procedure,
 seven out of these eight parameters were allowed to vary.
 The parameter $c_5$ was fixed to the conventional
 value $c_5\equiv c_4$, since the cross-section dependence on the
 proton beam momentum cannot be addressed by the present HARP data-set,
 which includes exclusively measurements taken at
 $p_{\hbox{\footnotesize {{beam}}}}=12.9$~\GeVc. In the $\chi^2$ minimization,
 the full error matrix was used. 

 Concerning the Sanford-Wang parameters estimation, the best-fit values of
 the Sanford-Wang parameters are reported in Table~\ref{tab:swpar_values_errors},
 together with their errors.
 The fit parameter errors are estimated
 by requiring $\Delta\chi^2\equiv
 \chi^2-\chi^2_{\hbox{\footnotesize {{min}}}}$=8.18,
 corresponding to the 68.27\% confidence level region for seven variable
 parameters. Significant correlations among fit parameters are found, as
 shown by the correlation matrix given in
 Table~\ref{tab:swpar_correlations}.

\begin{table}[tb]
\centerline{
\begin{tabular}{ c c } \hline
{\bf Parameter} & {\bf Value} \\ \hline
$c_1$      & $(4.4\pm 1.3)\cdot 10^2$ \\
$c_2$      & $(8.5\pm 3.4)\cdot 10^{-1}$ \\
$c_3$      & $(5.1\pm 1.3)$ \\
$c_4=c_5$  & $(1.78\pm 0.75)$ \\
$c_6$      & $(4.43\pm 0.31)$ \\
$c_7$      & $(1.35\pm 0.29)\cdot 10^{-1}$ \\
$c_8$      & $(3.57\pm 0.96)\cdot 10^1$ \\
\end{tabular}
}
\caption{\label{tab:swpar_values_errors}
Sanford-Wang parameters and errors obtained by fitting the
  dataset. The errors refer to the 68.27\% confidence
 level for seven parameters ($\Delta\chi^2=8.18$).
}
\end{table}

\begin{table}[tb]
\centerline{
\begin{tabular}{ c r r r r r r r }
\hline
{\bf Parameter} & $c_1$  & $c_2$  & $c_3$  & $c_4=c_5$  & $c_6$  & $c_7$  & $c_8$  \\
\hline
$c_1$           &  1.000 &        &        &            &        &        &        \\
$c_2$           & -0.056 &  1.000 &        &            &        &        &        \\
$c_3$           & -0.145 & -0.691 &  1.000 &            &        &        &        \\
$c_4=c_5$       & -0.322 & -0.890 &  0.831 &  1.000     &        &        &        \\
$c_6$           & -0.347 &  0.263 & -0.252 & -0.067     &  1.000 &        &        \\
$c_7$           & -0.740 &  0.148 & -0.067 &  0.077     &  0.326 &  1.000 &        \\
$c_8$           &  0.130 & -0.044 &  0.205 & -0.040     & -0.650 &  0.189 &  1.000 \\
\end{tabular}
}
\caption{\label{tab:swpar_correlations}
Correlation coefficients among the Sanford-Wang parameters, obtained
 by fitting the data.
}
\end{table}

\begin{figure}[tbp!]
\begin{center}
\includegraphics[width=0.8\textwidth]{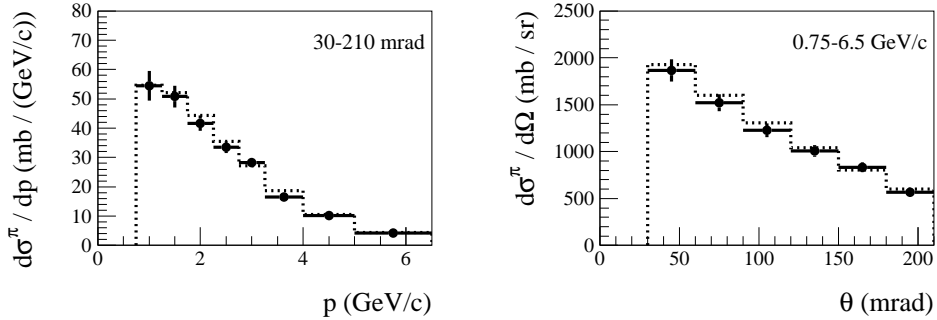}
\caption{\label{fig:harpswfit_1d} 
  Projections of the differential
  light hadron production cross-section as a function of $p$
  integrated over the range  $30 \le \theta < 210$~mrad
  (left panel), and production cross-section as a function of $\theta$
  in the range $0.75 \le p < 6.5$~\GeVc (right panel).
 The points show the HARP measurements, the dotted curve the best-fit
 Sanford-Wang parametrization. 
}
\end{center}
\end{figure}

 The HARP cross-section measurement is compared to the best-fit
 Sanford-Wang parametrization of Table~\ref{tab:swpar_values_errors} in
 Figs.~\ref{fig:harpswfit_2d} and~\ref{fig:harpswfit_1d}. 

The goodness-of-fit of the Sanford-Wang parametrization
hypothesis for the HARP results can be assessed by considering the best-fit $\chi^2$ value of
 $\chi^2_{\hbox{\footnotesize {{min}}}}$=305 for 41 degrees of freedom,
 indicating a very poor fit quality. We note that the goodness-of-fit strongly
 depends on the correlations among the HARP cross-section uncertainties
 in different $(p,\theta)$ bins, and therefore cannot be inferred
 from Fig.~\ref{fig:harpswfit_2d} alone. If these uncertainties were
 (incorrectly) treated as completely uncorrelated, the best-fit $\chi^2$ value
 would decrease from 305 to 57. A more comprehensive study of $\pi^+$
 production at various beam momenta and from various nuclear targets in HARP
 is planned and will follow in a subsequent publication, and should hopefully
 shed more light on the cause of the
 poor quality of the Sanford-Wang hypothesis reported here.

\section{Comparison with existing forward pion production data
 on aluminium}
\label{sec:compare}
Finally the HARP results are compared with existing $\pi^+$ production
data available in the literature directly from aluminium 
targets~\cite{ref:sugaya98,ref:vorontsov88,ref:vorontsov83,ref:abbott92}. The
comparison is restricted to proton beam momenta between 10 and 15~\GeVc 
(close to the K2K beam momentum of 12.9~\GeVc), and for pion
polar angles below 200 mrad (the range measured by HARP and of
relevance to K2K).

The comparison is based on the HARP Sanford-Wang parametrization
rather than on the HARP data points themselves, in order to match pion
momenta and angles measured in past Al experiments. Furthermore, a
correction to rescale the HARP Sanford-Wang parametrization at 12.9~\GeVc 
beam momentum to the 10--15~\GeVc beam momenta of the past Al
datasets is applied~\cite{ref:cho}.

\begin{figure}[tbp!]
\centerline{
\includegraphics[width=14.cm]{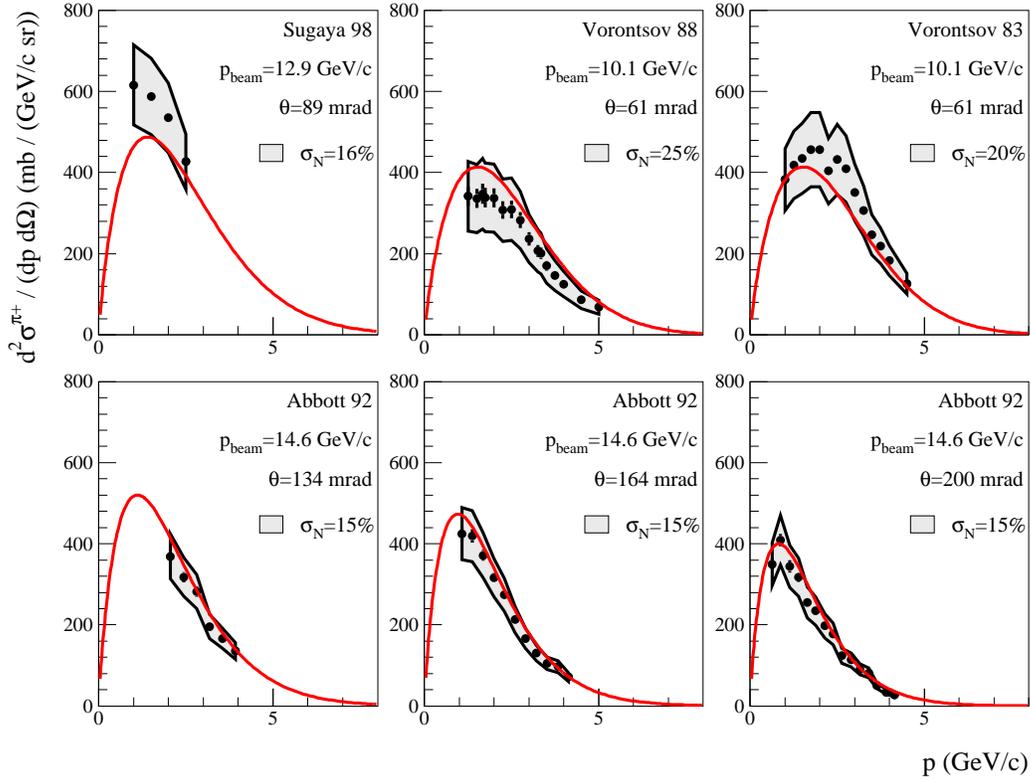}
}
\caption{\label{fig:comparisonwithaldata}
Comparison of the double-differential pion production cross-section
measured in HARP, and the one measured in past experiments using an
aluminium target and 10-15~\GeVc momentum beam protons.  The points
are the data from past experiments, and the shaded area reflect
their normalization uncertainty. The solid line is the 
HARP Sanford-Wang parametrization
rescaled to the beam momentum of past experiments, as discussed in the
text.  }
\end{figure}

Given these model-dependent corrections, it was found that the HARP results
 are consistent
 with Ref.~\cite{ref:vorontsov88} and
Refs.~\cite{ref:vorontsov83}, agree rather well with~\cite{ref:abbott92}
 and are somewhat lower than, but still marginally consistent with,
 Ref.~\cite{ref:sugaya98}. 
Figure~\ref{fig:comparisonwithaldata} shows the comparison between
HARP and the above datasets.

\section{HARP results as input to the K2K far--to--near neutrino
 flux ratio prediction}
\label{sec:far-near}

The main application of the measurement presented in this paper, the
double-differential $\pi^+$ production cross-section,
$d^2\sigma\ (p+\hbox{Al}\to\pi^+ +X)/(dpd\Omega )$, is to predict the
far--near ratio, $R$, for the muon neutrino disappearance search in the K2K
experiment.  

 As discussed in Sec.~\ref{sec:intro},
the determination of the far--near ratio is the leading
energy-dependent systematic error in the K2K
analysis~\cite{k2k,k2k-last}. To compute this quantity
a Monte Carlo program simulating all relevant beam-line
geometry and materials, and all relevant physics processes, is used.
In this simulation, the neutrino flux prediction
 uncertainty is dominated by the uncertainties in the forward $\pi^+$
 production arising from the interactions of the 
12.9~\GeVc protons in the aluminium
 target material. 
Therefore, it is instructive to recompute a
 prediction for the K2K far--to--near flux ratio prediction based on the
 new experimental information presented in this paper. 
The HARP-based prediction has been obtained by
 substituting the original
 $\pi^+$ production cross-section assumed in the K2K beam Monte Carlo hadronic
 model with the HARP Sanford-Wang parametrization discussed in Section
 \ref{sec:results}, while keeping unchanged all other ingredients 
of the K2K beam
 MC simulation, such as primary beam optics, pion re-interactions in the 
 aluminium target, pion focusing, pion decay, etc. More details on the default
 K2K beam MC assumptions can be found in Ref. \cite{k2k}.

\begin{figure}[htbp]
  \begin{center}
  \epsfig{file=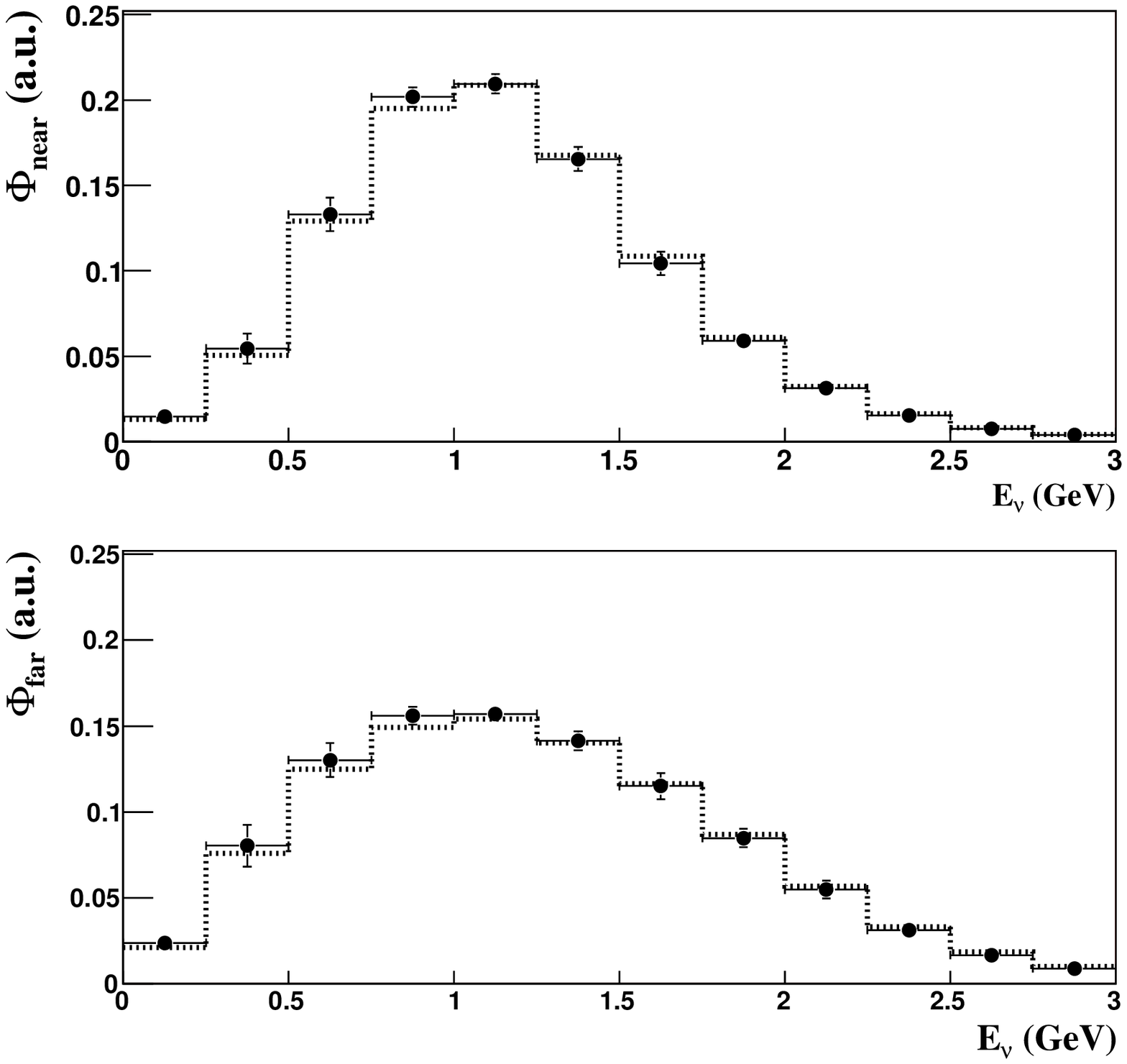,width=7.9cm} 
  \epsfig{file=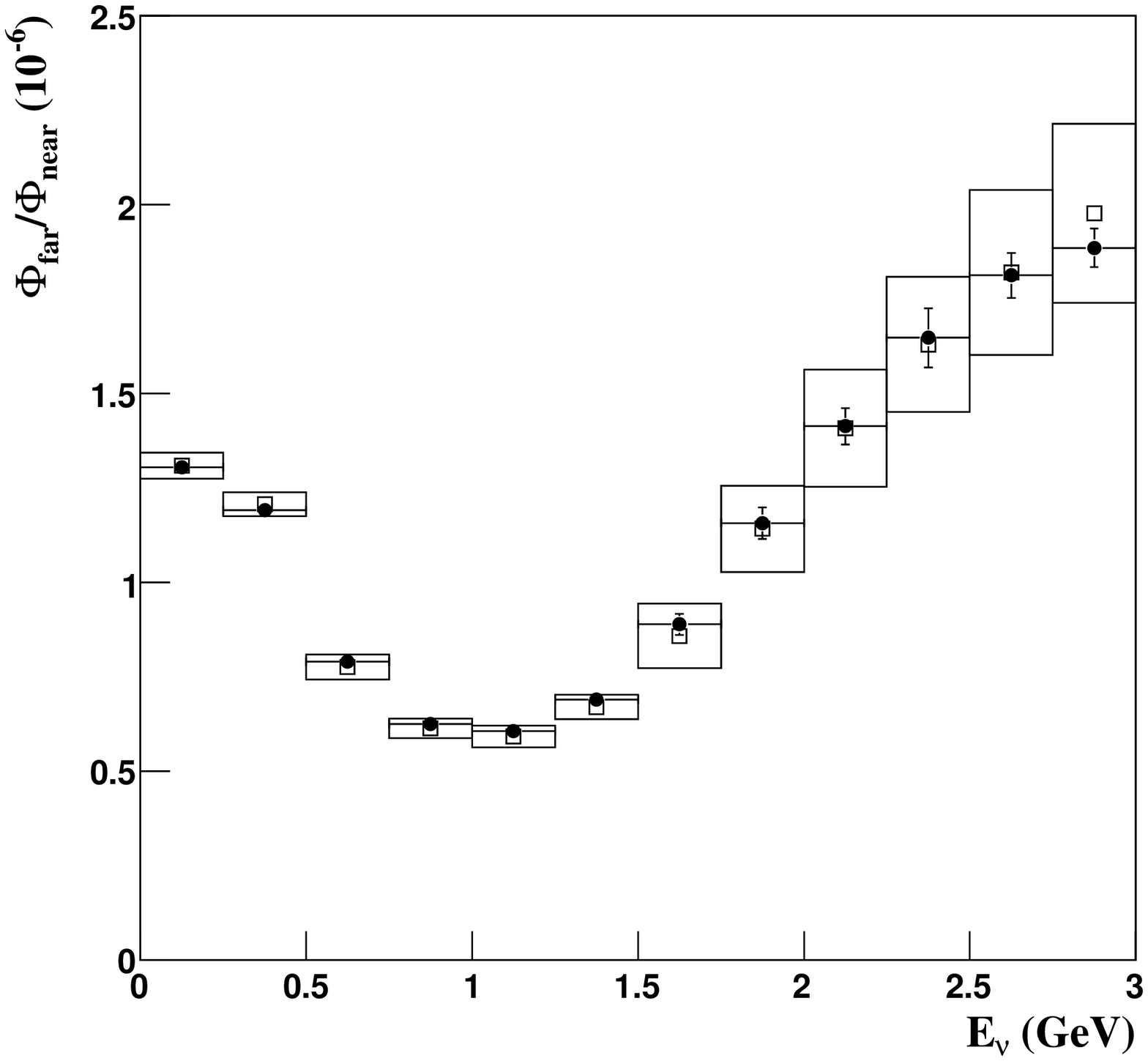,width=7.9cm}  

\caption{\label{fig:k2kfluxpredictions}
Muon neutrino fluxes in the K2K experiment as a function of
 neutrino energy $E_{\nu}$, as predicted by the default hadronic model
 assumptions in the K2K beam Monte Carlo simulation (dotted histograms),
 and by the HARP $\pi^{+}$ production measurement (full circles with
 error bars).
 Left panel shows unit-area normalized flux predictions at the
 K2K near (top) and far (bottom) detector locations, $\Phi_{\hbox{near}}$ and
 $\Phi_{\hbox{far}}$, respectively, while right panel shows
 the far--to--near flux ratio $\Phi_{\hbox{far}}/\Phi_{\hbox{near}}$
 (open squares with error boxes show K2K model results). 
}
  \end{center}
\end{figure}

The result of this exercise is shown in Fig.~\ref{fig:k2kfluxpredictions}.
The left panel shows 
muon neutrino fluxes in the K2K experiment as a function of
neutrino energy $E_{\nu}$, as predicted by the default hadronic model
assumptions in the K2K beam Monte Carlo simulation (dotted histograms),
and by the HARP $\pi^{+}$ production measurement (filled circles).
The plots on that panel show unit-area normalized flux predictions at the
K2K near (top) and far (bottom) detector locations, $\Phi_{\hbox{near}}$ and
$\Phi_{\hbox{far}}$, respectively. 
Right panel shows
the far--to--near flux ratio $\Phi_{\hbox{far}}/\Phi_{\hbox{near}}$
obtained from K2K Monte Carlo (empty squares with error boxes) and
from HARP measurament (filled dots with error bars).  
The fluxes predicted by HARP and the present
K2K model are in good agreement within the errors.
This is reflected also in a good agreement in $R$, in particular in the
oscillation region (below 1.5~\GeV). Finally, it is worth noting that
the error on $R$ associated with the HARP measurement (including
statistical and systematic errors) 
is of the order of 1\%, since most errors on the cross-section cancel
in the ratio. The current systematic error attached to $R$ in the K2K analysis
is of the order of 7\%. Thus, although the result presented here does not
yet represent a new measurement of $R$ (which requires a full evaluation 
of other systematic errors independent of the HARP measurement but
associated with the K2K beam line setup), it clearly shows the
considerable improvement that can be achieved by K2K by using this new
measurement. 
In addition, the data taken with the replica of the K2K target will be
valuable to study the effect of reinteractions which
needs to be taken into account in the beam simulation.

\section{Summary and conclusions}
\label{sec:summary}

In this paper we present a measurement of the double-differential production
cross-section 
${{d^2 \sigma^{\pi^+} }}/{{d p d\Omega }}$,
for positively charged pions.
The incident particles are protons of 
12.9~\GeVc momentum hitting a thin aluminium target of 5\%~$\lambda_{\mathrm{I}}$.
The measurement of this cross-section has a direct application
to the calculation of the neutrino flux of the K2K experiment.
The data were taken in 2002 in the T9
beam of the CERN PS.
Out of 4.7 million triggers processed, 3.4 million incoming
protons were selected.
After cuts, around 210~000 secondary tracks reconstructed in the forward
spectrometer were used in this analysis.
These high statistics results were corrected for measurement
resolutions.
These data were fitted with a Sanford-Wang parametrization.
The results are given for positively charged pions within a momentum range from
0.75~\GeVc to 6.5~\GeVc, and within an angular range from 30~\mrad to
210~\mrad. 
The average statistical error is 1.6\% per point.
The absolute normalization was performed using prescaled beam
triggers.
The overall efficiency for track reconstruction and particle
identification is known to better than 6\%, while the average
point-to-point error is 8.2\%.

\section{Acknowledgements}

We gratefully acknowledge the help and support of the PS beam staff
and of the numerous technical collaborators who contributed to the
detector design, construction, commissioning and operation.  
In particular, we would like to thank
G.~Barichello,
R.~Broccard,
K.~Burin,
V.~Carassiti,
F.~Chignoli,
G.~Decreuse,
C.~Detraz,  
M.~Dwuznik,   
F.~Evangelisti,
B.~Friend,
A.~Iaciofano,
I.~Krasin, 
J.-C.~Legrand,
M.~Lobello, 
F.~Marinilli,
J.~Mulon,
R.~Nicholson,
A.~Pepato,
P.~Petev, 
X.~Pons,
I.~Rusinov,
M.~Scandurra,
E.~Usenko,
and
R.~van der Vlugt,
for their support in the construction of the detector.
The collaboration is indebted to 
M.~Baldo~Ceolin,
P.~Binko,
E.~Boter,
M.~Doucet,
D.~D\"{u}llmann,
V.~Ermilova, 
A.~Pullia and A.~Valassi
for their contributions to the experiment.

We are indebted to the K2K collaboration who made available their
beam-line simulation for the calculation of the predicted
far--to--near neutrino flux ratio.

 The experiment was made possible by grants from
the Institut Interuniversitaire des Sciences Nucl\'eair\-es and the
Interuniversitair Instituut voor Kernwetenschappen (Belgium), 
Ministerio de Educacion y Ciencia, Grant FPA2003-06921-c02-02 and
Generalitat Valenciana, grant GV00-054-1,
CERN (Geneva, Switzerland), 
the German Bundesministerium f\"ur Bildung und Forschung (Germany), 
the Istituto Na\-zio\-na\-le di Fisica Nucleare (Italy), 
INR RAS (Moscow) and the Particle Physics and Astronomy Research Council (UK).
We gratefully acknowledge their support.


\end{document}